%% file: ms.tex
\documentclass[twocolumn]{aastex62}

\newcommand{\airwave}{\lambda_{\rm air}}

\newcommand{\Teff}{{T}_{\rm eff}}
\newcommand{\kms}{{km\,s}$^{-1}$}
\newcommand{\FeH}{{\rm [Fe/H]}}
\newcommand{\vmac}{v_{\rm mac}}

\newcommand{\vinst}{v_{\rm inst}}
\newcommand{\vbroad}{v_{\rm b}}
\newcommand{\lambdac}{\lambda_{\rm c}}
\newcommand{\fsyn}{f_{\rm syn}}
\newcommand{\fobs}{f_{\rm obs}}
\newcommand{\fobsorg}{f_{\rm obs,0}}
\newcommand{\dsyn}{d_{\rm syn}}
\newcommand{\dsyndagger}{d_{\rm syn\dagger}}
\newcommand{\dobs}{d_{\rm obs}}

\newcommand{\dsynstardagger}{d^*_{\rm syn\dagger}}

\newcommand{\alphasyn}{\alpha_{\rm syn}}
\newcommand{\alphaobs}{\alpha_{\rm obs}}
\newcommand{\RGC}{R_{\rm GC}}

\usepackage{booktabs}
\usepackage{multirow}
\usepackage{soul}

\received{October 20, 2019}
\revised{November 25, 2019}
\accepted{November 25, 2019}
\submitjournal{ApJS}

\shorttitle{Identification of lines of heavy metals in $YJ$ bands}
\shortauthors{Matsunaga et al.}

\begin{document}

\title{Identification of Absorption Lines of Heavy Metals in the Wavelength Range 0.97--1.32\,{$\mu$}m}

\correspondingauthor{Noriyuki Matsunaga}
\email{matsunaga@astron.s.u-tokyo.ac.jp}

\author{Noriyuki Matsunaga}
\affil{Department of Astronomy, School of Science, The University of Tokyo,
7-3-1 Hongo, Bunkyo-ku, Tokyo 113-0033, Japan}
\affil{Laboratory of Infrared High-resolution spectroscopy (LiH), Koyama Astronomical Observatory, \\ Kyoto Sangyo University, Motoyama, Kamigamo, Kita-ku, Kyoto 603-8555, Japan}

\author{Daisuke Taniguchi}
\affil{Department of Astronomy, School of Science, The University of Tokyo,
7-3-1 Hongo, Bunkyo-ku, Tokyo 113-0033, Japan}

\author{Mingjie Jian}
\affil{Department of Astronomy, School of Science, The University of Tokyo,
7-3-1 Hongo, Bunkyo-ku, Tokyo 113-0033, Japan}

\author{Yuji Ikeda}
\affil{Photocoding, 460-102 Iwakura-Nakamachi, Sakyo-ku, Kyoto 606-0025, Japan}
\affil{Laboratory of Infrared High-resolution spectroscopy (LiH), Koyama Astronomical Observatory, \\ Kyoto Sangyo University, Motoyama, Kamigamo, Kita-ku, Kyoto 603-8555, Japan}

\author{Kei Fukue}
\affil{Laboratory of Infrared High-resolution spectroscopy (LiH), Koyama Astronomical Observatory, \\ Kyoto Sangyo University, Motoyama, Kamigamo, Kita-ku, Kyoto 603-8555, Japan}

\author{Sohei Kondo}
\affil{Kiso Observatory, Institute of Astronomy, School of Science, The University of Tokyo, 10762-30 Mitake, Kiso-machi, Kiso-gun, Nagano 397-0101, Japan}
\affil{Laboratory of Infrared High-resolution spectroscopy (LiH), Koyama Astronomical Observatory, \\ Kyoto Sangyo University, Motoyama, Kamigamo, Kita-ku, Kyoto 603-8555, Japan}

\author{Satoshi Hamano}
\affil{National Astronomical Observatory of Japan, 2-21-1 Osawa, Mitaka, Tokyo 181-8588, Japan}

\author{Hideyo Kawakita}
\affil{Laboratory of Infrared High-resolution spectroscopy (LiH), Koyama Astronomical Observatory, \\ Kyoto Sangyo University, Motoyama, Kamigamo, Kita-ku, Kyoto 603-8555, Japan}
\affil{Department of Physics, Faculty of Science, Kyoto Sangyo University, Motoyama, Kamigamo, Kita-ku, Kyoto 603-8555, Japan}

\author{Naoto Kobayashi}
\affil{Kiso Observatory, Institute of Astronomy, School of Science, The University of Tokyo, 10762-30 Mitake, Kiso-machi, Kiso-gun, Nagano 397-0101, Japan}
\affil{Institute of Astronomy, School of Science, The University of Tokyo, 2-21-1 Osawa, Mitaka, Tokyo 181-0015}
\affil{Laboratory of Infrared High-resolution spectroscopy (LiH), Koyama Astronomical Observatory, \\ Kyoto Sangyo University, Motoyama, Kamigamo, Kita-ku, Kyoto 603-8555, Japan}

\author{Shogo Otsubo}
\affil{Laboratory of Infrared High-resolution spectroscopy (LiH), Koyama Astronomical Observatory, \\ Kyoto Sangyo University, Motoyama, Kamigamo, Kita-ku, Kyoto 603-8555, Japan}

\author{Hiroaki Sameshima}
\affil{Institute of Astronomy, School of Science, The University of Tokyo, 2-21-1 Osawa, Mitaka, Tokyo 181-0015}

\author{Keiichi Takenaka}
\affil{Department of Physics, Faculty of Science, Kyoto Sangyo University, Motoyama, Kamigamo, Kita-ku, Kyoto 603-8555, Japan}

\author{Takuji Tsujimoto}
\affil{National Astronomical Observatory of Japan, 2-21-1 Osawa, Mitaka, Tokyo 181-8588, Japan}

\author{Ayaka Watase}
\affil{Department of Physics, Faculty of Science, Kyoto Sangyo University, Motoyama, Kamigamo, Kita-ku, Kyoto 603-8555, Japan}

\author{Chikako Yasui}
\affil{National Astronomical Observatory of Japan, 2-21-1 Osawa, Mitaka, Tokyo 181-8588, Japan}
\affil{Laboratory of Infrared High-resolution spectroscopy (LiH), Koyama Astronomical Observatory, \\ Kyoto Sangyo University, Motoyama, Kamigamo, Kita-ku, Kyoto 603-8555, Japan}

\author{Tomohiro Yoshikawa}
\affil{Edechs, 17203 Iwakura-Minami-Osagi-cho, Sakyo-ku, Kyoto 606-0003, Japan}



\begin{abstract}

Stellar absorption lines of heavy elements can give us various insights into
the chemical evolution of the Galaxy and nearby galaxies. 
Recently developed spectrographs for the near-infrared wavelengths
are becoming more and more powerful for producing a large number of
high-quality spectra, but identification and characterization of
the absorption lines in the infrared range remain to be fulfilled.
We searched for lines of the elements heavier than
the iron group, i.e., those heavier than Ni, in the $Y$ 
(9760--11100\,{\AA}) and $J$ (11600--13200\,{\AA}) bands.
We considered the lines in three catalogs, i.e.,
Vienna Atomic Line Database (VALD), the compilation by R.~Kurucz, and 
the list published in 1999 by Mel\'endez \& Barbuy. 
Candidate lines were selected based on synthetic spectra and
the confirmation was done by using WINERED spectra of
13 supergiants and giants within FGK spectral types (spanning
4000--7200\,K in the effective temperature).
We have detected lines of \ion{Zn}{1}, \ion{Sr}{2}, \ion{Y}{2}, \ion{Zr}{1},
\ion{Ba}{2}, \ion{Sm}{2}, \ion{Eu}{2}, and \ion{Dy}{2},
in the order of atomic number.
Although the number of the lines is small, 23 in total, they are
potentially useful diagnostic lines of the Galactic chemical evolution,
especially in the regions for which interstellar extinction hampers
detailed chemical analyses with spectra in shorter wavelengths.
We also report the detection of lines whose presence was not predicted 
by the synthetic spectra created with the above three line lists.

\end{abstract}

\keywords{line: identification --- techniques: spectroscopic --- stars: abundances --- supergiants --- infrared: stars}


\section{Introduction} \label{sec:intro}
The identification of stellar absorption lines in the near-infrared range is
not complete compared to the established lists of lines
in the optical \citep[see, e.g.,][]{Andreasen-2016}.
In this work, we focus on the elements with the atomic number $Z \geq 29$,
i.e., those heavier than the iron group elements of which the heaviest is Ni.
Those heavy elements are useful for studying the detailed chemical evolution of the Galaxy 
\citep{McWilliam-1997,Sneden-2008,DelgadoMena-2017},
although their absorption lines are rather limited.
The so-called neutron-capture ($n$-capture) elements usually stand for
the elements with $Z\geq 31$ \citep[e.g.,][]{Sneden-2008}, and 
they are mostly synthesized 
in the presence of excessive neutrons.
Depending on the density of neutrons
and the resultant time scale of the $n$-capture process, compared with 
the time scale of the $\beta$ decay, the $n$-capture nucleosynthesis has
two main modes,
$s$-process (slow) and $r$-process (rapid),
and they occur at different astronomical sites.
In contrast, Cu ($Z=29$) and Zn ($Z=30$) are often included in the ``iron peak'' elements
and not in the $n$-capture elements; the origins of these two elements are, however,
considered to be multiplex and complicated \citep{DelgadoMena-2017}.
We will further discuss, in the Discussion (Section~\ref{sec:implication}),
the groups of the heavy elements
and the insights that can be yielded by the elements
with absorption lines confirmed in this study.

We here investigate the absorption lines that appear in near-infrared
spectra of FGK-type stars, in particular, supergiants.
We have two main reasons for considering supergiants:
(i)~many absorption lines are expected to be strong in stars with low surface gravity
as we see below, and (ii)~their high luminosities are advantageous 
as stellar tracers of the Galaxy and nearby galaxies.
In particular, Cepheid variable stars
are supergiants within the range of FGK types, aged at 10--300\,Myr, and they are useful for studying
the Galactic disk because their distances and ages can be accurately estimated
\citep[see, e.g., the review by][]{Matsunaga-2018}.
For example, \cite{daSilva-2016} investigated abundances of
five $n$-capture elements (Y, La, Ce, Nd, and Eu)
in hundreds of Cepheids, of which 73 were observed by the authors themselves.
They used optical high-resolution spectra for measuring
six \ion{Y}{2}, six \ion{La}{2}, three \ion{Ce}{2}, six \ion{Nd}{2},
and two \ion{Eu}{2} lines which are located between 4500
and 8000\,{\AA} \citep[see also][]{Lemasle-2013}.
In addition, measurements of the heavy elements in Cepheids and supergiants,
based on the line list of \citet{Kovtyukh-1999},
are found in a series of papers by Luck
and collaborators \citep[][and references therein]{Luck-2011a,Luck-2011b,Luck-2018}.
The number of the lines used in the previous studies is not so large, 
and yet they provide us with valuable information on 
the chemical enrichment of the Galactic disk. 
Identification and characterization are less advanced for absorption lines in the infrared range,
but many efforts have gradually increased the lines available for measuring stellar abundances.

Mel\'endez \& Barbuy (\citeyear{Melendez-1999}; hereinafter referred to as MB99)
used the solar spectrum \citep[Kitt Peak Solar Atlas;][and references therein]{Wallace-1996}
to compile the oscillator strengths, {$\log gf$}, and
other parameters of the 978 lines that they identified
between 10000 and 13400\,{\AA}, a part of the $YJ$ bands,
in addition to the 1240 lines in the $H$ band. We use this MB99 line list
as a starting point of our line selection in addition to 
Vienna Atomic Line Database \citep[VALD;][]{Ryabchikova-2015}
and the compilation by R.~Kurucz (KURZ; updated at
his web site\footnote{http://kurucz.harvard.edu/linelists/gfnew/}).
The latter two give more general compilations of absorption lines
including those purely based on theoretical calculations.

In the recent decade, high-resolution spectrographs 
covering infrared wavelengths have been used to explore 
absorption lines including weak ones.
For example, dozens of absorption lines of the heavy elements
($Z\geq 29$) have been found
\citep[e.g.,][]{Hasselquist-2016,Cunha-2017,Afsar-2018,BocekTopcu-2019,Chojnowski-2019} 
with high-quality spectra from the APOGEE survey \citep{Majewski-2017}
and IGRINS \citep{Park-2014}, although they do not cover the $YJ$ bands
of our interest.
In contrast, GIANO \citep{Oliva-2012} covers the $YJ$ bands
in addition to $H$ and $K$.
\citet{Origlia-2013,Origlia-2016} measured Sr abundances 
using one or two lines with GIANO $Y$-band spectra.
These two studies by Origlia {et~al.} targeted red supergiants in stellar clusters. 
In contrast, \citet{Caffau-2016} investigated 
GIANO spectra of FG-type dwarfs and
found three \ion{Sr}{1} lines in the $Y$ band.
Some warmer stars provide us with
unique data useful for identifying
various absorption lines.
\citet{Hubrig-2012}
reported one \ion{Sr}{2} line and
one \ion{Dy}{2} line in the $Y$ band
in one or more
peculiar A-type stars by
investigating the spectra from CRIRES \citep{Kaeufl-2004}.

The studies mentioned in the preceding paragraph mark important progress
in the identification and characterization of absorption lines
in the near infrared, but the current list of confirmed lines is clearly
incomplete. 
In this paper, we investigate the lines of the heavy elements
with $Z\geq 29$ seen in
the observed spectra, covering the $YJ$ bands, collected with
the near-infrared spectrograph, WINERED.\footnote{http://merlot.kyoto-su.ac.jp/LIH/WINERED/}
It is a high-resolution echelle spectrograph 
covering the wavelength range of 0.90--1.35\,{$\mu$}m
with the resolving power of 28000 or higher \citep{Ikeda-2016,Otsubo-2016}.
We use WINERED spectra of supergiants and giants
with various effective temperatures, $\Teff$, between 4000 and 7200\,K. 
The dependency on the temperature varies from line to line,
and the set of the WINERED spectra enables us to check whether
the absorption at the wavelength of 
each candidate line depends on $\Teff$ as expected for the particular line.
The purpose of this study is to confirm the absorption lines of the heavy elements that
are seen in the $YJ$-band spectra of FGK-type stars.

\section{Spectral data} \label{sec:data}
\subsection{Observational spectra} \label{sec:data-obs}
We investigate $YJ$-band spectra of 13 stars, supergiants and giants.
Their spectra were obtained in 2015 and 2016 
with WINERED attached to the 1.3\,m Araki Telescope at
Koyama Astronomical Observatory in Kyoto, Japan.
The spectral resolution of WINERED with the setting of
the WIDE mode and the 100\,{$\mu$m} slit is around 28000 \citep{Ikeda-2016}. 
We observed dozens of supergiants, giants, and dwarfs
whose temperatures were derived making use of
the line-depth ratios by \citet{Kovtyukh-2003,Kovtyukh-2006}
and \citet{Kovtyukh-2007}.
The entire set of the spectra for these objects will be considered
for discussing the line-depth ratios (Jian et~al., in preparation),
but a part of the spectra of supergiants and giants are used
in this study (Table~\ref{tab1}).
We use the stellar parameters, including $\Teff$,
taken from \citet{Luck-2014} and \citet{Hekker-2007}
because they obtained the four necessary parameters,
i.e., $\Teff$, $\log g$, $\FeH$, and the microturbulence ($\xi$), altogether.
Their measurements are based on high-resolution spectra ($R\geq 40000$),
and the obtained parameters are precise enough for our purpose;
in fact, the spectra synthesized with the given parameters reproduce
the observed spectra at around the 1\,\% level as we see below.
Figure~\ref{fig1} presents
the distribution of the 13 objects on the $(\Teff, \log g)$ plane.

All the observed spectra were reduced by following the standard procedure adopted
in the WINERED pipeline (Hamano et~al., in preparation)
that was established using PyRAF,\footnote{PyRAF is a product of the Space Telescope Science Institute, which is operated by AURA for NASA.}
which calls IRAF tasks\footnote{IRAF is distributed by the National Optical Astronomy Observatories, which are operated by the Association of Universities for Research in Astronomy, Inc., under cooperative agreement with the National Science Foundation.}
from Python scripts.
The reduction steps include sky subtraction, scattered light subtraction,
flat-fielding (using a halogen lamp with an integrating sphere),
geometric transformation, aperture extraction, and wavelength calibration
based on Th--Ar lamp spectra. In addition, small wavelength
shifts\footnote{The wavelength shifts are, at least partly, due to the instability of the instrument caused by varying ambient temperature, and we made an instrumental upgrade to minimize such shifts in late 2016.}
between individual exposures were corrected, if necessary, before they were
combined to give an averaged one-dimensional spectrum for each target.
Throughout this paper, we use air wavelengths rather than vacuum wavelengths.
Then, the one-dimensional spectrum was normalized at the continuum level,
and the telluric correction was performed 
as described in \citet{Sameshima-2018a} using the WINERED spectrum of
an A0\,V star taken at the same night. 

\input{table1.tex}

\begin{figure}
\includegraphics[clip,width=0.98\hsize]{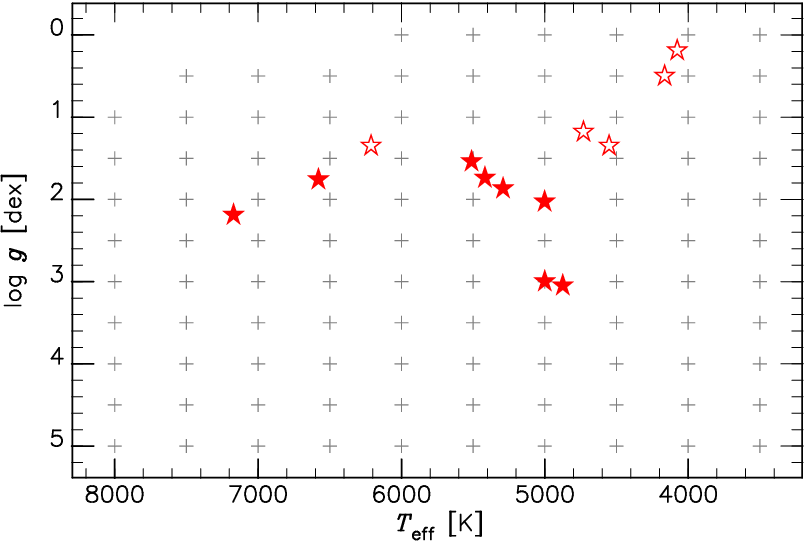}
\caption{
Stellar parameters used for spectral synthesis.
The effective temperatures, $\Teff$,
and the surface gravities, $\log g$,
of our targets (Table~\ref{tab1})
are indicated by star symbols, 
the open symbol for those with $\log g < 1.5$ and
the filled symbol for those with $\log g > 1.5$. 
The gray `$+$' symbol indicates the grid of
$(\Teff, \log g)$ used for a set of synthetic spectra
(Section~\ref{sec:data-syn}).
\label{fig1}}
\end{figure}

It is worthwhile to add remarks on some objects.
Because they are bright, there were
studies with optical spectra dating back decades ago,
e.g., \citet{Steel-1945} for HD\,20902 ($\alpha$~Per).
In particular, HD\,194093 (or $\gamma$~Cyg) is an interesting supergiant.
Since the early discovery by \citet{Adams-1926,Adams-1927},
it is known that the spectrum of $\gamma$~Cyg is rich in rare-earth lines,
e.g., Eu and Dy
\citep[see, also,][]{Roach-1942}.
The star $\gamma$~Cyg is located close to 
the so-called $\gamma$-Cygni supernova remnant, but \citet{Johnson-1975}
found that they are not connected with each other.

\subsection{Synthetic spectra} \label{sec:data-syn}
We used MOOG \citep[the version released in February 2017][]{Sneden-2012}
to create synthetic spectra for selecting and confirming  
absorption lines of the heavy elements.
We adopted the 1D plane-parallel atmosphere models
compiled by R.~Kurucz.\footnote{We adopted the files named amXX.dat or apXX.dat where XX is to be replaced by two digits indicating the abundances, not the original ones available in \citet{Kurucz-1993}, from http://kurucz.harvard.edu/grids/.}
The local thermal equilibrium (LTE) is assumed in
the models we used and any part of the spectral synthesis.
We considered two sets of stellar parameters.
Those for the first set are listed in Table~\ref{tab1} and
used for making direct comparisons between the synthetic
and observed spectra.
We estimated the broadening of individual objects
($\vbroad$, in the full width at half maximum)
including the instrumental resolution based on several
isolated lines in the observed spectra.
For the second set, we used the
$\Teff$ and $\log g$ values on the grid indicated by the `+' points in Figure~\ref{fig1}
together with the following parameters fixed:
the microturbulence $\xi=2$\,{\kms}, the broadening $\vbroad=10.7$\,{\kms},
and the chemical abundances to be solar \citep{Asplund-2009}.
The fixed microturbulence and broadening are not optimal 
for stars with broad ranges of $\Teff$ and $\log g$.
The $\xi$ of 2\,{\kms} is typical for red giants, but $\xi$ tends to be
larger for stars with higher luminosity \citep{Gray-2001}.
The macroturbulence ($\vmac$), or broadening in general,
also shows the dependency on the luminosity class \citep{Gray-1986}.
Likewise, both $\xi$ and $\vmac$ depend on $\Teff$
\citep[e.g.,][]{Ryabchikova-2016}.
This infers that
the predicted line strengths in the second set of the synthetic spectra
may be biased especially for dwarfs and supergiants. Nevertheless, 
those synthetic spectra useful to predict where
absorption lines in question get significant or deep
in the parameter space of $(\Teff, \log g)$.
Incidentally, the absorption lines we discuss here are mostly shallow
and not strongly saturated;
therefore, $\xi$ does not affect our predictions so much.

For both of the sets mentioned above, we assumed the solar abundance ratios of \citet{Asplund-2009},
$\log \epsilon _{\rm Fe,\odot}=7.50$~dex for ${\rm [Fe/H]}=0$ and other `Photosphere' abundances
listed in their Table~\ref{tab1}, except the carbon abundance.
We changed [C/H] by some amounts, between $-0.31$ and 0.28\,dex,
in order to get
better agreements of CN lines between the synthetic and observed spectra.
The observed spectra in the lower temperature range show many CN lines, and
those lines are not well reproduced by the synthetic spectra of many stars  
without the changes. This may be caused by the surface abundance ratios
modified by the first dredge-up \citep{Luck-1985,Takeda-2013,Lyubimkov-2015}.
The purpose of this adjustment is merely to get the synthetic spectra
that match better with the observed ones. The [C/H] values we used
are not necessarily accurate estimates of the carbon abundances;
therefore, we do not give the [C/H] used for individual objects.
In contrast, [C/H] is fixed to be solar in the second set
of synthetic spectra for the grid of $(\Teff, \log g)$.

In order to assess the absorption of target lines
even if they are contaminated by other lines,
we consider four kinds of synthetic spectra:
(i)~normal ones with all lines included,
(ii)~those with lines of only one species (i.e., each atom or atomic ion) included,
(iii) those with only one target line included,
and (iv)~those with the target lines excluded.
We consider the {$\log gf$} values and other parameters 
of a given target line in each of the three lists. 
For other atomic lines, we adopted the lines and their parameters in KURZ
when we considered target lines in the KURZ list, but
we used the VALD line list for considering target lines in VALD or those in MB99.
The KURZ list we used is the version compiled on 2017 Oct 8, while we downloaded
the VALD3 list to use on 2019 August 27.
For molecular lines, we used those listed in VALD, in which
lines of C$_2$, CH, CN, CO, and OH molecules are included
within the wavelength range of our interest.
Only CN lines appear significant
in the temperature ranges we investigate. In addition,
we adopted FeH lines compiled by Plez (private communication;
see also \citealt{Onehag-2012}).
We adopted the FeH's dissociation energy of 1.59\,eV,
at 273.15\,K, from \citet{Schultz-1991}.
The FeH lines appear only in a few objects with the lowest temperatures.

In comparing the observed and synthetic spectra of the 13 objects,
we noticed dozens of absorption lines
which are well visible in the observed spectra
but not predicted in the synthetic ones.
We list such lines and discuss their characteristics
in Appendix~\ref{sec:unknown}.

\section{Detection of absorption lines} \label{sec:lines}
\subsection{Line selection} \label{sec:selection}
We searched for absorption lines of 
elements heavier than the iron group elements
in the three line lists, KURZ, VALD, and MB99.
Combining the three lists, our target elements are 
from Cu to Th, i.e., $29\leq Z\leq 90$, with some gaps.
We limited ourselves to the ionization states between
I (neutral) and III (doubly ionized). We also limited the wavelength ranges to
9760--11100 and 11600--13200\,{\AA}, 
corresponding to the orders of 51--57\,th ($Y$ band) and 43--48\,th ($J$ band)
of WINERED.

In the given wavelength ranges,
MB99 lists 19 lines of 6 species in total,
i.e.,
\ion{Zn}{1}, \ion{Ge}{1}, \ion{Sr}{2}, \ion{Y}{2}, \ion{La}{2}, and \ion{Eu}{2}.
MB99 covers the wavelengths longer than
10000\,{\AA} only. 
MB99 actually lists 20 lines of our target species
between 10000 and 13400\,{\AA}, but
one of the lines, \ion{Ge}{1}~11125.12, falls within
the gap of the target in the wavelength range. 
There is dense and strong telluric absorption around the
\ion{Ge}{1} line \citep[see, e.g., Fig.~4 in][]{Sameshima-2018a},
and we neglected this line.
We consider the 19 lines in MB99 in the next selection step.

In contrast to MB99, VALD and KURZ list large numbers of absorption lines for
various elements between Cu
and Th. Hundreds of lines are listed
for some elements, but we cannot expect to see so many
lines of these heavy elements in real stellar spectra
as we confirm below.
Before we compare the observed spectra with
the synthetic ones, we limited the number of candidate lines
for each species using the synthetic spectra.
We considered the spectra
for the grid of $(\Teff, \log g)$
with lines of only one species included
at each time (see Section~\ref{sec:data-syn}).
We picked up lines that become deeper than 0.02 in depth
in at least one of the synthetic spectra at all the grid points.
67 lines of 9 species in KURZ and 63 lines of 14 species in VALD
were found to be candidates.

There are many overlaps between the lines selected
with the three line lists, and there are 108 different lines
of 14 species in total;
40, 35, and 5 lines are listed only in KURZ, VALD, and MB99, respectively.
All of the 5 lines selected from MB99 only are of \ion{Ge}{1}.
In the following,
we consider whether these lines appear significant
in the observed spectra
or not and, if detected, examine their temperature dependence
in comparison with the synthetic spectra.

\subsection{Measurements} \label{sec:measurements}
A significant fraction of the lines we investigate are
expected to be blended with other absorption lines
even if they exist,
and it is difficult to measure the equivalent widths. 
We therefore consider depths, i.e., $d \equiv 1-f(\lambdac)$, where
$f(\lambdac)$ is the flux at the line center, $\lambdac$, in normalized spectra.

In order to make direct comparisons between the depths in the observed
and synthetic spectra, we perform the normalization to adjust 
the continuum levels of the two kinds of spectra.
The reduction performed by the standard pipeline software
for WINERED includes the continuum normalization
\citep[see, e.g., Section~3.1 in][]{Taniguchi-2018}.
Although the synthetic spectra we created for individual objects show
reasonable agreements between the observed and synthetic spectra, 
there are noticeable deviations especially within absorption lines.
Such deviations can be attributed to inaccurate {$\log gf$} values
\citep{Andreasen-2016,Kondo-2019} and maybe to inaccurate abundances assumed.
Moreover, narrow flat parts 
outside apparent absorption lines are found to be lower than the unity
in some spectral regions of the synthetic spectra,
but the normal continuum procedure applied to the observed spectra
adjusts such regions to be around 1.
We therefore match the continuum levels of
the observed and synthetic spectra around each target line as follows.

First, we searched for velocity offsets to fix
small offsets in the wavelength scale, if any, by minimizing
the residuals between the observed and synthetic spectra.
For each target line at the wavelength of $\lambdac$,
the residuals were calculated for spectral parts around the line center.
We used the width of $\pm 300$\,{\kms} for
the objects with $\Teff < 5200$\,K and the width of $\pm 1000$\,{\kms} 
for the others for this wavelength adjustment.
The large width for the warmer stars was necessary to accommodate
many absorption lines which enable us to estimate
the offset precisely.
When we compare the observed and synthetic spectra,
the synthetic spectra were pixelized to match the pixels
of the observed spectra by integrating the normalized flux
within the wavelength range
of each pixel of the latter, and $\fsyn$ indicates
the counts of thus digitized spectra.
We corrected the observed spectra produced by the pipeline
for the velocity offsets; $\fobsorg$ indicates
the counts of the observed spectra after this correction.

Second, we considered 
the pixel-by-pixel ratio between each observed spectrum and the corresponding
synthetic one, $r_{\rm org}\equiv \fobsorg / \fsyn$,
for adjusting the continuum level of the former to that of the latter.
For the normalization around each target line, we used
the adjacent parts on both sides of the line,
i.e., $\lambdac-\Lambda_2<\lambda<\lambdac-\Lambda_1$
and $\lambdac+\Lambda_1<\lambda<\lambdac+\Lambda_2$,
where $\Lambda_1$ and $\Lambda_2$ are the wavelength steps 
corresponding to the velocity of 15\,{\kms} and 100\,{\kms},
at around $\lambdac$, respectively. The ratio shows
deviations from 1 caused by various errors including
the offset and slope in the pre-normalized continuum level
of $\fobsorg$ and imperfect reproduction of the stellar spectrum
in $\fsyn$.
The ratio, as a function of pixel, of the adjacent parts around each target line
was fitted as a linear trend, $a(\lambda-\lambdac)+b$,
by minimizing
\begin{eqnarray}
\chi^2 = \sum_{\lambda_i} w_{f} w_{\lambda} \left\{ r_{\rm org}(\lambda_i) - a(\lambda_i -\lambdac)-b\right\}^2, 
\end{eqnarray}
where the summation is taken over the pixels, $\lambda _i$,
within the adjacent ranges mentioned above.
The pixels within {$\pm$}15\,\kms around the center were
not included. 
We introduced two weight terms, $w_{f}$ and $w_{\lambda}$,
\begin{eqnarray}
w_{f} &=& [{\fsyn}(\lambda_i)]^2, \label{eq:wf} \\
w_{\lambda} &=& {\rm exp}\left(-\frac{|\lambda_i-\lambda_c|}{\Lambda_2}\right). \label{eq:wl}  
\end{eqnarray}
The former depends on the count of $\fsyn$;
the pixels with flux closer to the continuum
are more weighted than the pixels within strong absorption. 
The latter, $w_{\lambda}$, gives higher weights to the pixels closer to
the target line in wavelength than to those more separated.
Admittedly, we have no solid mathematical background to use
the weights as given in Equations~(\ref{eq:wf}) and (\ref{eq:wl}).
Nevertheless, after some experiments, we found that
they tend to give reasonable results even if the observed spectra show deviations
from the synthetic one such as lines with clearly different depths.
Next, we divided the original observed spectra by the fitted linear trend 
to obtain the re-normalized spectra, 
$\fobs (\lambda) = \fobsorg (\lambda) / [a(\lambda - \lambdac)+b]$. 
We also calculated the weighted standard deviation of the residual
between the observed and synthetic spectra,
\begin{eqnarray}
e = \sqrt{\frac{\sum \left\{w_f w_\lambda \left[ \fobs (\lambda_i) - \fsyn (\lambda_i) \right]^2 \right\}}{\sum w_f w_\lambda}} \, . 
\end{eqnarray}
This $e$ serves as the error in depth when we compare the depths of the line in
the observed and synthetic spectra. 
The residual is attributed to
various factors including pixel-to-pixel errors in the observed spectra,
the inconsistency between the observed and synthetic spectra,
and the error in the normalization.
The median of the $e$ values calculated for individual 
ranges around the target lines is 
{$\sim$}0.01 or smaller for most of the 13 objects.
For the two objects with $\Teff < 4200$\,K, however,
the median $e$ values are nearly 0.02.
We note that the error estimates described here do not include the uncertainty concerning
the target lines themselves because the normalization 
and the calculations of $e$ were 
done without considering the pixels within $\Lambda_1 = 15$\,{\kms} around the lines.
The residuals at the target wavelengths, if significant, would tell us
the inconsistency between the depths observed and
those predicted with the line lists.

Then, we measured the depths, $d$,
at the line wavelength, $\lambda_c$, by linearly interpolating
the two adjacent pixels, $\lambda_i$ and $\lambda_{i+1}$,
with $\lambda_i \leq \lambda_c < \lambda_{i+1}$;
$\dobs$ and $\dsyn$ indicate the depths measured with
$\fobs$ and $\fsyn$, respectively.
While the two depths of each target line
can be directly compared with each other for each object,
the depths for different objects are affected by
differences in various stellar parameters. 
Here we consider the indicator, $d^* = \gamma d$, 
i.e., the depths multiplied by the factor $\gamma$ which compensates
the effects of stellar metallicity and line width,
\begin{eqnarray}
\gamma \equiv 10^{-\FeH} \left(\frac{\vbroad}{\vinst}\right), 
\end{eqnarray}
where the first term compensates the metallicity effect and
the second term converts the depths in spectra broadened by
the width of $\vbroad~(\simeq \sqrt{ \vmac^2 + \vinst^2})$ into
those in the spectra merely broadened by the instrumental line width, 
$\vinst=10.7$\,{\kms}, of the WIDE mode of WINERED
(i.e., with the broadening by macroturbulence, $\vmac$, ignored).
The $\gamma$ factor for each target is listed in Table~\ref{tab1}.
It should be noted that $d^*$ does not necessarily agree with
the depth of a line in the spectrum of a star with $\FeH = 0$\,dex
broadened by the instrumental width if the line is blended with other lines.
Nevertheless, the conversion from $d$ to $d^*$ makes it easier
to compare the depths of 13 targets directly and examine the dependence on $\Teff$.
The $d^*$ may still be affected by
$\log g$ and $\xi$. 
We will take into account the effect of
$\log g$ by considering the depths
in the synthetic spectra obtained for the grid of $(\Teff, \log g)$
that was described in Section~\ref{sec:data-syn}.
On the other hand, $\xi$ has little impact on shallow lines 
for which we need precise measurements for confirming the line identification.

Finally, we evaluate the indicator that measures the absorption of
the target lines themselves. As mentioned in Section~\ref{sec:data-syn}, 
we synthesized the spectra with the target lines excluded.
Their depths at the wavelengths in question are
also measured and denoted as $\dsyndagger$. They are expected to be zero
if there is no blending line, but they are often not zero.
By subtracting $\dsyndagger$ from the normal depth, we can estimate
the contribution of a target line to absorption. We calculated
this indicator, $\alpha$, for the depths in both observed and synthetic spectra;
$\alphaobs \equiv \dobs - \dsyndagger$ and $\alphasyn \equiv \dsyn - \dsyndagger$.
If $\alphaobs$ is large with respect to $e$, 
the absorption by the target line is suggested to be significant.
When we compare the measurements for the 13 objects,
we consider the conversion by considering 
the effects of metallicity and broadening,
$\alpha^* \equiv d^* - \dsynstardagger = \gamma (d-\dsyndagger)$,
where $\alpha^*$ and $d^*$ are considered for both the observed and synthetic spectra.
In addition, we use $\beta \equiv \dsyndagger/\dsyn$ as
the estimate of the blend; $\beta$ varies from 0 (no blend) to
1 (fully contaminated). We define $\beta = 1$ when
$\dsyn = 0$, but we are not interested in
such cases because our purpose is to confirm or reject predicted lines. 

\subsection{Results} \label{sec:results}
\begin{figure}[!tb]
\includegraphics[clip,width=0.98\hsize]{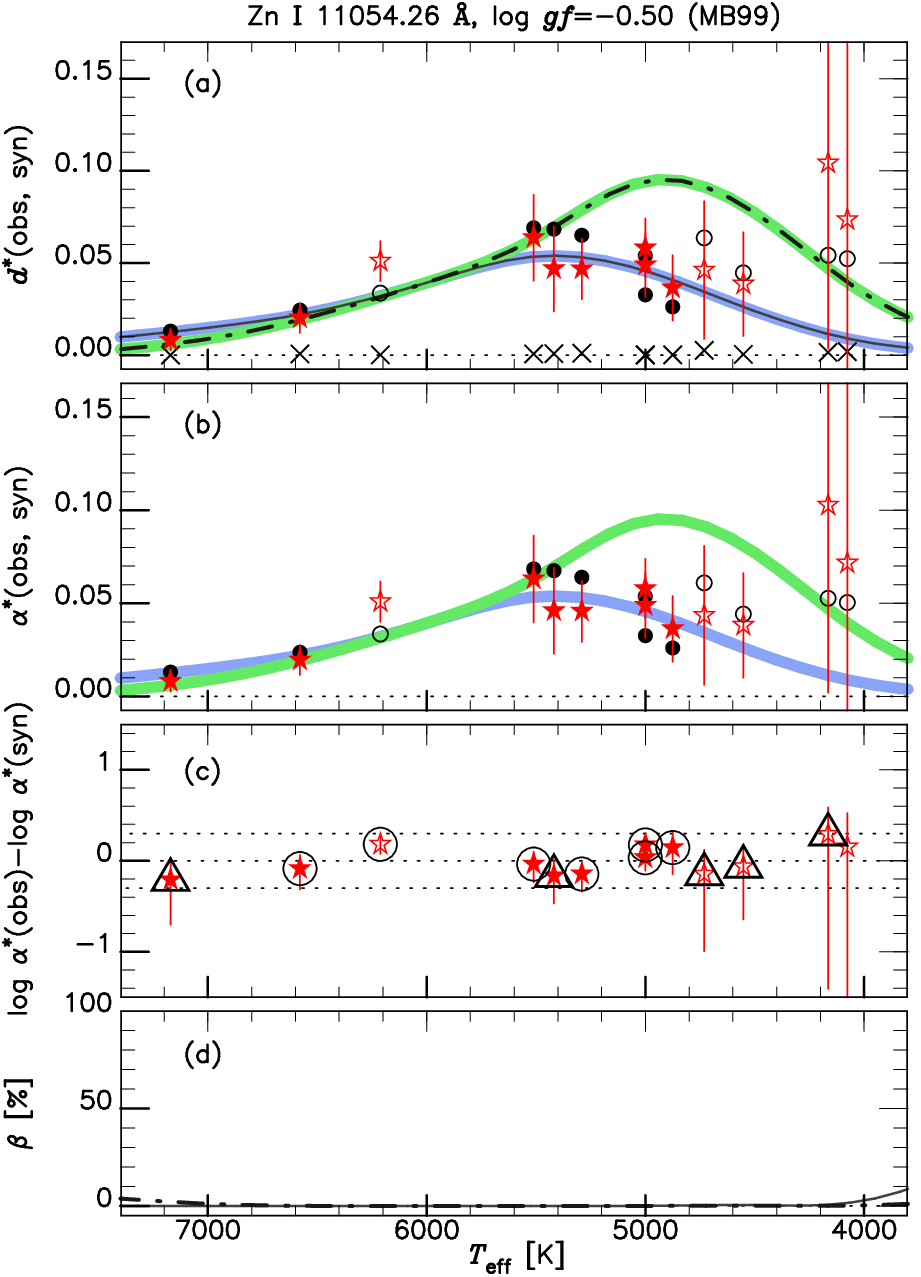}
\caption{
The temperature dependency of the line depths and blends.
The species and wavelength of the line 
together with the {$\log gf$} and its source (KURZ, VALD, or MB99)
are labeled at the top.
The top two panels plot the depths, $d^*$, and the
contribution of the target lines themselves, $\alpha^*$,
against $\Teff$. Both $d^*$ and $\alpha^*$ are
after the correction of metallicity and broadening
(see Section~\ref{sec:measurements}). Star symbols indicate
the measurements with the observed spectra of 13 objects;
the open symbol for those with $\log g < 1.5$ and
the filled symbol for those with $\log g > 1.5$. 
Open and filled circles indicate the values obtained with
the synthetic spectra for the objects with
$\log g<1.5$ (open) and those with $\log g>1.5$ (filled).
The `$\times$' symbol in the top panel indicates the depths 
in the synthetic spectra with the target lines excluded.
The green and blue curves in the panels~(a) and (b)
indicate the $\Teff$ trend of the target line itself
expected for stars with 0.5 and 2.5, respectively, in $\log g$,
while the dot--dashed and solid curves in the panel (a) indicate
the corresponding trends of the depth including blends.
In the third panel, we plot 
$\log \alphaobs^* - \log \alphasyn^* \simeq 0$
and the error against $\Teff$ for the 13 objects.
The star symbols in the panel~(c) are
accompanied by circles or triangles
if we found significant ($\alphaobs/e > 2$) or 
marginal ($1<\alphaobs/e<2$) detection.
The last panel shows the blends, $\beta$, for
the synthetic spectra of stars with
$\log g=0.5$ (dot--dashed) and $2.5$ (solid).
The complete figure set (23 images) is
available in the online journal.
\label{fig2}}
\end{figure}

\begin{figure*}
\includegraphics[clip,width=0.98\hsize]{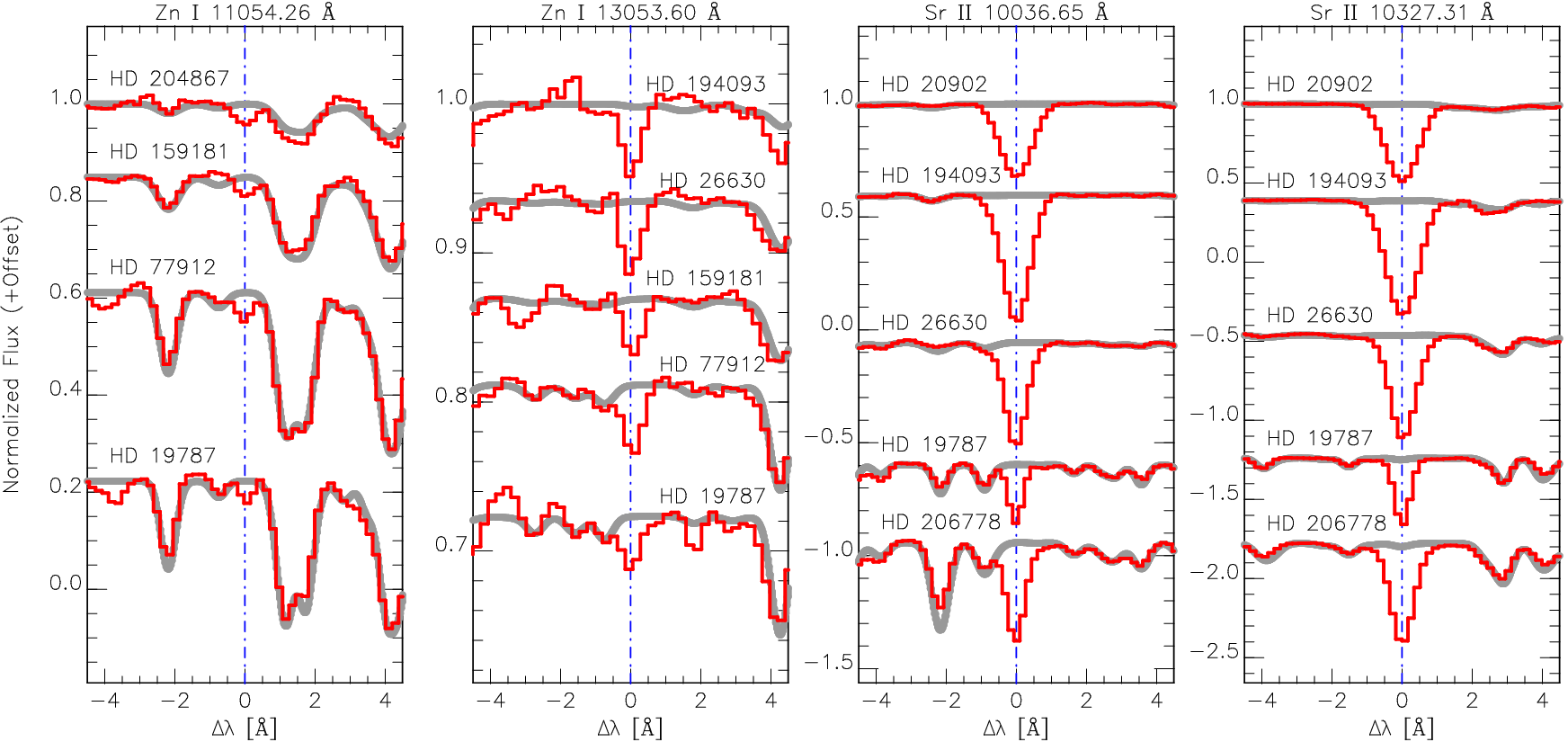}
\caption{
Spectra around the absorption lines detected.
The species and the air wavelength, indicated by the vertical line,
of each line are labeled at the top of each panel.
The red and gray curves respectively indicate the observed spectra and the spectra
synthesized without the target line included. 
The complete figure set (23 panels) is
available in the online journal.
\label{fig3}}
\end{figure*}

We measured the depths, $\dobs$ and $\dsyn$, and also the errors, $e$,
for the 108 candidate lines, selected in Section~\ref{sec:selection},
in the spectra of the 13 objects. 
While many lines were not seen, there are dozens of lines
showing large $\alphaobs/e$ values. We considered the lines with
$\alphaobs/e > 2$ to be significant unless they are
affected by spurious noises too much.
Then, we examined the $d^*$, $\alpha^*$, and $\beta$
plotted against $\Teff$ (Figure~\ref{fig2})
and checked if the dependence 
on $\Teff$ is consistent with the expectation from the synthetic spectra.
For some lines, the gravity effect on the depths measured for
the targets with different $\log g$ is significant, and
we took it into account by comparing the $\Teff$ trends
expected for the two $\log g$ values, 0.5 and 2.5, illustrated in Figure~\ref{fig2}.
In addition, we checked the appearance of the spectra,
comparing the synthetic spectra with the target lines removed
with the observed spectra
(see Figure~\ref{fig3}),
as well as how
the telluric absorption could have left noises after the correction.

Among the 108 lines we examined in the observed spectra
(Section~\ref{sec:selection}), 
we detected 23 lines (Table~\ref{tab2}).
Figure~\ref{fig3} presents the spectra
of some stars with the line detected.
We here summarize the characteristics of the lines we identified,
but more details on the lines of individual species are given in Appendix~\ref{sec:remarks}. 
We detected two \ion{Zn}{1}, three \ion{Sr}{2}, and five \ion{Y}{2} lines listed in all the three lists, KURZ, VALD, and MB99,
but no other lines of these species were confirmed.
We also detected one MB99 line of \ion{Eu}{2} (10019.52\,{\AA})
together with \ion{Eu}{2}~9898.27 and 10165.56 that are not included in MB99.
In addition, among the candidate lines which are not in MB99,
we identified four \ion{Zr}{1}, one \ion{Ba}{2}, two \ion{Ce}{2},
two \ion{Sm}{2}, and one \ion{Dy}{2} lines.

A few lines have been already reported in previous papers,
besides MB99, based on high-resolution spectra.
Although the temperature of the object is different from ours,
\citet{Sameshima-2018b} identified the two \ion{Sr}{2} lines
at 10327.31 and 10914.89\,{\AA} in 
a WINERED spectrum of an A0\,V-type star.
The same lines were used for measuring abundances of red supergiants
by \citet{Origlia-2013}, 
and the one at 10914.89\,{\AA} was also identified
by \citet{Hubrig-2012} in peculiar A-type stars.
In addition to these \ion{Sr}{2} lines, we also confirmed the one at 10036.65\,{\AA} (a strong line listed in MB99).
The \ion{Dy}{2} line at 10835.94\,{\AA} reported by \citet{Hubrig-2012}
with spectra of peculiar A-type stars
was not confirmed in any of our targets.

The 85 lines 
We could not confirm 85 candidate lines in the observed spectra.
They include
some lines listed in MB99: \ion{Zn}{1}~13196.62, \ion{La}{2}~11874.19, and six \ion{Ge}{1} lines (see the notes on the individual elements
in the Appendix).
We found that there may be absorption by some of the unconfirmed lines,
but rejected them either because the lines are not deep enough to be confirmed ($\alphaobs/e<2$; e.g., \ion{Ce}{2}~9774.63)
or the telluric absorption seems to leave spurious noises
(e.g., \ion{Sm}{2}~9788.96).
In addition, we found quite strong absorption at the wavelength of \ion{Dy}{2}~10305.36,
but did not conclude that the line was confirmed because the observed $\Teff$ trend is not
consistent with the prediction. The \ion{Dy}{2} line, if exists, is at least contaminated by
an unidentified line.

We examined $\log \alphaobs^* - \log \alphasyn^*$ and
judged if the {$\log gf$} values given in the line lists
reproduce the depths of significant lines
($\log \alphaobs^* - \log \alphasyn^* \simeq 0$) or not.
The superscript $c$ in Table~\ref{tab2} indicates that 
the measured depths differ from those predicted with
the given {$\log gf$} by {$\sim$}0.3~dex or more.
We note, however, that our conclusions on the {$\log gf$} cannot be final
because we only made comparisons with the synthetic spectra
created with the solar [X/Fe] values assumed.

\input{table2.tex}

\section{Discussion} \label{sec:discussion}
\subsection{Astronomical implications} \label{sec:implication}
In the following, we discuss the implications that
individual elements with the detected line(s) may give
to the chemical evolution of the Galaxy and nearby galaxies.
We here focus on the chemical imprints 
one would find in Galactic-disk stars of around the solar metallicity.
For stars with different metallicities or 
those in different systems, 
some elements may be explained by different origins.
For example, Ba is an $s$-process element at around
the solar metallicity but
the contribution of the $r$-process gets stronger 
at the low-metallicity range \citep{Burris-2000}.
Moreover, there are proposed $n$-capture processes that show patterns different
from those of the main $s$- and $r$-processes
\citep[e.g.,][]{Burris-2009,Hampel-2016}.
Our main targets, supergiants, are young and expected to
be relatively metal rich, and previous explanations
on the origins of the heavy elements in the Sun and
the Solar system \citep{Burris-2000,Sneden-2008} gives
at least an approximate idea on the origins of
individual elements in the supergiants. 

Zn is often included in the iron peak elements
and considered as the heaviest one of this group. 
In fact, in a broad range of the metallicity down to $\FeH \simeq -2$\,dex,
Zn in stars in the solar neighborhood shows a concentration around
[Zn/Fe]$=0$ and thus seems to be created along with Fe and other iron peak elements
\citep{Sneden-1991}.
However, it has become clear that systematic differences in the [Zn/Fe] trends
are found in different systems (thin and thick disks, bulge, halo, and dwarf galaxies)
thanks to the efforts of various authors
\citep[][and references therein]{Duffau-2017,Ji-2018,Hirai-2018}.
The origins of Zn remain still elusive and
its implication to the Galactic chemical evolution
may be unique compared with other better-understood elements
\citep[see, e.g.,][]{Tsujimoto-2018}.

The $s$-process elements are created mainly in
low- and intermediate-mass Asymptotic Giant Branch (AGB) stars.
Good reviews on the process are found, e.g., 
in \citet{Busso-1999} and \citet{Karakas-2014}.
Sr, Y, and Zr are grouped as light $s$-process elements,
although they show slightly different
trends from each other in disk stars \citep{DelgadoMena-2017}.
Ba and Ce are often called heavy $s$-process elements together with Nd.
The abundance trends of the three heavy $s$-process
elements are, however, not necessarily common.
For example, \citet{Andrievsky-2013,Andrievsky-2014} reported
that the radial gradient of Ba has a negligible slope,
while \citet{Lemasle-2013} and \citet{daSilva-2016} found significant
slopes of the gradient for the other heavy s-process elements,
Ce and Nd, as well as for Y and Zr, i.e., light $s$-process elements.
As noted by \citet{Andrievsky-2014} and \citet{Luck-2014}, 
only strong Ba lines were used with non-LTE calculations, which may leave
significant uncertainties. Readers are referred to other studies
such as \citet{DOrazi-2009}, \citet{Bensby-2014},
and references therein
concerning the Ba abundances by using the strong lines in the optical,
but their targets are dwarfs and giants. 
\ion{Ba}{2}~13058.01 found in this study
is, in contrast, expected to be weak; therefore,
this line may add an important constraint on the Ba abundances.
On the other hand, the three \ion{Sr}{2} lines in the $Y$ band are very strong
and the same problem of the non-LTE effect may prevent us from obtaining accurate Sr abundances.

Eu is representative of $r$-process elements.
Sm and Dy are considered to be mainly formed
through the $r$-process 
around the solar metallicity \citep{Burris-2000}.
The contribution of the $s$-process to Sm can be larger,
especially for some isotopes, but the trend of
Sm seen in the disk stars with about the solar metallicity 
is similar to that of Eu \citep{Battistini-2016}.
After the discovery of the gravitational wave and 
the electromagnetic counterpart of GW\,170817
\citep{Abbott-2017}, it has become
more evident that the neutron star mergers accompanied by
kilonovae are major contributors to $r$-process elements
\citep{Smartt-2017,Tanaka-2017}.
However, explaining the chemical evolution of
$r$-process abundances seen in disk stars requires
further investigations concerning some complications
due, e.g., to the delay-time distributions
of the neutron star mergers \citep{Hotokezaka-2018},
the initial mass functions \citep{Tsujimoto-2019},
and potential contribution from other origins
\citep{Siegel-2019}. New observational constraints
on the $r$-process enrichment 
can be expected if the near-infrared diagnostic lines of Eu and Dy
are used for measuring
the abundances of stars in unexplored regions
of the Galactic disk behind severe interstellar extinction.

\subsection{Expected targets} \label{sec:expected}
\begin{figure}[!tb]
\includegraphics[clip,width=0.98\hsize]{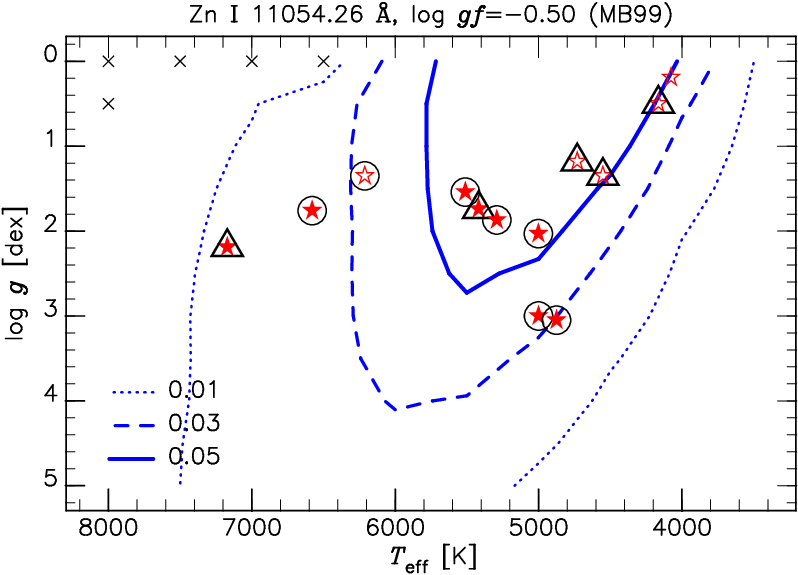}
\caption{
Contours of the predicted depths of the lines we detected.
The species and wavelength of the line 
together with the {$\log gf$} and its source (KURZ, VALD, or MB99)
are labeled in the top of each panel.
The contours of $\alphasyn$ for the levels indicated within each panel
are created based on the synthetic spectra 
with the solar abundance and the broadening $\vbroad = 10.7$\,{\kms}
for the grid of $(\Teff, \log g)$ illustrated in Figure~\ref{fig1}.
We note that the grid lacks 
the points indicated by the `$\times$' symbol.
The star symbols (open if $\log g<1.5$ and filled if $\log g>1.5$)
indicate the observed targets in this study,
and they are accompanied by circles or triangles
if we found significant ($\alphaobs/e > 2$) or 
marginal ($1<\alphaobs/e<2$) detection of the line.
The complete figure set (23 images) is
available in the online journal.
\label{fig4}}
\end{figure}

The lines we report here will be useful diagnostic lines 
of detailed chemical abundances available in the $YJ$ bands. 
In order to give a rough idea on the $(\Teff, \log g)$ range
in which the detected lines are seen, 
we present in Figure~\ref{fig4} the contours of $\alphasyn = \dsyn - \dsyndagger$
based on the synthetic spectra obtained for
the $(\Teff, \log g)$ grid (see Figure~\ref{fig1}).
Some lines are blended and may be difficult to measure the abundances
even if $\alpha$ gets significant.
Also, note that the depths refer to those expected for
the spectra with the WINERED resolution, $R=28000$, for stars with 
the solar metallicity, $\FeH=0$\,dex, and sufficiently small broadening.
Nevertheless, the contours will be useful, e.g., for selecting
the targets to study the abundances discussed in this paper
and making the observational plans to achieve the detection of the lines.

Only two species (\ion{Zn}{1} and \ion{Zr}{1})
among the 9 with the lines detected
are neutral. In contrast, the others are
singly ionized. The first ionization potentials of the elements
with the detected lines are low, 5.2--6.2\,eV, compared to Zn (9.4\,eV) and Zr (6.6\,eV).
The lines of the ionized species 
are sensitive to $\log g$ and strong in low-gravity stars
as expected. Furthermore, Figure~\ref{fig4} indicates that
supergiants with $\Teff \sim 5000$\,K and $\log g \lesssim 1.5$\,dex
show the strongest absorption of the lines of
\ion{Sr}{2}, \ion{Y}{2}, \ion{Ba}{2}, \ion{Sm}{2}, \ion{Eu}{2},
and \ion{Dy}{2}. 
The contours of \ion{Zn}{1} are similar to 
those of the above species to some extent,
the peak being around 5000\,K and
stronger towards the lower surface gravity,
but the sensitivity to the gravity is not so high as
the ionized species.
In contrast, the lines of \ion{Ce}{2} require 
both low temperature, $\Teff < 4000$\,K,
and low surface gravity, $\log g < 1$, 
to be significant. In the case of \ion{Zr}{1}, 
the line strengths grow rapidly towards low $\Teff$
but are insensitive to surface gravity.

Cepheids have intermediate temperatures, 4500--6500\,K,
and low surface gravities, $\log g \lesssim 1.5$\,dex
\citep[see, e.g.,][]{Genovali-2014}.
Their locations on the Hertzsprung-Russell diagram (HRD)
thus agree with the peaks of the strengths for most of
the lines reported in this paper, although 
\ion{Zr}{1} may be seen only in Cepheids with the lowest $\Teff$
among this group of pulsating stars in the Cepheid instability strip.
Therefore, $YJ$-band spectra of Cepheids would allow us to measure
various heavy elements including both $s$- and $r$-process elements
and also those with mixed origins such as Zn.

Cepheids are young, 10--300~Myr old, and provide valuable information
on chemical abundances of the present Galaxy. It is known that
their metallicities are anti-correlated with the distances
from the Galactic center ($\RGC$), known as the abundance gradient
\citep{Genovali-2014}. Observing Cepheids in a wide range of $\RGC$
thus allows investigating the abundance patterns from the metal-rich end
([Fe/H]$\sim +0.2$ at $\RGC \leq 6$\,kpc) to the metal-poor end
([Fe/H]$\lesssim -0.3$ at $\RGC \geq 14$\,kpc) of the current Galactic disk.
The youth of Cepheids offers a crucial advantage in
studying the disk chemical evolution. The radial migration of
the disk \citep{Sellwood-2002} affects the current abundance distribution of 
various elements for stars in the disk
\citep[e.g., see a recent study on the $r$-process abundance in the solar neighborhood by][]{Tsujimoto-2019}.
Cepheids are, however, considered
to be located almost at their birth positions in terms of $\RGC$
and tell us, at least approximately, the current abundances of star-forming gas.
Recent large-scale surveys have found a large number of new Cepheids
spread in a large range of $\RGC$
\citep{Udalski-2018,Chen-2019,Dekany-2019,Skowron-2019},
and some of them are heavily reddened in the Galactic disk.
Near-infrared spectroscopic observations are desired to
obtain their abundances.

Bright red giants, roughly $\Teff \lesssim 5000$\,K and
$1 \lesssim \log g \lesssim 2$\,dex, are often targeted
for investigating the heavy elements in nearby dwarf
galaxies \citep[e.g.,][]{McWilliam-2013,Ji-2016,Ji-2018}.
\ion{Sr}{2} and \ion{Zr}{1} lines are expected to be strong,
while \ion{Y}{2} lines are probably weak, $\lesssim 0.05$ in depth.
Some \ion{Eu}{2} lines may also be seen but the abundance
measurements would require high signal-to-noise ratios
with the typical depths expected to be
around {$\sim$}0.02 or smaller. Observing the lines of
other heavy elements would be even more challenging.
Bright red giants are important because they are usually
the easiest targets in stellar systems like the Galactic bulge
and dwarf galaxies which are dominated by old stars.
Measuring their Eu abundances, if possible,
would be very useful for studying the chemical evolution
of those systems together with the $s$-process
abundances traced with Sr and Zr. 

At the lower parts of the Hertzsprung-Russell diagram,
$\log g \lesssim 3$, only a small fraction of 
the lines reported in this study are expected to be visible.
The three \ion{Sr}{2} lines are expected to be still strong,
$\gtrsim 0.2$, for a wide range of $\Teff$.
In fact, \citet{Caffau-2016} detected the three lines
and estimated the abundances by taking into account
the non-LTE effect. 
The two \ion{Zn}{1} lines may be seen in stars with
intermediate temperatures around 5500\,K,
while \ion{Zr}{1}~9822.56 is expected to be visible
only in late-type stars with $\Teff < 5000$\,K.
None of the other species is probably significant unless
the abundances are highly enhanced. 

\section{Summary} \label{sec:summary}
We identified 23 lines of 9 elements heavier than the iron group elements,
i.e., Zn, Sr, Y, Zr, Ba, Ce, Sm, Eu, and Dy,
using the $YJ$-band WINERED spectra
of 13 supergiants and giants.
Besides the lines we detected,
significantly more lines of the targeted heavy elements
were selected based on KURZ and/or VALD but not detected.
We also found lines that are clearly present in the observed spectra
but are not predicted in the synthetic counterparts (Appendix~\ref{sec:unknown}).
It is vital to establish the list of 
lines of various elements in the infrared range.

Although we have identified the absorption lines
of rare heavy elements, including newly detected lines,
there remains a lot to be done.
Further searches for new lines should be performed.
Our search was done by using WINERED spectra 
with the resolution of 28000 with the typical S/N around 200--300.
It would be useful to make a more complete survey of
the relevant absorption lines by using
spectra with higher resolution and/or higher quality, although 
the resolution of 28000 is high enough to resolve 
the intrinsic line profiles of many supergiants. 
Solid estimates of {$\log gf$} values are also important
for abundance measurements with these lines
especially because the numbers of lines are small.
For this purpose, our targets and dataset are not optimal.
For example, relatively metal-poor stars with enhanced abundances of
$s$-process or $r$-process elements \citep[e.g.,][]{Sneden-2008,Bisterzo-2011}
would be good calibrators because
the lines of the enhanced elements can be measured with
reduced line contamination.

\acknowledgments

We thank Andy McWilliam for useful comments on the analysis and the manuscript.
We are grateful to the staff of the Koyama Astronomical Observatory for
their support during our observation. This study is financially supported by
JSPS KAKENHI (grant No.~16684001, 20340042, 21840052, 26287028, and 18H01248)
and the MEXT Supported Program for the Strategic Research Foundation at
Private Universities, 2008--2012 (No.~S0801061) and 2014--2018 (No.~S1411028).
HS acknowledges the JSPS grant No.~19K03917. 
This study has made use of
the SIMBAD database, operated at CDS, Strasbourg, France,
and also the VALD database, operated at Uppsala University,
the Institute of Astronomy RAS in Moscow, and the University of Vienna.
We thank the referee, Prof. Robert Kurucz, for comments that helped us to improve this paper.

\software{
WINERED pipeline (Hamano et al., in preparation), IRAF \citep{Tody-1986,Tody-1993}, PyRAF (Science Software Branch at STScI 2012), MOOG \citep[February 2017 version;][]{Sneden-2012}.
}

\appendix

\section{Identification of lines of individual elements}
\label{sec:remarks}
Here we describe the identification
(or the failure of confirmation) of
the lines of the heavy elements using the WINERED spectra of 13 objects
(Table~\ref{tab1}). We examined 108 lines of 14 species in total.
Table~\ref{tab2} lists the 23 detected lines.

\begin{itemize}
\item 
Zn (Zinc, $Z=30$) --- 
We investigated four \ion{Zn}{1} lines using the observed spectra.
The line at 13150.5\,{\AA} (13150.464\,{\AA} in KURZ and 13150.533\,{\AA} in VALD)
is not included in MB99, 
but the other three are included.
We clearly detected two lines at 11054.26 and 13053.60\,{\AA},
which are free from blends at most of
the $\Teff$ range we investigated.
For each line of them, the {$\log gf$} values in KURZ and VALD are consistent
within 0.1\,dex, but the {$\log gf$} in MB99 is different by 0.15--0.3\,dex.
The {$\log gf$} values in MB99 seem to give 
better agreements between the predicted and measured depths for
both of the lines although the differences between the three lists
are not large.
Among the two lines that were not confirmed,
\ion{Zn}{1}~13150.5 is dominated by
the strong blending line, \ion{Al}{1}~13150.75.
The other one, \ion{Zn}{1}~13196.6, also suffers from significant blends
especially at lower $\Teff$, if exists, and it tends to be affected by
telluric lines too.
KURZ and VALD list slightly different wavelengths, within 0.1\,{\AA},
for the two undetected lines but these lines are not detected
in the observed spectra in any case.

\item 
Ga (Gallium, $Z=31$) --- 
We investigated two \ion{Ga}{1} lines, 
included in both KURZ and VALD, using the observed spectra.
Neither of them is included in MB99 although their wavelengths
are longer than 10000\,{\AA}.
They were predicted to be significant, $\gtrsim 0.02$,
only at the lowest temperatures, $\Teff \lesssim 4000$\,K.
Moreover, both of them are strongly blended with other lines
($\gtrsim 80\,\%$),
and we could confirm none of the \ion{Ga}{1} lines.
Around the wavelength of \ion{Ga}{1}~11949.23, 
\ion{Ti}{1}~11949.55 gives strong absorption at lower $\Teff$
and \ion{Ca}{2}~11949.74 gives strong absorption at higher $\Teff$.
Around \ion{Ga}{1}~12109.85, \ion{Si}{1}~12110.66 gets
stronger towards the lower $\Teff$.

\item 
Ge (Germanium, $Z=32$) --- 
We investigated eight \ion{Ge}{1} lines using the observed spectra.
All of them are located at longer than 10000\,{\AA}.
One line at 11714.75\,{\AA} is listed in all the three lists.
In addition, three lines are listed only in KURZ and VALD,
while four lines only in MB99.
The \ion{Ge}{1} lines are expected to get strongest
at around 4500\,K, but most of the lines we investigated 
do not reach 0.05 in $\alphasyn^*$ according to the {$\log gf$}
values available in the lists. 
Some lines are expected to be strongly blended while others are not, but
we could not confirm any of the \ion{Ge}{1} lines.
The $\alphaobs^*$ we obtained are consistent with zero for
almost all of the objects expected to show the Ga absorption,
which indicates that the {$\log gf$} values are at least smaller than expected.

\item 
Sr (Strontium, $Z=38$) --- 
We investigated nine \ion{Sr}{2} lines using the observed spectra.
Three of them (10036.65, 10327.31, and 10914.89\,{\AA}) are listed in all the three line lists.
These lines are strong and there are almost no blends.
The {$\log gf$} values in the three lists are
more-or-less consistent with each other, within 0.2\,dex,
and they give 
reasonable agreements between the observed and synthetic spectra.
We note that these three lines are very deep and
the departures from the predicted depths may well be caused by
the factors, other than inaccurate {$\log gf$} values,
that were not taken into account in our simple spectral synthesis
(e.g., the non-LTE effect).
In contrast, we could confirm no other \ion{Sr}{2} lines.
The reduced spectra of a couple of objects present
a hint of \ion{Sr}{2}~12974.38, but telluric absorption lines 
around this wavelength seem to give spurious noise on the spectra.
We also found non-zero $\alphaobs/e$ for \ion{Sr}{2}~12013.96 in a few objects,
but the dependency of the depth on $\Teff$ and $\log g$ is not consistent with predicted for this line. 
There is one \ion{Fe}{1} line at 12013.88\,{\AA},
and the $\alphaobs$ we measured could be explained by  
the error in oscillator strength of this line with a {$\log gf$} larger
than listed.
For \ion{Sr}{2}~13121.96, two entries with different {$\log gf$} values,
$0.692$ and $-0.609$\,dex, are listed in VALD with the same wavelength
and the same excitation potential (EP).
Including or excluding the shallower one with $\log gf=-0.609$ does not change
the synthetic spectra significantly and, in any case, the absorption was not
confirmed in the observed spectra.

\item 
Y (Yttrium, $Z=39$) --- 
We investigated two \ion{Y}{1} and five \ion{Y}{2} lines using the observed spectra. The \ion{Y}{1} lines are not included in MB99,
while it lists all the \ion{Y}{2} lines.
The \ion{Y}{1} lines were predicted to be visible at very low temperatures, around 3500\,K, but still shallow, {$\sim$}0.03;
we could not confirm them in the observed spectra.
In contrast, we detected all the five \ion{Y}{2} lines.
All of them except \ion{Y}{2}~10605.15 are significantly blended 
at least at $\Teff \lesssim 5000$\,K, but the \ion{Y}{2} lines are expected to get
deepest at 5000--5500\,K and this dependency on $\Teff$ 
can be well traced with our measurements.
The {$\log gf$} values in MB99 reproduce the observed depths better for
\ion{Y}{2}~10186.46, while 
the {$\log gf$} values in the three lists are at least roughly
consistent with each other and
they predict reasonable depths for the other four lines.

\item 
Zr (Zirconium, $Z=40$) --- 
We investigated 40 \ion{Zr}{1} lines using the observed spectra.
Two lines are listed in both KURZ and VALD, while
the other 38 lines are found in KURZ only.
None of them is listed in MB99. 
We confirmed four lines in a few objects at low $\Teff$
without significant blends. One of them is listed in both KURZ and VALD,
while the others are given in KURZ only.
The {$\log gf$} values in the KURZ list give reasonable
predictions of the depths observed where the detection is significant.
We note that the observed and synthetic spectra show poor agreements
around \ion{Zr}{1}~10654.18 (see the online material for Figure~\ref{fig3}).
The absorption line observed at around 10657.4\,{\AA} is listed as an unknown one
in Table~\ref{tab3}. In addition, the observed spectra show broad and stronger absorption than
the synthetic ones at around 10652\,{\AA},
which indicates that more than one lines contribute to
the observed absorption but the line lists used for the synthetic spectra
give too low {$\log gf$} values or completely miss necessary lines.
A part of the inconsistency can be explained by the low {$\log gf$} values,
approximately $-2.8$\,dex, of \ion{Fe}{1}~10652.24 (EP$=$5.478\,eV)
in KURZ and VALD, whose {$\log gf$} in MB99 is significantly higher, $-1.79$\,dex. 
At least one other line is necessary to explain the wide absorption observed,
and CN~10651.796 but with {$\log gf$} higher than given in VALD
would probably explain the absorption in combination with the \ion{Fe}{1} line.

\item 
Ba (Barium, $Z=56$) ---
We investigated one \ion{Ba}{2} line using the observed spectra.
This line is listed in KURZ and VALD but not in MB99.
The wavelengths in the former two catalogs are slightly different,
13057.716\,{\AA} in KURZ and 13058.015\,{\AA} in VALD.
Synthetic spectra indicate: (i)~The line gets deepest at around 5300\,K.
(ii)~It becomes shallower and shallower with increasing $\log g$, and 
the $d^*$ never reaches 0.01 at $\log g\gtrsim 2.5$\,dex.
(iii)~There are strong blends throughout the temperature range
from 4000 to 8000\,K; 30\,\% or higher (reaching almost 100\,\%)
at $\lesssim 4,500$\,K and $\log g=0.5$\,dex, and
50\,\% or higher at $\log g=2.5$\,dex.
The absorption by this \ion{Ba}{2} line is supported by 
the observed spectra of a few objects with $\Teff > 5000$\,K
where the blends are moderate.
The overall result indicates that the detection of this line is solid,
but the confirmation based on more spectra is desirable. 
The observed depths, where significant, are larger than
predicted based on the synthetic spectra by a factor of {$\sim$}3,
although this may be because we assumed the solar abundance ratio
of Ba to the metallicity, i.e., [Ba/Fe]$=0$.

\item 
La (Lanthanum, $Z=57$) --- 
We investigated one \ion{La}{2} line, 11874.19\,{\AA},
using the observed spectra.
This line is included in both VALD and MB99 but not in KURZ.
Based on the synthetic spectra, the line is expected to be
as deep as 0.05 in many objects at $\Teff \lesssim 6000$\,K and
reach {$\sim$}0.2 in depth in the two lowest-$\Teff$ objects.
However, the depths we measured are consistent with zero
for most objects and we could not confirm this line
in any of the observed spectra.

\item 
Ce (Cerium, $Z=58$) --- 
We investigated 22 \ion{Ce}{2} lines using the observed spectra.
All of these are at $\lambda < 10000$\,{\AA} and listed in VALD, but
not included in KURZ or MB99. Most of the lines are expected to be
significantly blended with other lines.
We detected two lines at 9805.49\,{\AA} and 9853.11\,{\AA};
they are also blended with other lines,
but clearly appear
in the shoulders of the contaminating lines. 
The observed depths of the two lines, where significant,
are larger than
predicted based on the synthetic spectra by a factor of {$\sim$}5.
There may be a couple of other \ion{Ce}{2} lines
visible in the observed spectra,
although we could not conclude that they are real. 
In the case of \ion{Ce}{2}~9774.63,
the spectra of the two objects with the lowest $\Teff$
show absorption,
but their $\alphaobs/e$ values, 1.5--1.8, suggest that
the detection is marginal. While no significant blend
of stellar lines is expected,
the telluric absorption may disturb the observed spectra
at around this line.
The $\alphaobs/e$ values for \ion{Ce}{2}~9889.47
are significant for a couple of objects with low $\Teff$,
but strong blending lines, 
\ion{Fe}{1}~9889.035 and CN~9890.1513,
prevent us from confirming it.
There seems to be an absorption line at 
the wavelength of \ion{Ce}{2}~9949.45 in a few objects,
but it looks more significant in stars with
intermediate $\Teff$, 5000--6500\,K.
Such $\Teff$ dependency is inconsistent with
the $\Teff$ trend of the \ion{Ce}{2} line
predicted with the synthetic spectra.

\item 
Sm (Samarium, $Z=62$) --- 
We investigated four \ion{Sm}{2} lines using the observed spectra.
They are listed in VALD, but KURZ includes none.
One of them is at $\lambda > 10000$\,{\AA} but not listed in MB99.
We found that all the four lines may be present but it is 
not easy to make firm conclusions.
The predicted depths of these lines are rather small;
$\alpha$ is expected to be smaller than {$\sim$}0.03 in our objects
except the two with the lowest $\Teff$, {$\sim$}4100\,K.
In the low-$\Teff$ objects, however, larger depths are predicted
to be accompanied by severe blends, and we could not confirm
any \ion{Sm}{2} line significantly.
In contrast, we made significant and marginal detections
of some lines in warmer objects.
For \ion{Sm}{2}~9850.67 and 9936.51 (included in Table~\ref{tab2}),  
the significant absorption was found in two objects
for the former and in three objects for the latter in addition to
marginal detections in a few more objects.
The $\alphaobs/e$ for \ion{Sm}{2}~10083.34 was {$\sim$}2.5 for
the coolest objects, HD\,52005, and marginal, 1--2,
in a few objects. However, this line is significantly blended
with \ion{Cr}{1}~10083.18, and the $\Teff$ trend of the $\alphaobs$
is similar to that of the contaminating line.
We could not confirm the presence of this \ion{Sm}{2} line.
We also found that $\alphaobs/e$ for \ion{Sm}{2}~9788.96
is not zero for some objects; 
larger than 2 for HD\,20902 and 1--2 for 
HD\,194093 and HD\,204867.
In this case, however, the dependency of the absorption on $\Teff$ and $\log g$
is inconsistent with the prediction based on the synthetic spectra.
In addition, this line is located 
where the telluric absorption is relatively strong
\citep{Sameshima-2018a}.
In this paper, we conclude that we detected the two lines,
\ion{Sm}{2}~9850.67 and 9936.51,
but further confirmation and characterization
based on spectra of higher resolution and/or higher quality
are desired for all the four candidate lines.

\item 
Eu (Europium, $Z=63$) --- 
We investigated five \ion{Eu}{2} lines using the observed spectra.
The one at 10019.52\,{\AA} is listed in MB99.
We detected this line in addition to two new lines at 9898.30 and 10165.56\,{\AA}.
For all of them,
the detection is clearest in HD\,194093, which is
enhanced in rare-earth elements as mentioned in Section~\ref{sec:data-obs}.
We also found that the three lines are significant or at least marginally
significant in HD\,204867, while they were marginally detected
in a few more objects.
For \ion{Eu}{2}~10019.52,
$\alphaobs/e$ are approximately 8 and 4 for HD\,194093 and HD\,204867, respectively.
The {$\log gf$} in MB99 seems to predict the observed depths slightly better
than the values in KURZ and VALD.
The synthetic spectra predict that
\ion{Fe}{1}~11019.79 and \ion{Si}{1}~10020.07 similarly contaminate
the \ion{Eu}{2} line for the two objects. While the \ion{Fe}{1} line
is included in both VALD and MB99 and its EP and {$\log gf$} are 
consistent in the two lists, the \ion{Si}{1} line is not included in MB99.
The latter is much weaker than suggested with the VALD list even if it exists,
and this explains the inconsistency between the observed and synthetic spectra
presented in Figure~\ref{fig3} (see the panel for \ion{Eu}{2}~10019.52 available as the online material).
For the other two lines detected, 9898.30 and 10165.56\,{\AA},
there are telluric lines that may disturb the detection of the \ion{Eu}{2} lines.
We examined the spectra before and after the telluric correction together with
the telluric spectra and also the target spectra of individual exposures,
and they suggest that the \ion{Eu}{2} lines are real. 
The other two lines at 10034.22 and 10142.99\,{\AA} were not detected.

\item 
Dy (Dysprosium, $Z=66$) --- 
We investigated four \ion{Dy}{2} lines,
listed in VALD only, using the observed spectra.
None of them are found in MB99 although three are located at $\lambda > 10000$\,{\AA}.
The KURZ list does not include those lines either.
We clearly detected the line at 10523.39\,{\AA} in several objects
although the overall fits between the observed and synthetic spectra tend to be poor
in the surrounding wavelength range. 
For this line, the {$\log gf$} in VALD predicts reasonable depths
that are comparable with the measured.
We found clear absorption at the wavelength of \ion{Dy}{2}~10305.36
in many objects. The absorption line looks reasonably isolated except
the objects with the lowest $\Teff$. However, 
the measured depths are significantly larger than predicted
by {$\sim$}0.8\,dex in the logarithm of depth.
Moreover, the temperature dependency of the depths does not follow
the prediction well. In fact, the trend of $\alphaobs^*$ is similar
to some of the unidentified lines, discussed in Section~\ref{sec:unknown},
that get stronger towards the lower temperature, while
\ion{Dy}{2} should get strongest at around 5000~K. 
Although this \ion{Dy}{2} line may exist, there seems to be an unidentified line,
at almost the same wavelength,
which makes it hard to confirm the \ion{Dy}{2} line.
We cannot include this line in the list of confirmed lines.
The other two lines, \ion{Dy}{2}~9763.05 and 10835.94, were not detected.

\item 
Er (Erbium, $Z=68$) --- 
We investigated one \ion{Er}{2} line using the observed spectra.
This line at 11059.56\,{\AA} is not listed in KURZ or MB99.
According to the prediction based on the synthetic spectra,
the line is shallow across the temperature range we investigated.
The depth may be as deep as {$\sim$}0.03 only in stars with
$\Teff \sim 5000$\,K and $\log g \lesssim 0.5$,
but such objects were not included in this study
and we could not detect this line in any object.

\end{itemize}

\section{Detection of unidentified lines} \label{sec:unknown}
\begin{figure*}
\includegraphics[clip,width=0.98\hsize]{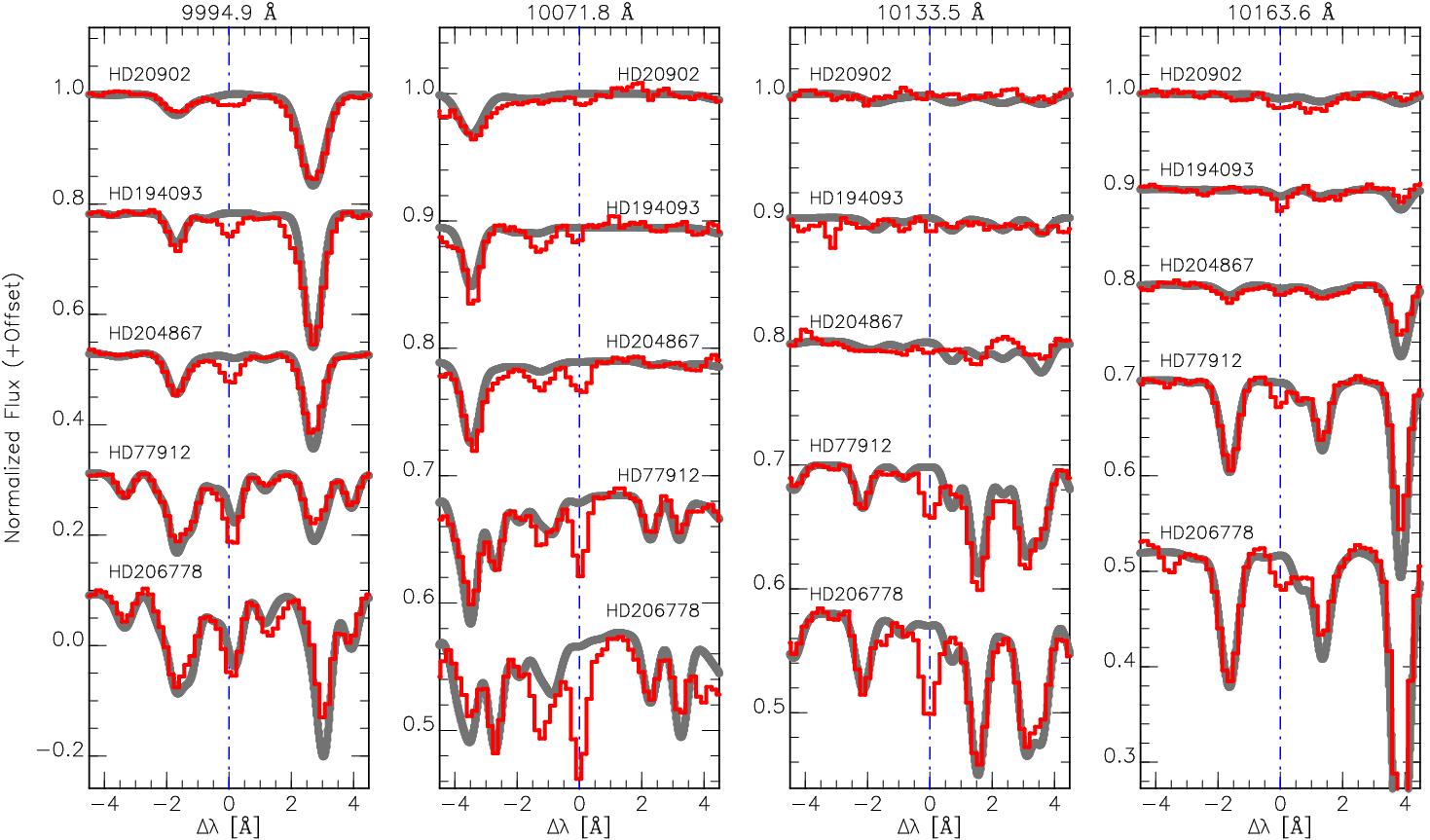}
\caption{
Absorption lines that were not predicted in the synthetic spectra
(Table~\ref{tab3}).
The red and gray curves indicate the observed and synthetic spectra, respectively.
The wavelength of each line is indicated by the vertical line
and labeled at the top of each panel.
The spectra of five stars with different $\Teff$ (increasing from top to bottom)
are presented, even if the given line is not seen in each star,
to show its dependence on $\Teff$.
The complete figure set (45 panels) is
available in the online journal.
\label{fig5}}
\end{figure*}

Some lines are clearly significant in the observed spectra
but not found in the synthetic spectra.
Table~\ref{tab3} lists such unidentified lines
whose depths are sufficiently large,
{$\sim$}0.05 or more, at least in a few objects and show systematic trends with $\Teff$.
Spectra of five objects, regardless of the significance in each object,
around the wavelengths of the unidentified lines are presented in Figure~\ref{fig5}
with which one can trace the $\Teff$ trend of
each line between 4000 and 7200\,K.
The KURZ list was used for the synthetic spectra in Figure~\ref{fig5},
but none of the KURZ, VALD, and MB lists has the unidentified lines.
Many of the lines get deeper towards the lower (or higher) end
of the $\Teff$ range we investigated, but some lines show a peak
within the temperature range of our targets
or a rather flat trend (e.g., 9994.9 and 12571.1\,{\AA}).

\input{table3.tex}

Table~\ref{tab3} does not include lines which appear both in the observed and synthetic spectra
consistently even if the predicted depths are significantly shallower than
those in the observed.
For example, the observed spectra show a clear line at
10516.1\,{\AA} which is almost invisible in the spectra synthesized
with KURZ or VALD.
This line is probably \ion{Ca}{1}~10516.14 (EP$=$4.74\,eV, $\log gf=-0.52$\,dex)
as listed in MB99.
VALD and KURZ also list this line but give
a significantly smaller {$\log gf$} value, $-1.438$\,dex.
If we use the {$\log gf$} in MB99 for the spectral synthesis,
the line observed is explained reasonably well.
Therefore, we do not include this line in Table~\ref{tab3}.
In fact, inconsistency in line strengths between the observed and synthetic spectra
is found in many lines, e.g., \ion{Mg}{1}~12456.935 and
a pair of \ion{N}{1}~12461.25 and CN~12460.83 that give
different contribution according to $\Teff$,
showing much stronger absorption
in the observed; they are outside the scope of this paper and
not included in Table~\ref{tab3}.

As long as one of the three lists (KURZ, VALD, and MB99) gives good identification
of the observed feature, Table~\ref{tab3} does not include such a line even if
the other two lists cannot explain the feature.
For example, we found that \ion{Fe}{1}~10452.75 (EP=3.88\,eV)
is listed in MB99 but not in KURZ and VALD.
While warmer stars ($\Teff \gtrsim 5000$\,K) show a \ion{C}{1} line
listed in all the line lists
almost at the same wavelength,
the absorption in low-$\Teff$ stars requires
the \ion{Fe}{1} line in MB99.
Moreover, this line was included
in the abundance measurements by \citet{Kondo-2019} and
gave [Fe/H] values within 0.03\,dex of the final results based on
dozens of \ion{Fe}{1} lines.
In the case of \ion{S}{1}~10635.97 listed in MB99, in contrast,
KURZ and VALD list this line with the wavelength of
10633.08\,{\AA}. 
The observed spectra show the absorption at around the wavelength given in MB99.
The EP and {$\log gf$} of this line are consistent 
in all the three lists, and the predicted depths as a function of $\Teff$
agree with those observed. 

For some of the unidentified lines in Table~\ref{tab3},
different lines at almost the same wavelength were confirmed
in both observed and synthetic ones at a particular $\Teff$ range
but the observed spectra show unexpected absorption with
a completely different trend against the temperature, which indicates 
the presence of the unidentified lines. At around 10434.0\,{\AA}, for example,
a CN line is seen at lower temperatures,
$\Teff \lesssim 5500$\,K, and
this line appears in both the observed and synthetic spectra. However,
we found another line, which is the unidentified line we report,
growing with increasing $\Teff$.

Some lines in Table~\ref{tab3} seem to have counterparts
in some line lists but at different wavelengths
(and no list gives the observed wavelength).
In the following, we give remarks on such lines that seem to give
the corresponding absorption,
in the observed spectra, at shifted wavelengths
and show the consistent $\Teff$ trends as predicted.

10133.5\,{\AA} --- The \ion{Fe}{1} line listed in KURZ and VALD
with the wavelength of 10134.21\,{\AA} roughly follows
the $\Teff$ trend of the absorption at 10133.5\,{\AA}
in the observed spectra. The observed spectra do not show
the absorption at 10134.21\,{\AA}.
The predicted $\Teff$ trend of the \ion{Fe}{1} line
is slightly different by itself from the observed at
the lowest temperatures around 4000\,K, but this deviation may
be explained by the blend with \ion{Ti}{1}~10133.39
and some molecular lines.
No line is listed in MB99 around this wavelength.

10476.5\,{\AA} --- The observed $\Teff$ trend 
may be explained by a CN line.
There is, in fact, one CN line at 10475.295\,{\AA}, listed in KURZ and VALD,
which appears rather shallow in the observed spectra even if it exists.
Maybe the wavelength in VALD should be replaced by 10476.5\,{\AA} or, instead,
there is another CN line which gives the observed wavelength and strengths. 
No line is listed in MB99 around this wavelength.

12658.9\,{\AA} --- Interestingly, the feature predicted with  
two \ion{Ni}{1} lines at 12655.38 and 12655.60\,{\AA}
shows the $\Teff$ trend consistent with the observed,
but no significant absorption was observed in our 13 targets at the given wavelengths. 
The two \ion{Ni}{1} lines, present in both KURZ and VALD,
seem to give similar contributions to the feature,
although MB99 lists only one line at 12655.60\,{\AA}.

\newpage

\begin{center}
Here presented are individual plots for Figure~\ref{fig2} 
which are available as the online material. 
\end{center}
\begin{tabular}{cc}
\includegraphics[clip,width=0.43\hsize]{fig2a.eps}  &
\includegraphics[clip,width=0.43\hsize]{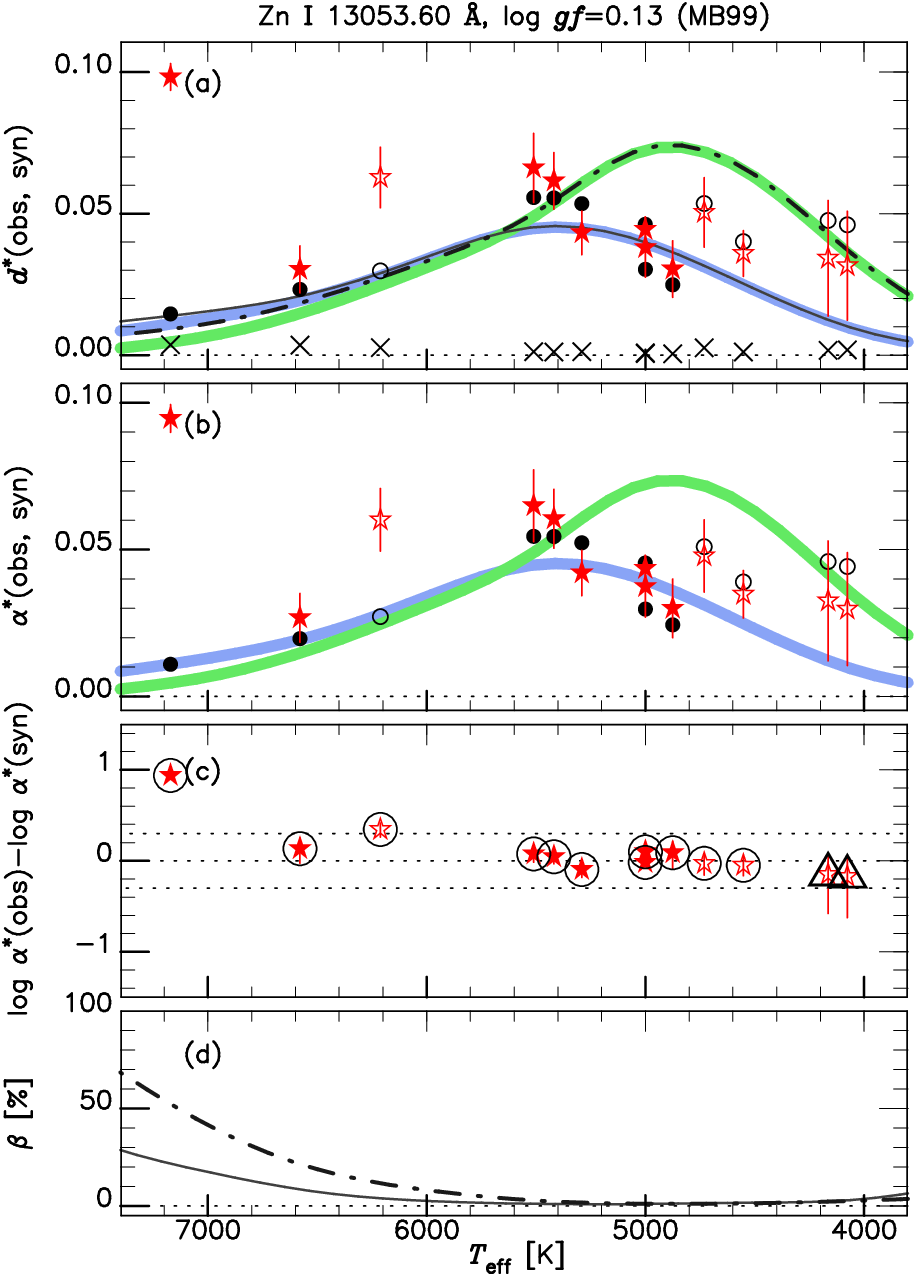}  \\
\includegraphics[clip,width=0.43\hsize]{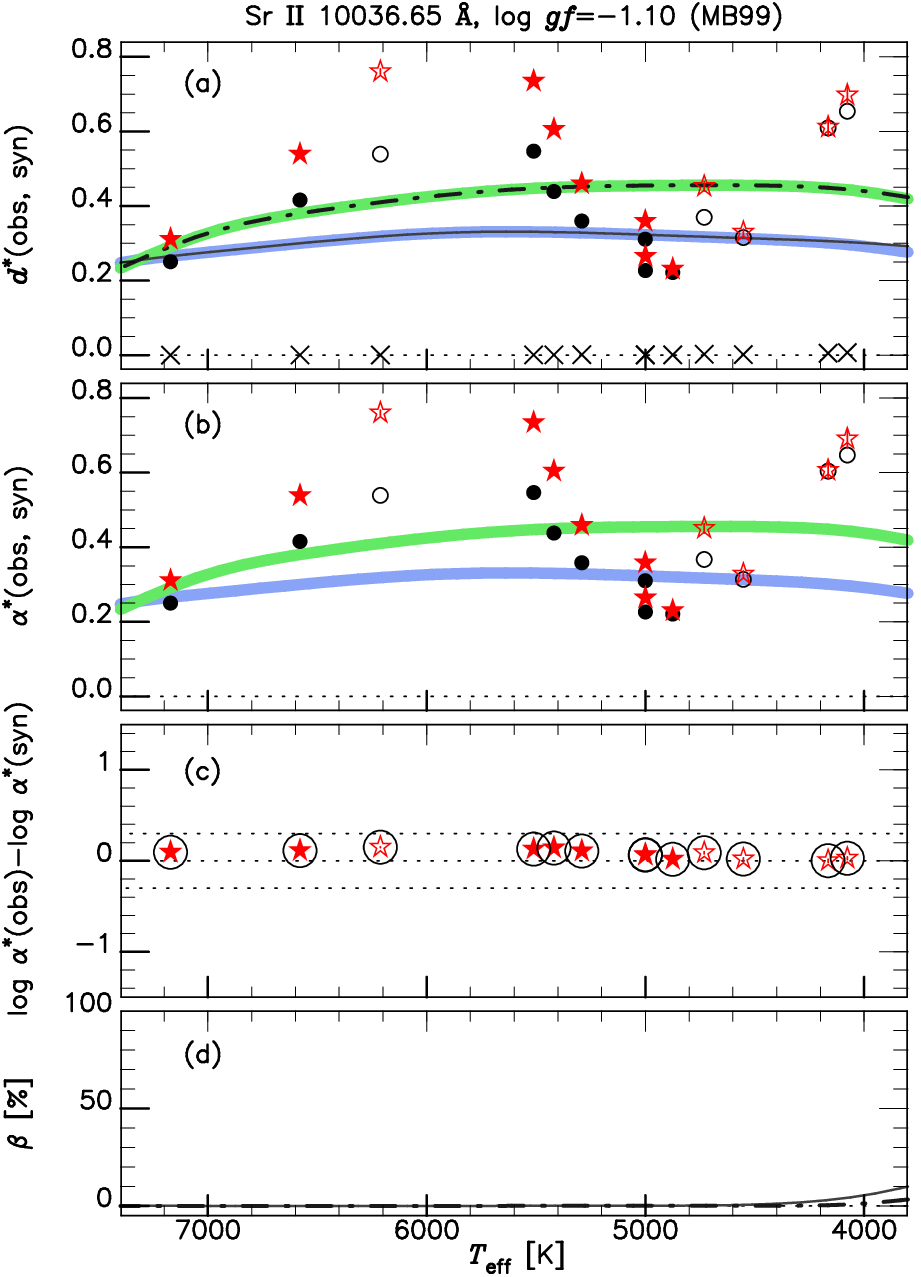}  &
\includegraphics[clip,width=0.43\hsize]{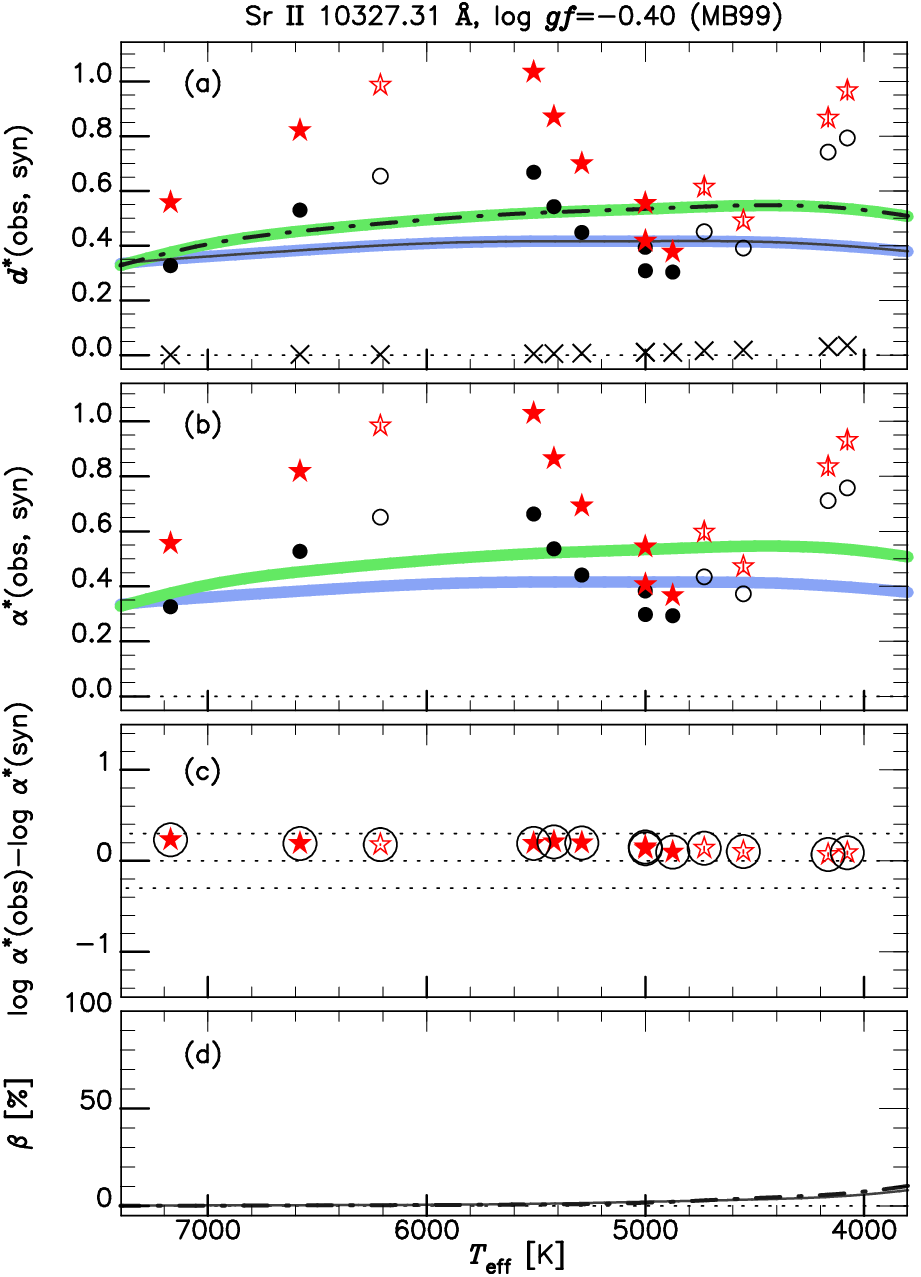}  
\end{tabular}

\newpage

\begin{center}
Here presented are individual plots for Figure~\ref{fig2} 
which are available as the online material. 
\end{center}
\begin{tabular}{cc}
\includegraphics[clip,width=0.43\hsize]{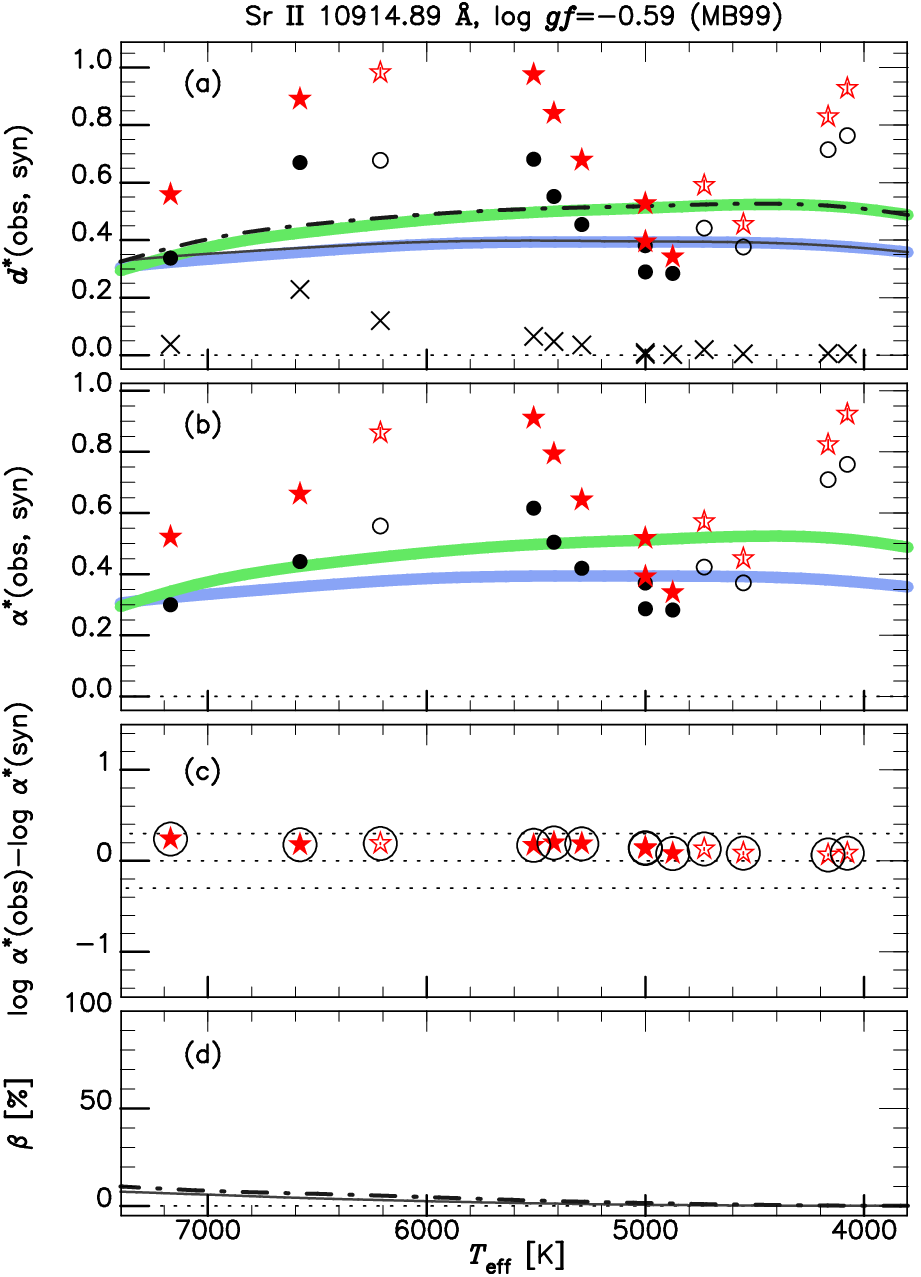}  &
\includegraphics[clip,width=0.43\hsize]{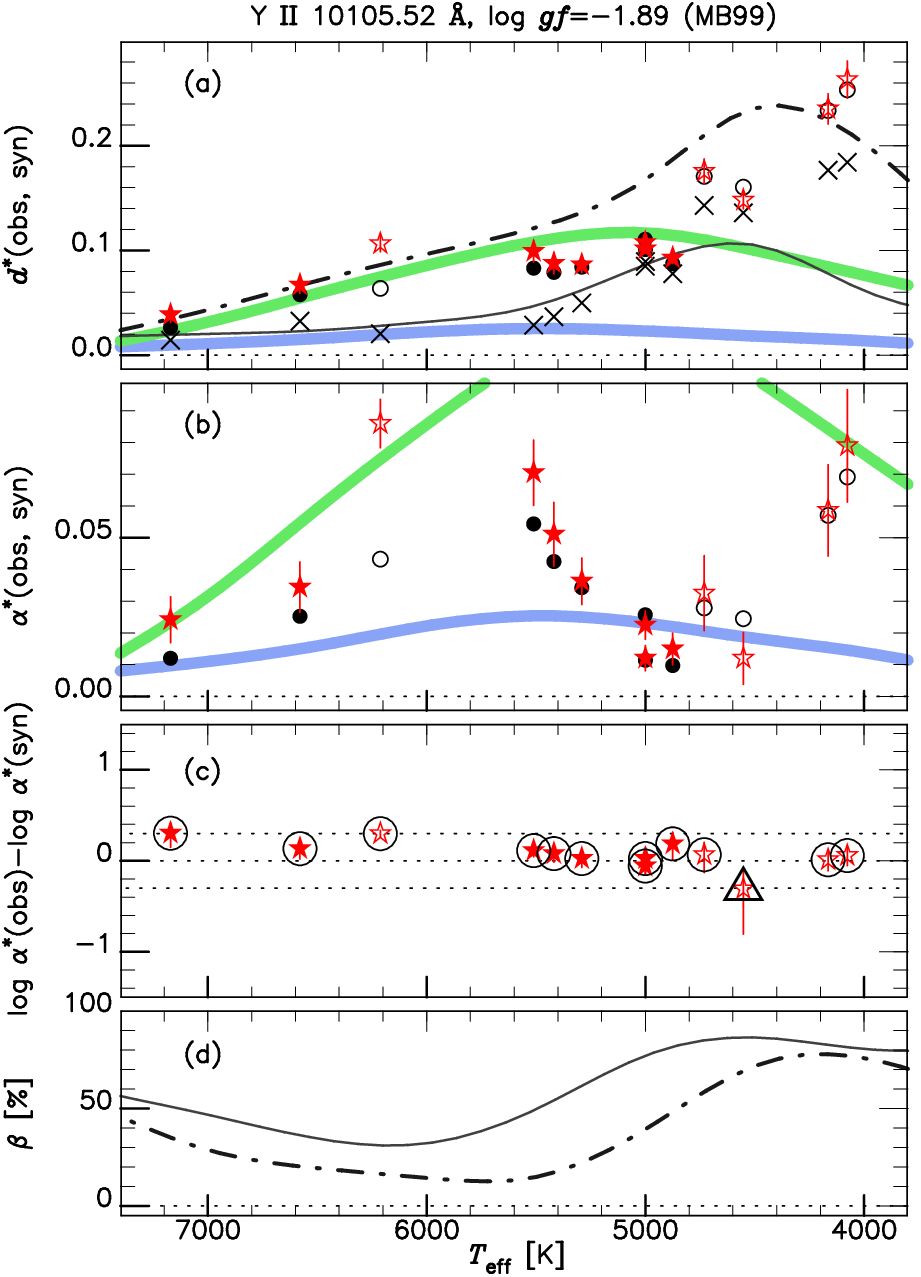}  \\
\includegraphics[clip,width=0.43\hsize]{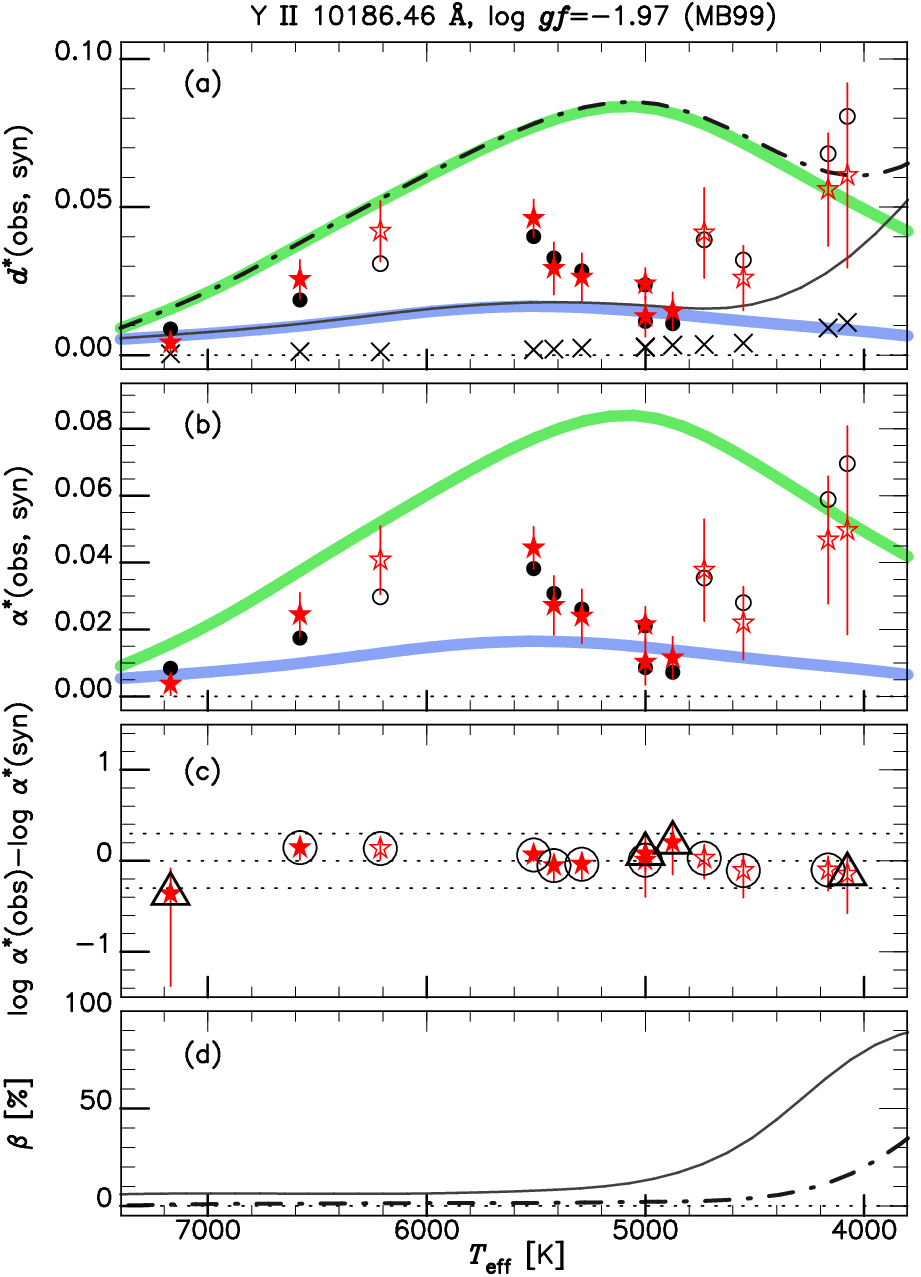}  &
\includegraphics[clip,width=0.43\hsize]{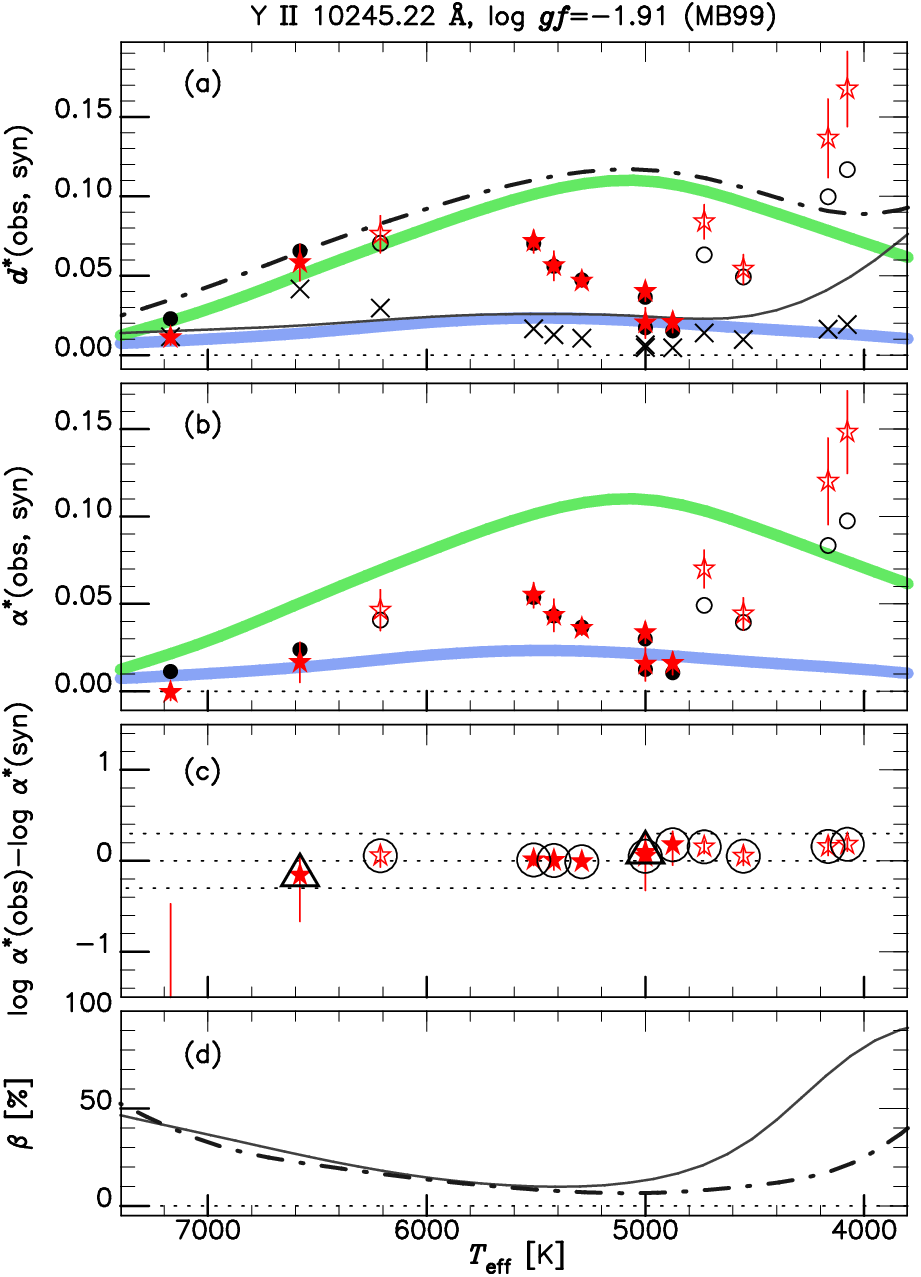}  
\end{tabular}

\newpage

\begin{center}
Here presented are individual plots for Figure~\ref{fig2} 
which are available as the online material. 
\end{center}
\begin{tabular}{cc}
\includegraphics[clip,width=0.43\hsize]{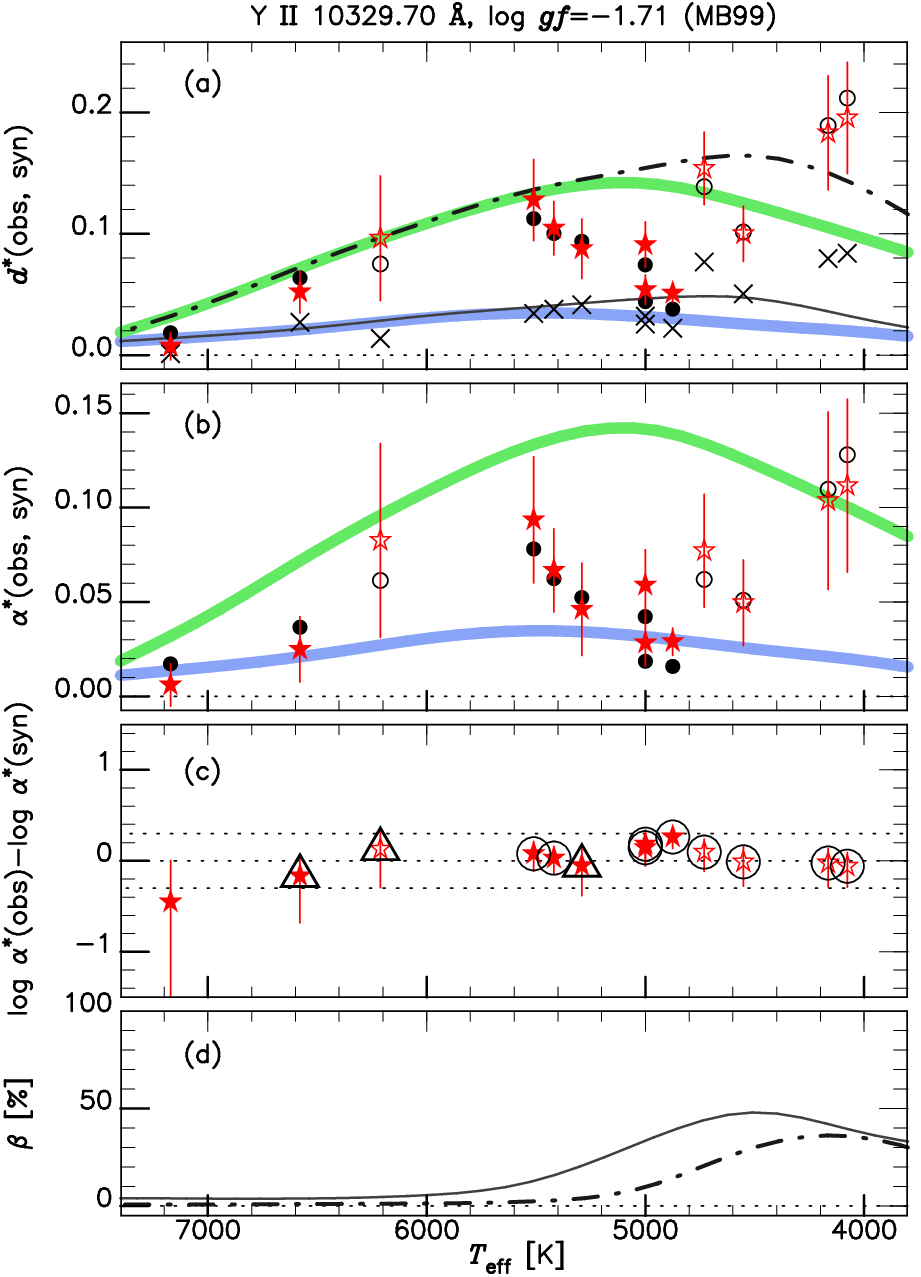}  &
\includegraphics[clip,width=0.43\hsize]{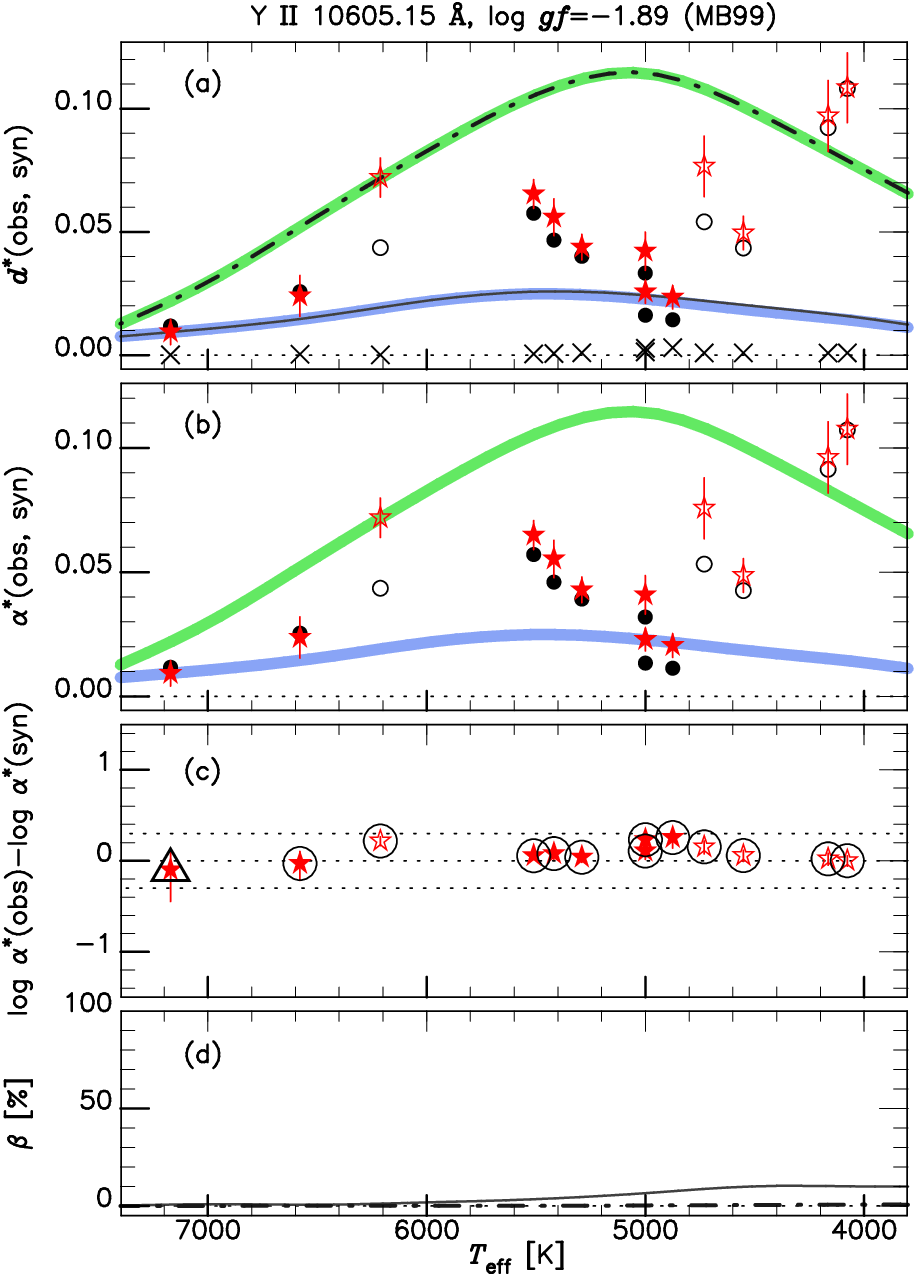}  \\
\includegraphics[clip,width=0.43\hsize]{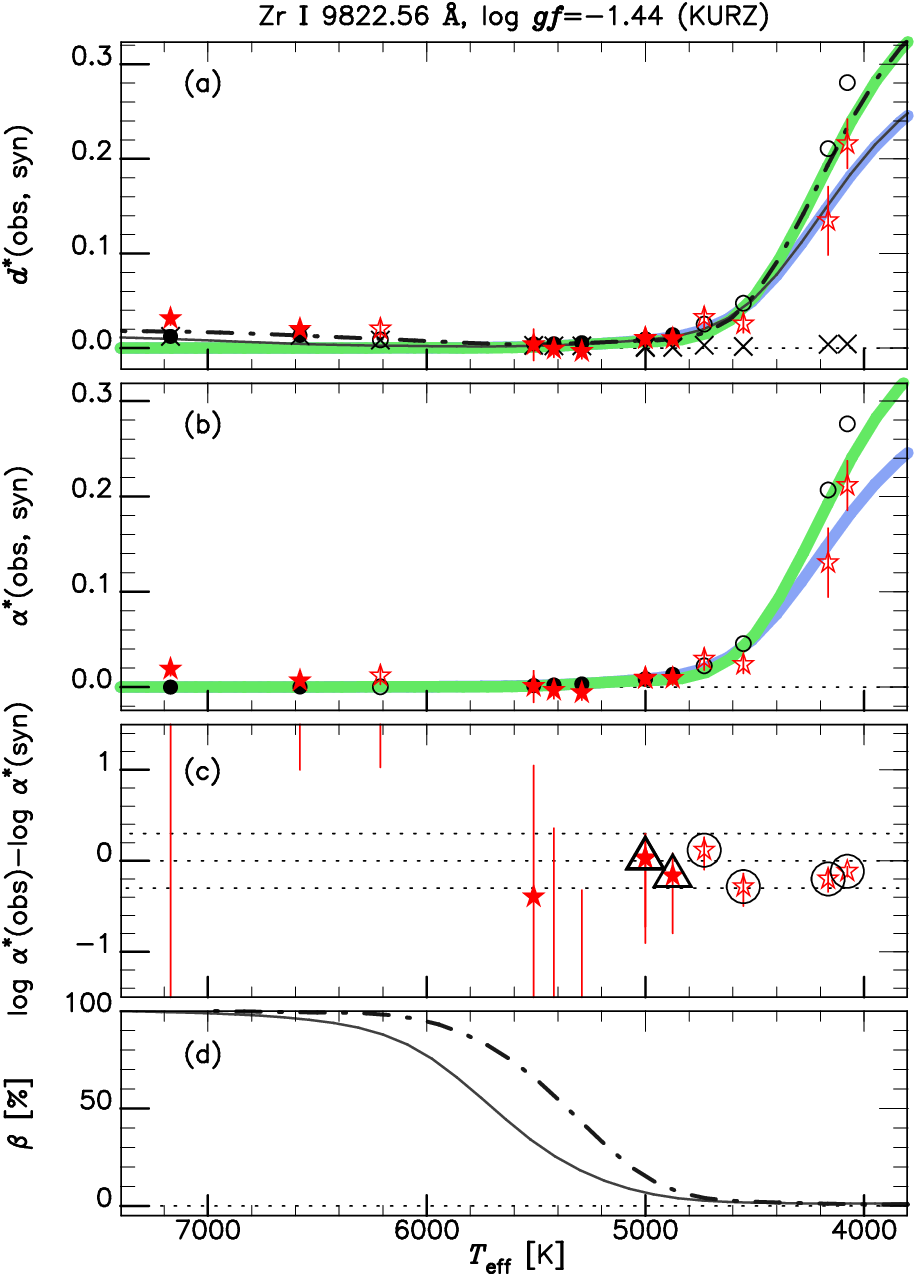}  &
\includegraphics[clip,width=0.43\hsize]{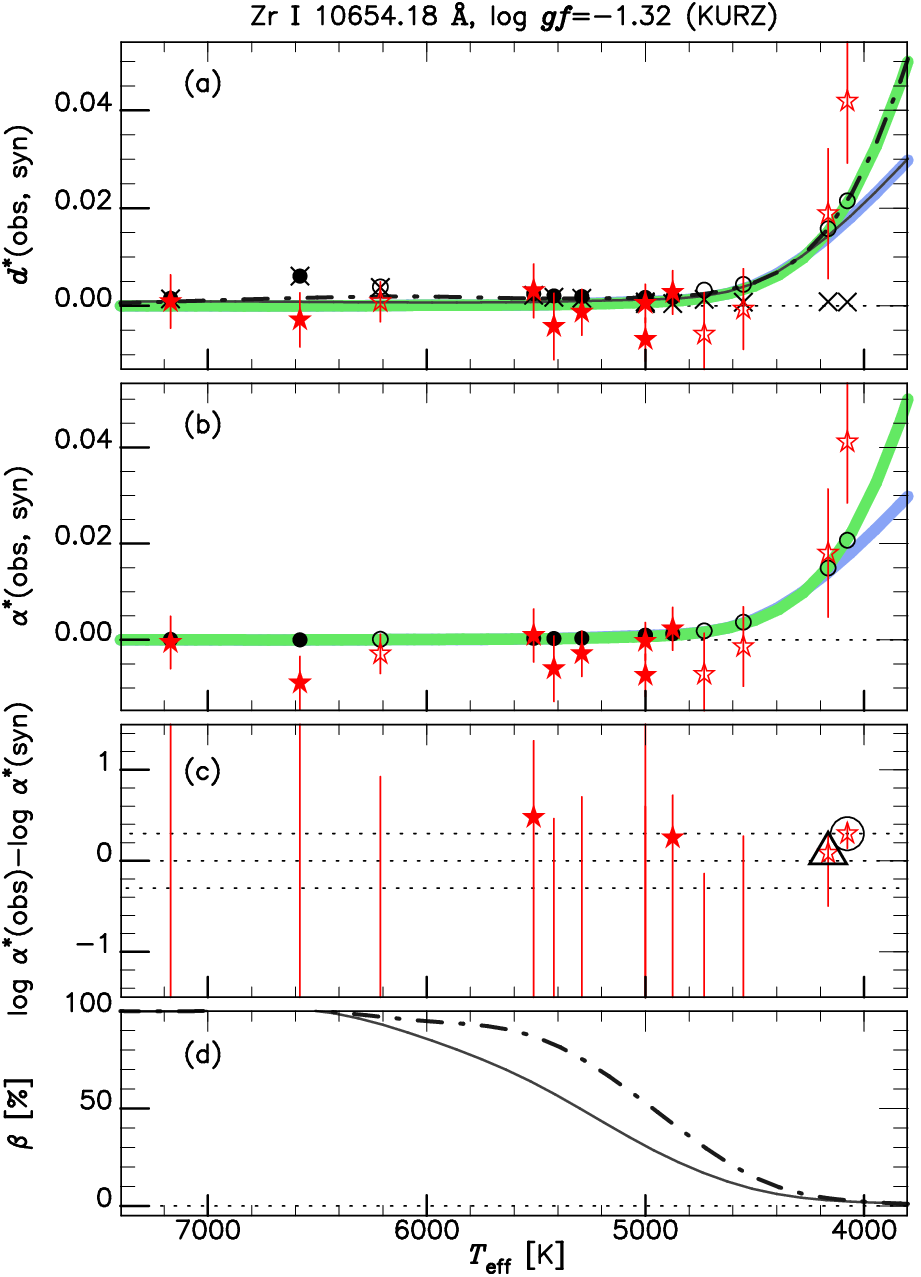}  
\end{tabular}

\newpage

\begin{center}
Here presented are individual plots for Figure~\ref{fig2} 
which are available as the online material. 
\end{center}
\begin{tabular}{cc}
\includegraphics[clip,width=0.43\hsize]{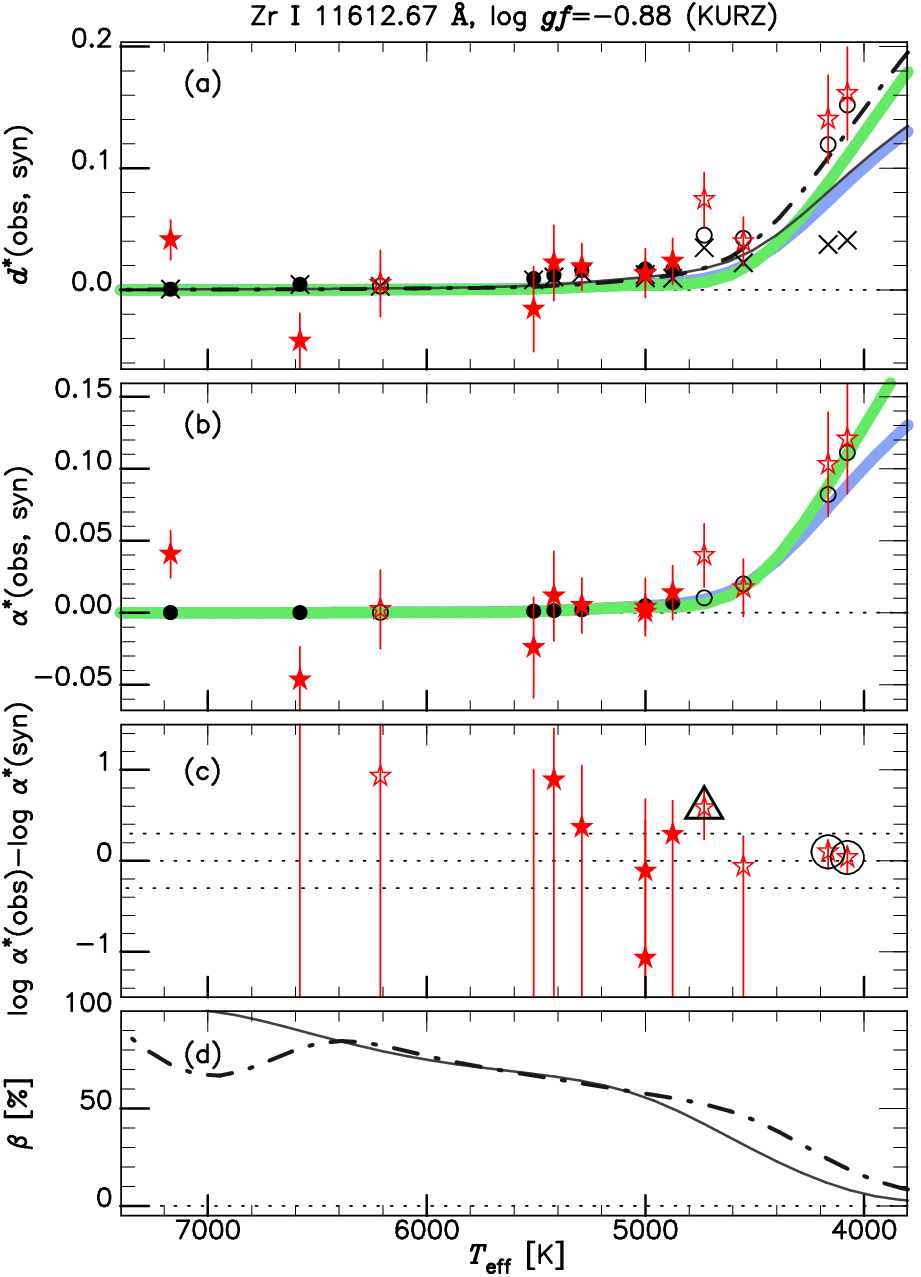}  &
\includegraphics[clip,width=0.43\hsize]{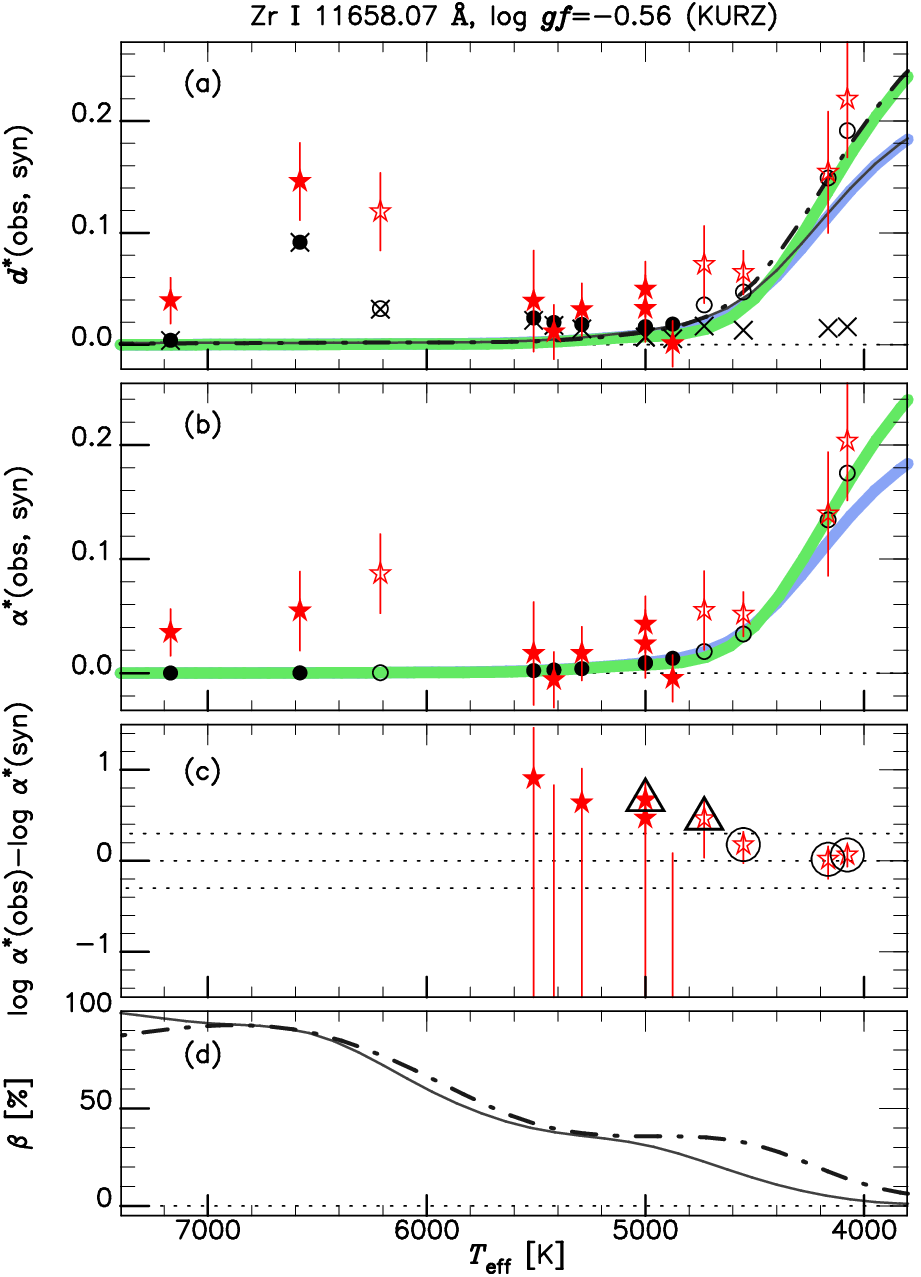}  \\
\includegraphics[clip,width=0.43\hsize]{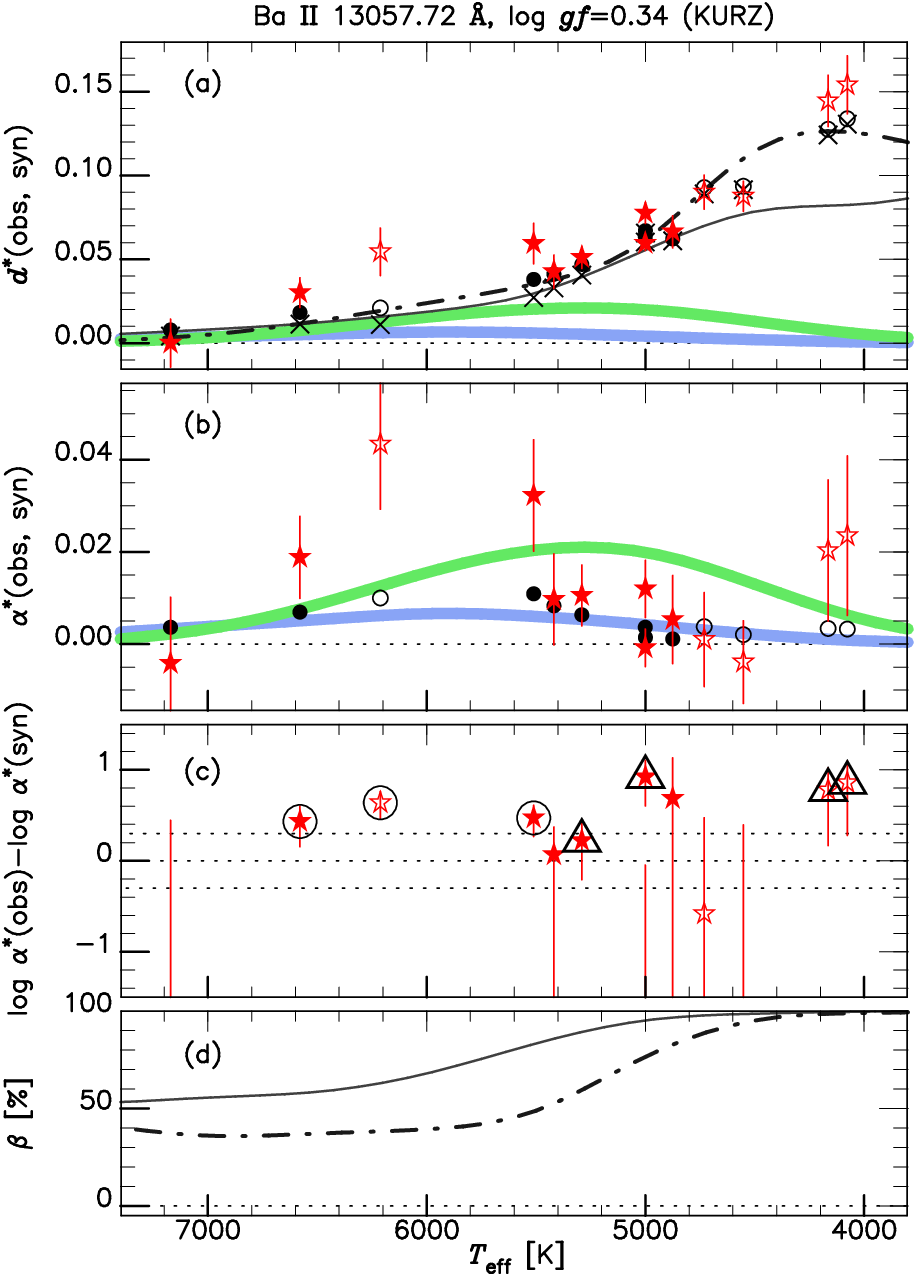}  &
\includegraphics[clip,width=0.43\hsize]{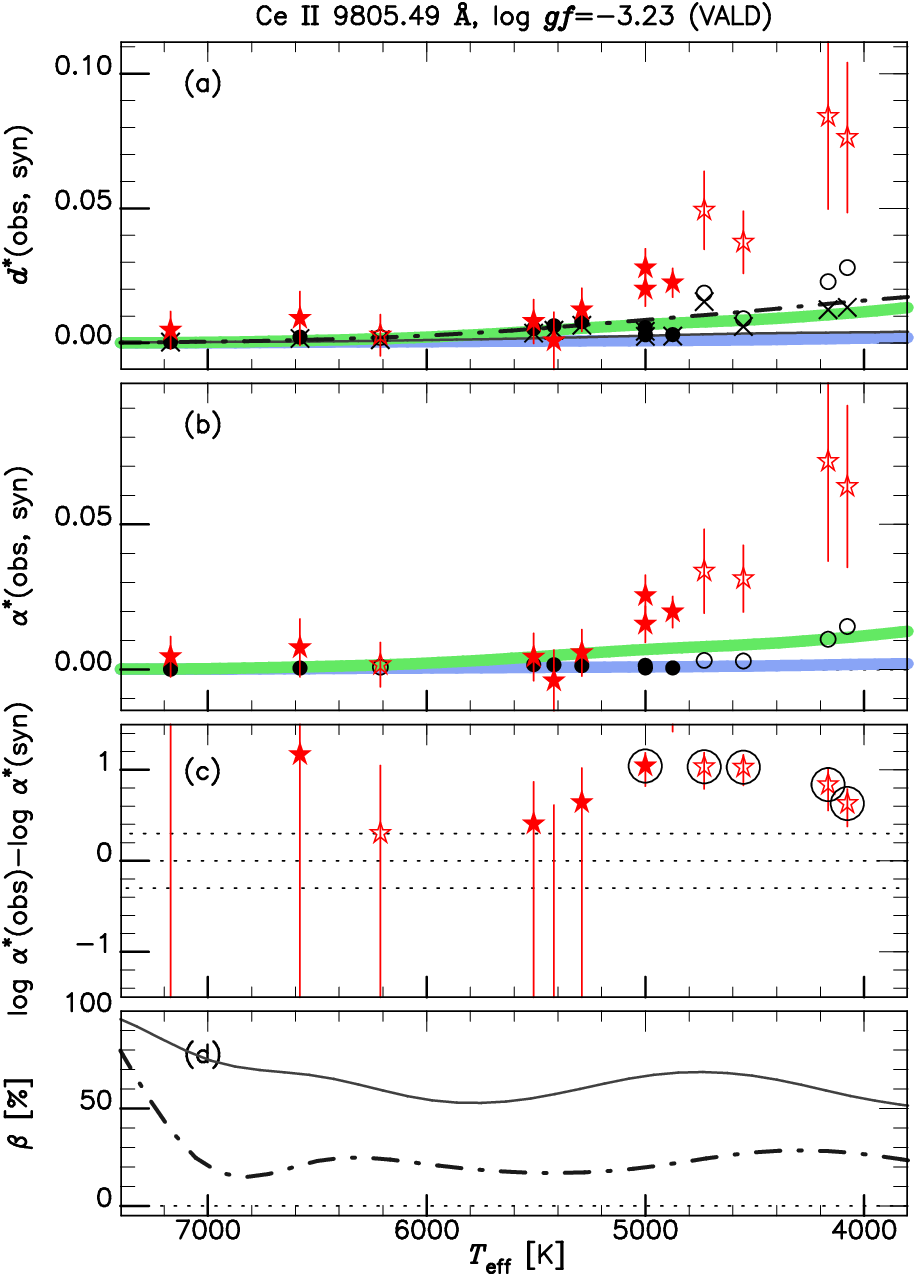}  
\end{tabular}

\newpage

\begin{center}
Here presented are individual plots for Figure~\ref{fig2} 
which are available as the online material. 
\end{center}
\begin{tabular}{cc}
\includegraphics[clip,width=0.43\hsize]{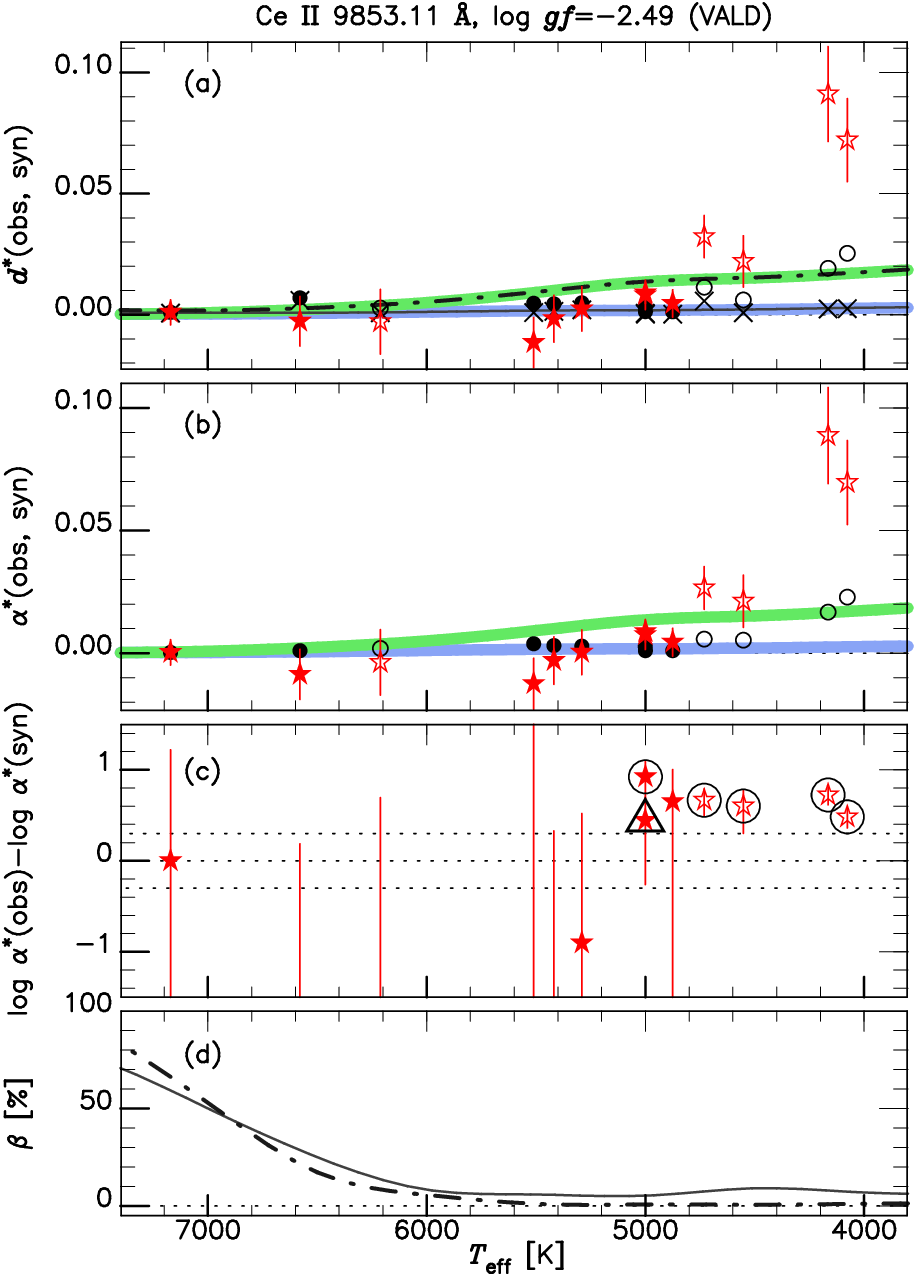}  &
\includegraphics[clip,width=0.43\hsize]{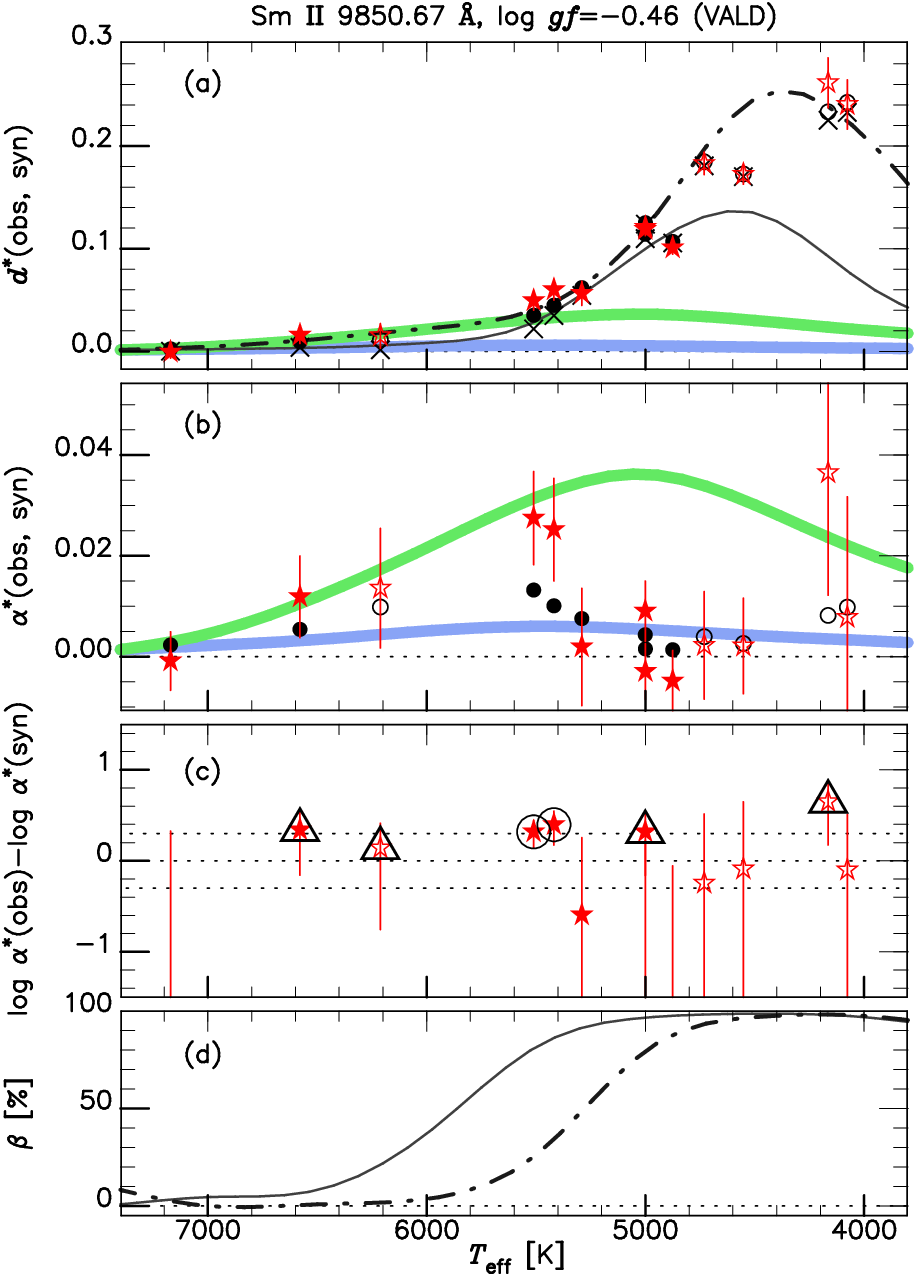}  \\
\includegraphics[clip,width=0.43\hsize]{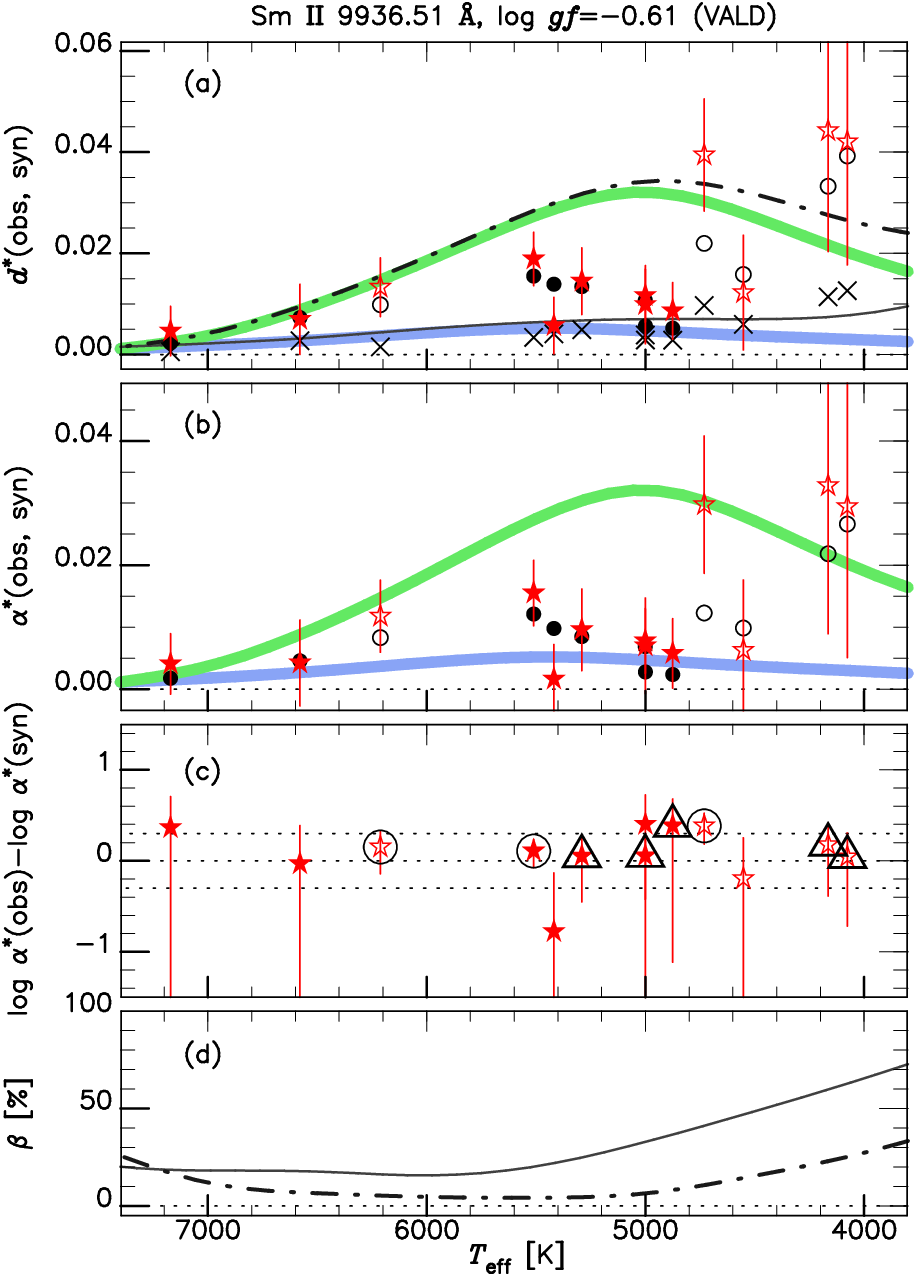}  &
\includegraphics[clip,width=0.43\hsize]{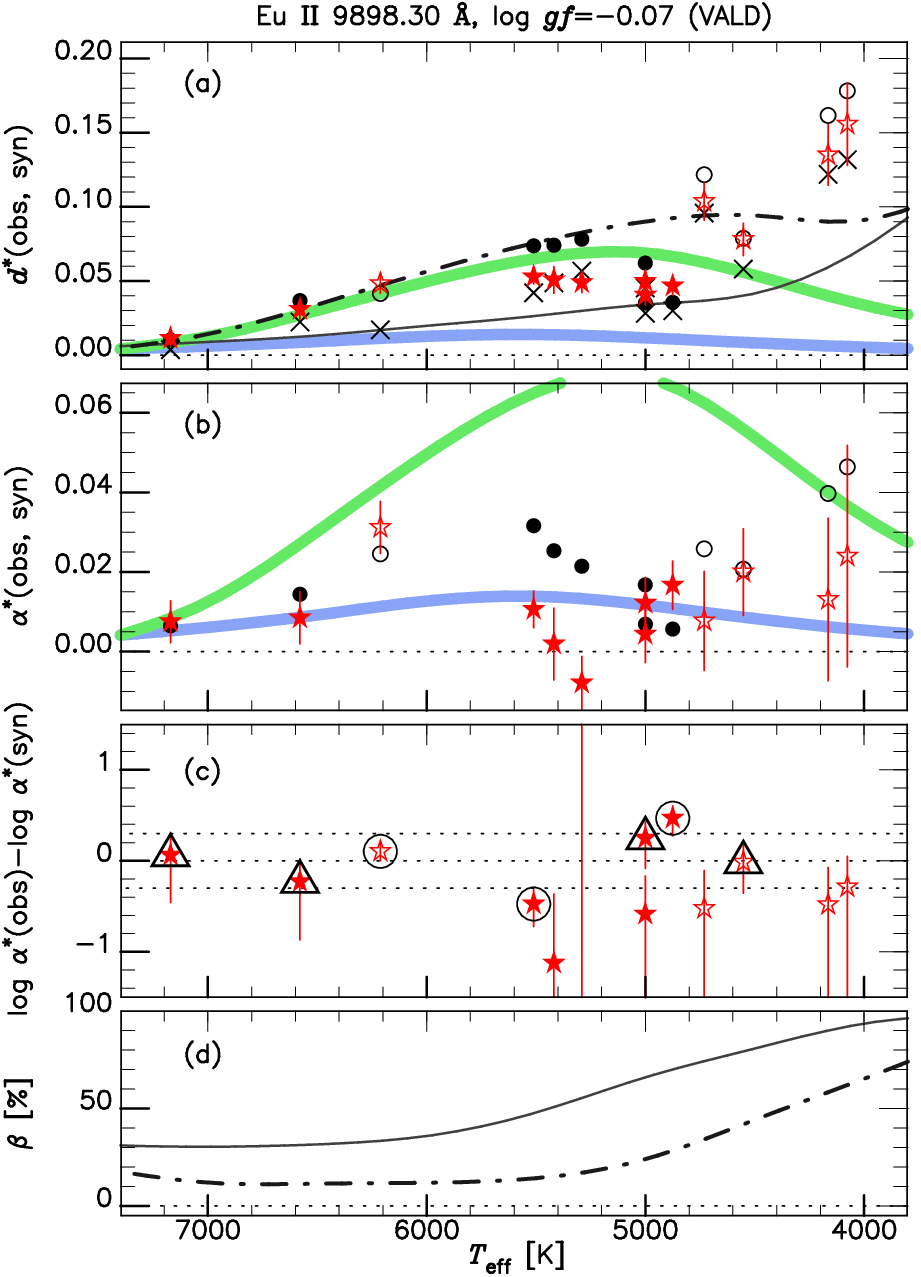}  
\end{tabular}

\newpage

\begin{center}
Here presented are individual plots for Figure~\ref{fig2} 
which are available as the online material. 
\end{center}
\begin{tabular}{cc}
\includegraphics[clip,width=0.43\hsize]{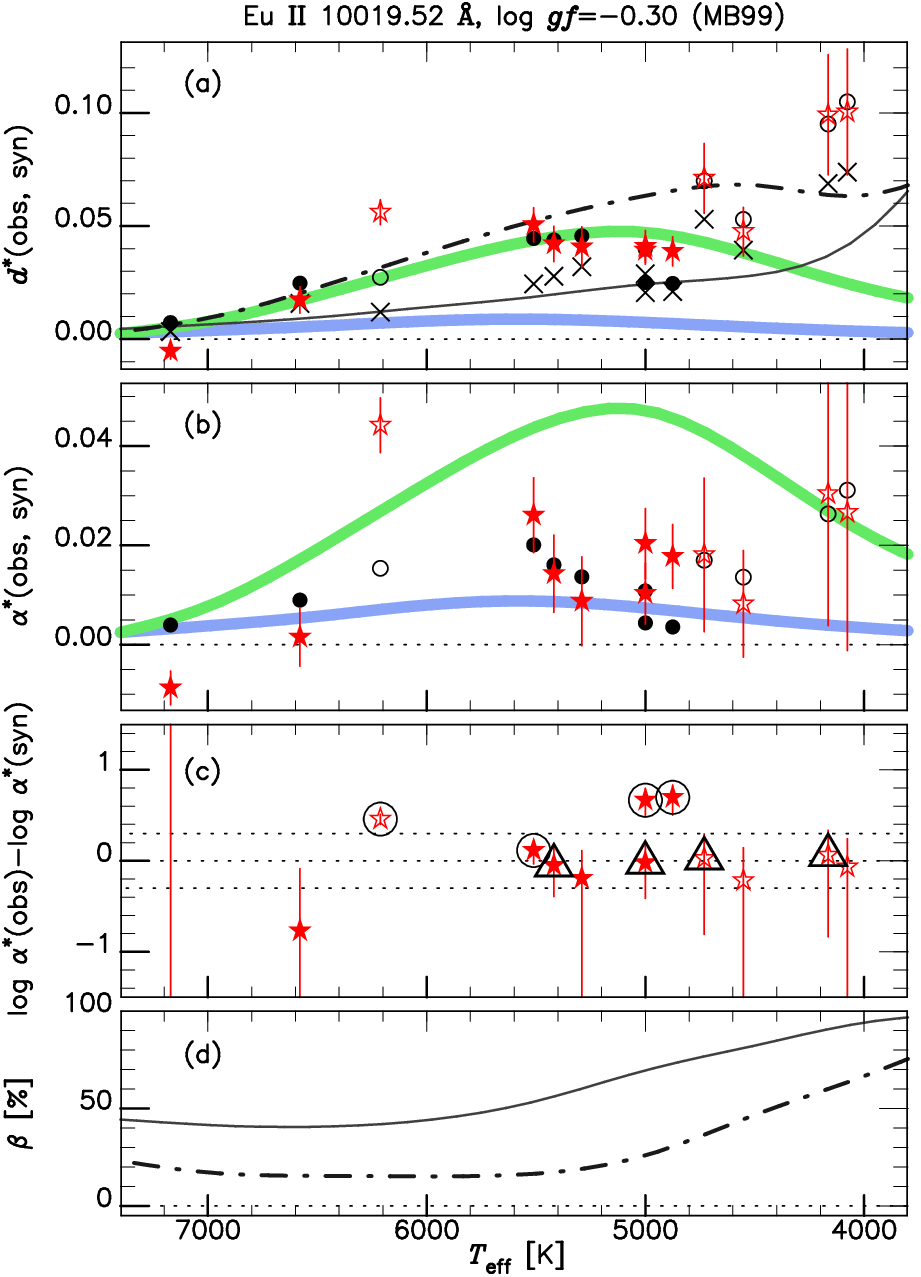}  &
\includegraphics[clip,width=0.43\hsize]{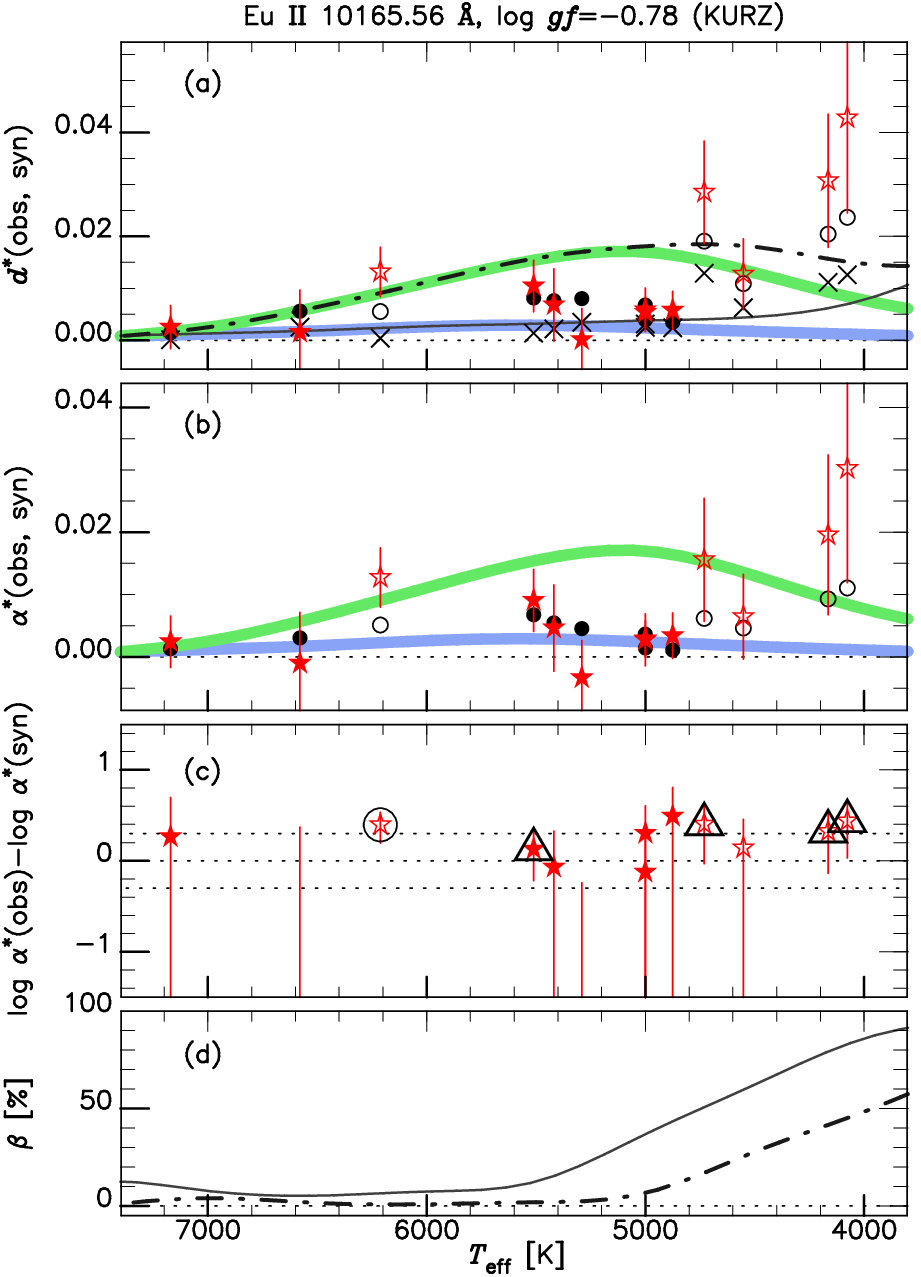}  \\
\includegraphics[clip,width=0.43\hsize]{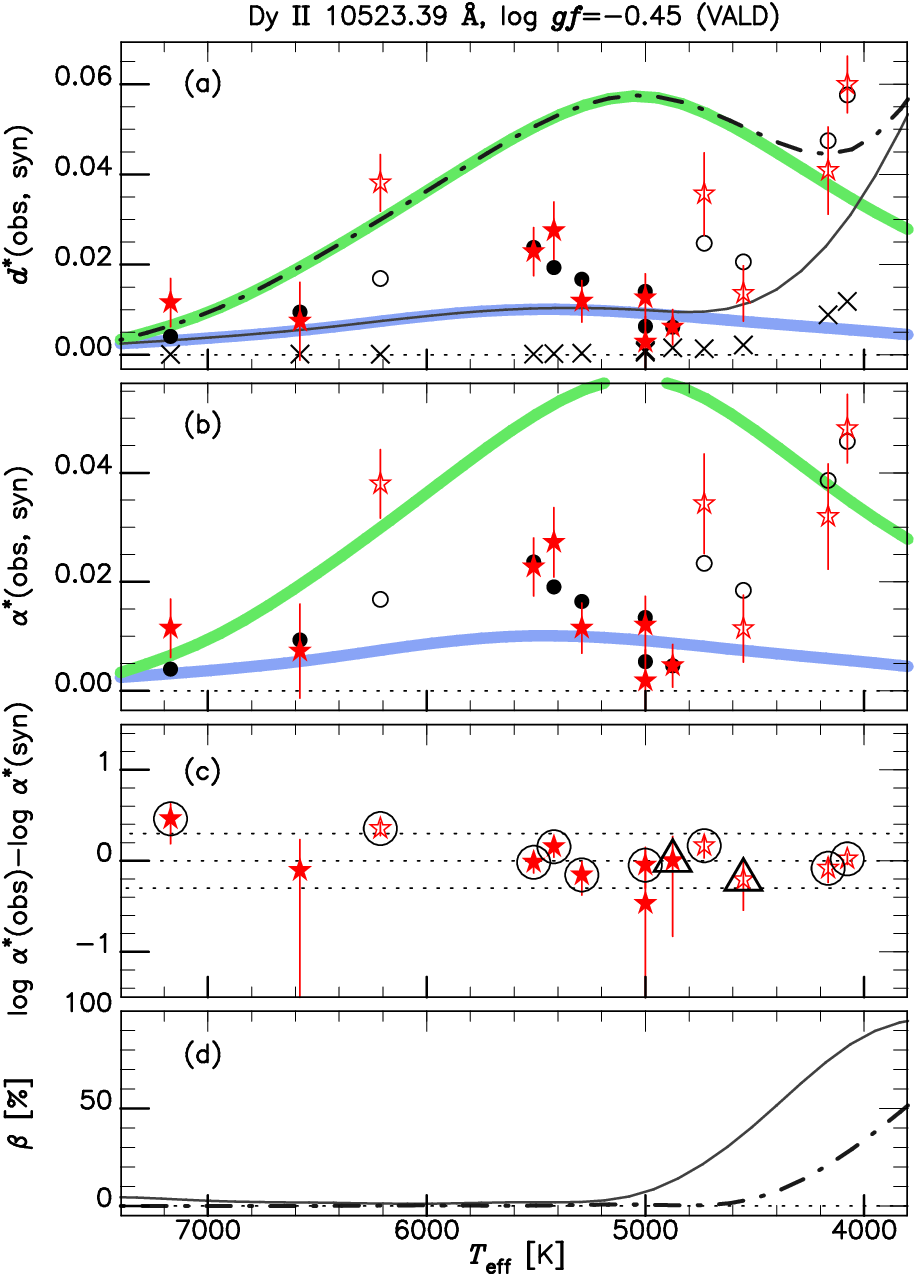}  &
\end{tabular}

\newpage

\begin{center}
Here presented are individual plots for Figure~\ref{fig3} 
which are available as the online material. 
\end{center}
\begin{tabular}{c}
\includegraphics[clip,width=0.98\hsize]{fig3a.eps}  \\
\includegraphics[clip,width=0.98\hsize]{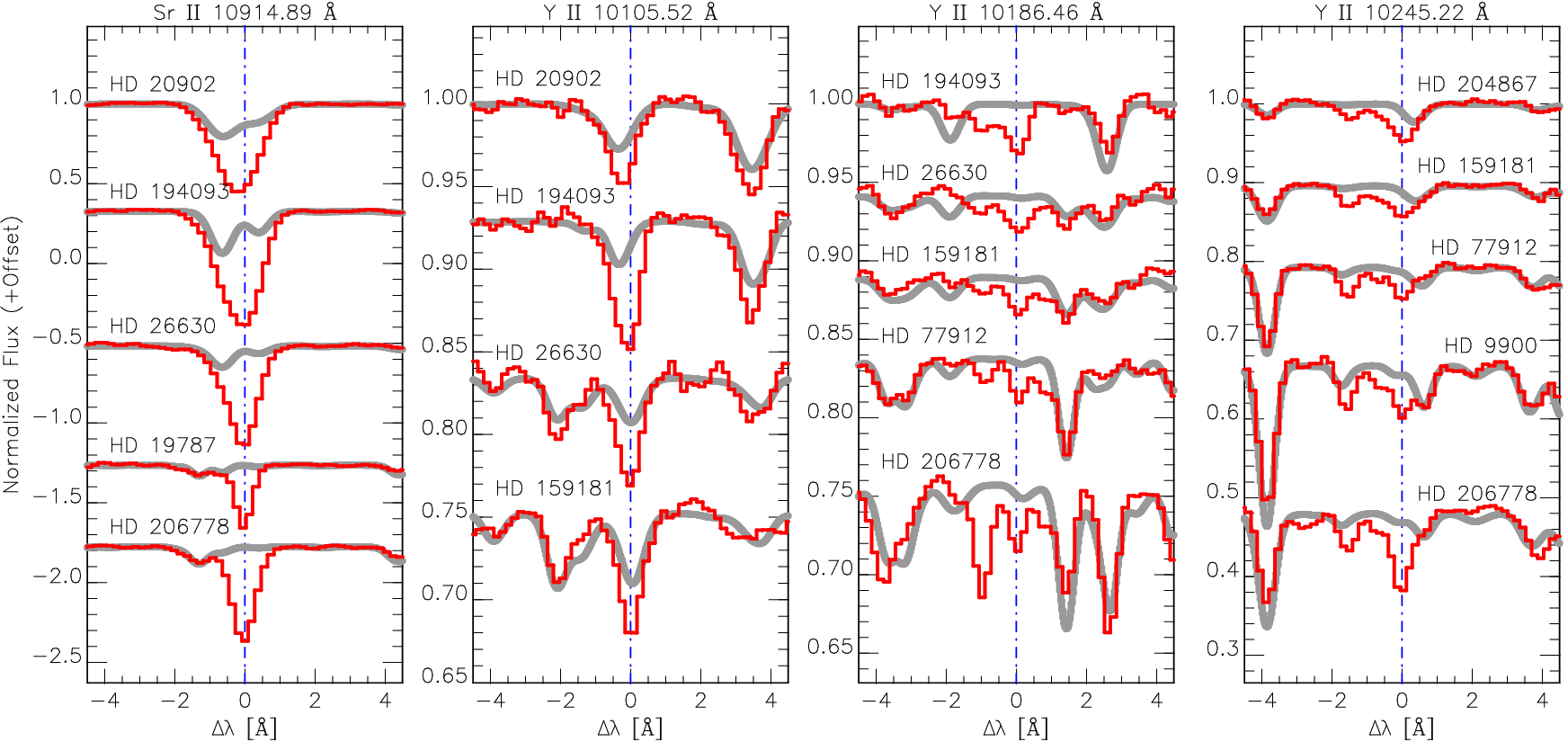} 
\end{tabular}

\newpage

\begin{center}
Here presented are individual plots for Figure~\ref{fig3} 
which are available as the online material. 
\end{center}
\begin{tabular}{c}
\includegraphics[clip,width=0.98\hsize]{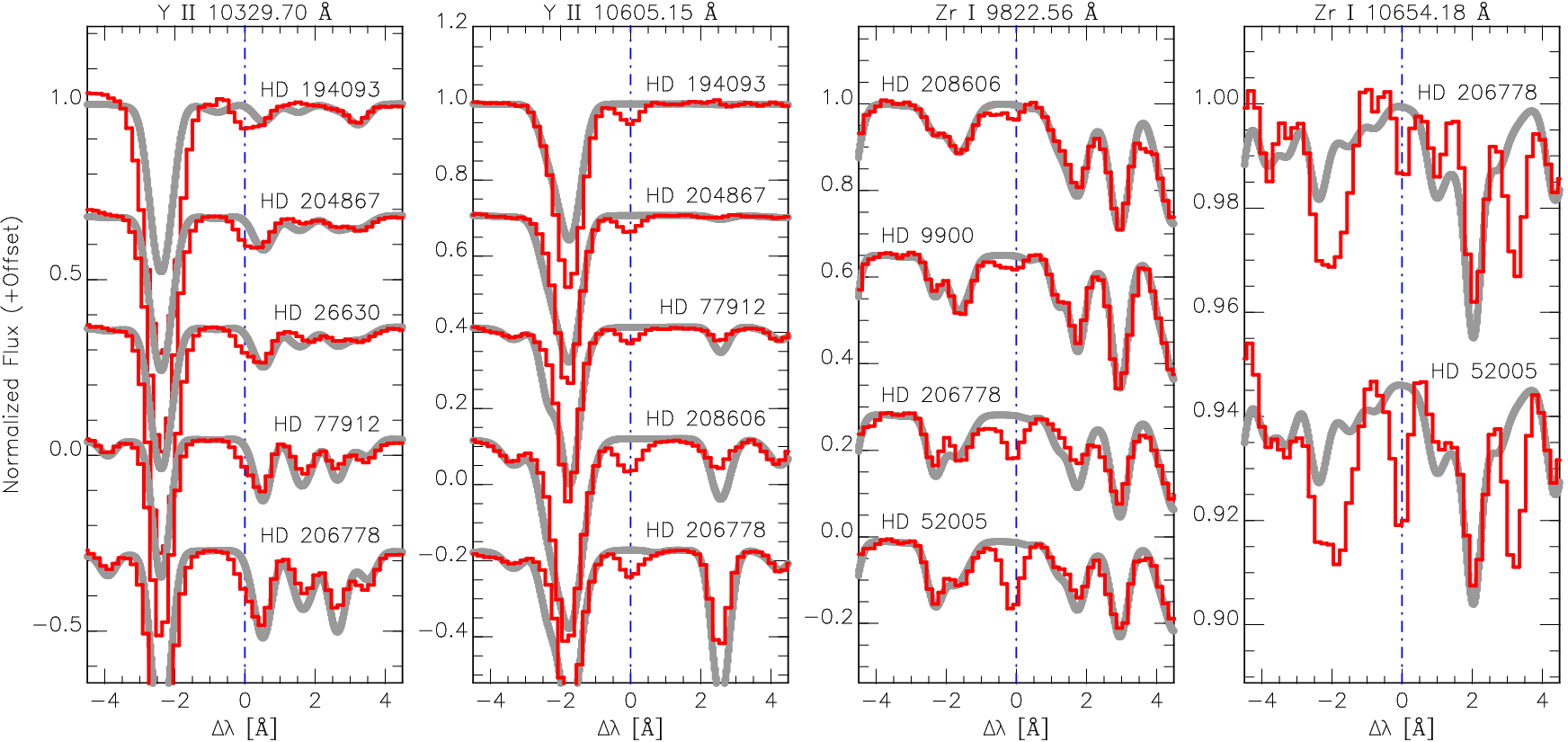}  \\
\includegraphics[clip,width=0.98\hsize]{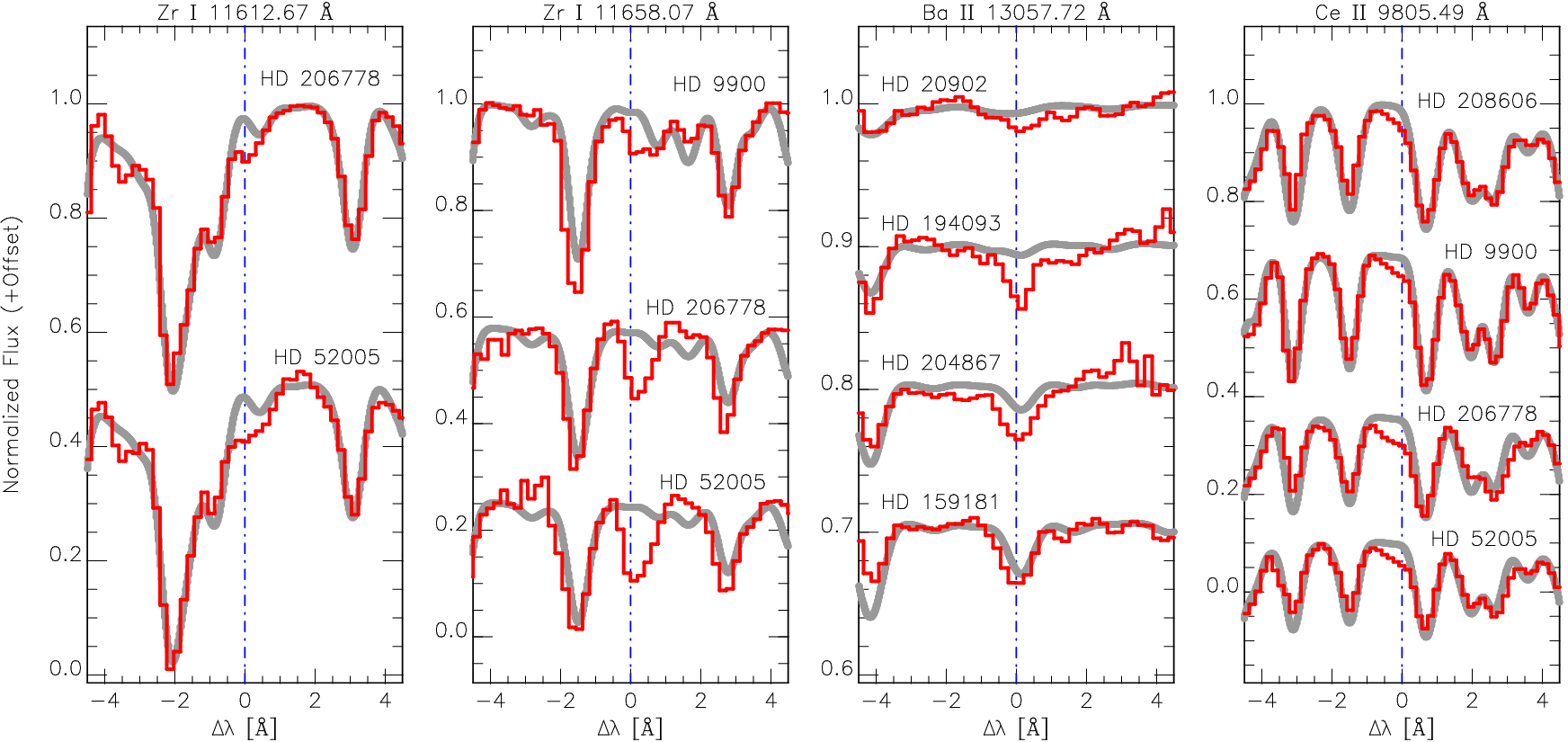} 
\end{tabular}

\newpage

\begin{center}
Here presented are individual plots for Figure~\ref{fig3} 
which are available as the online material. 
\end{center}
\begin{tabular}{c}
\includegraphics[clip,width=0.98\hsize]{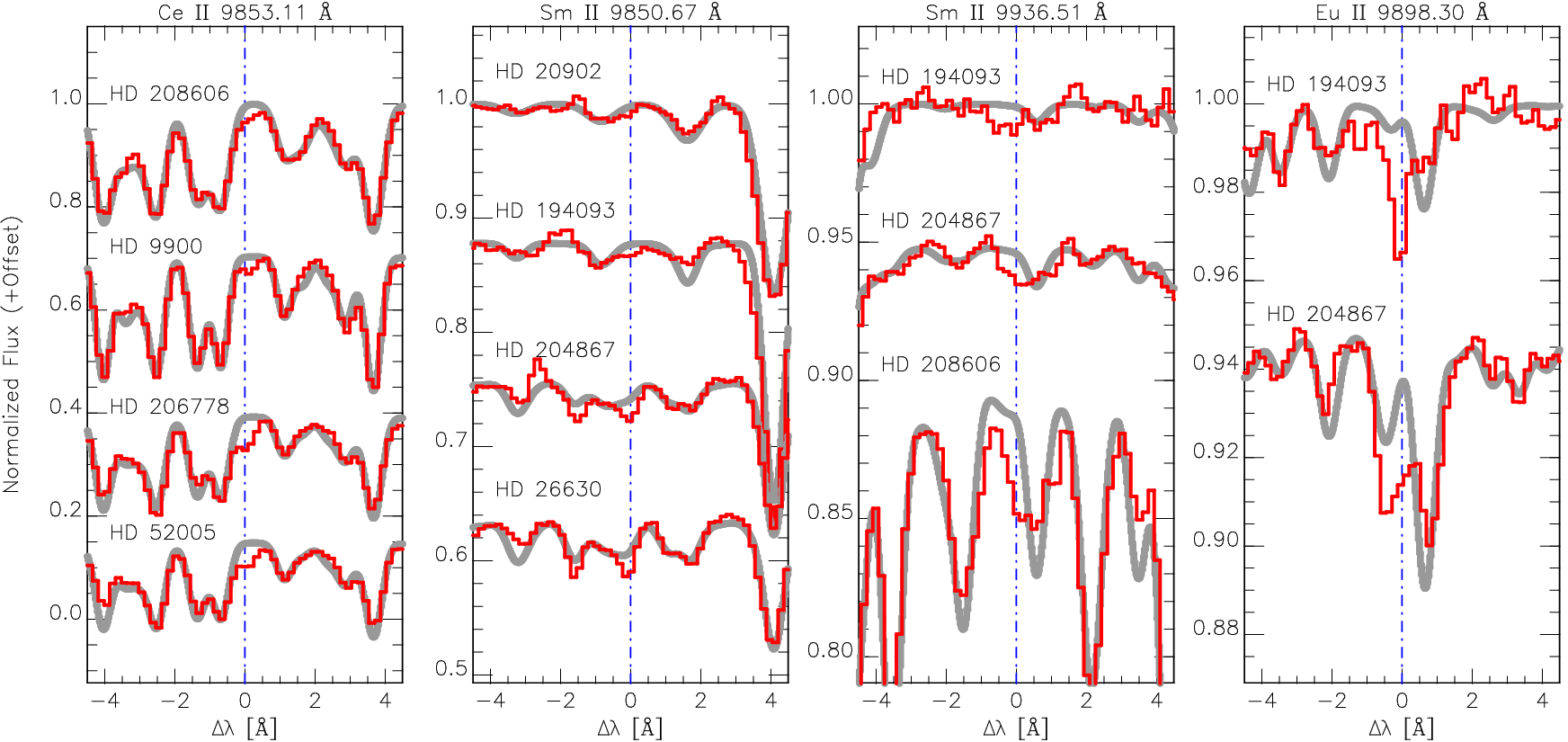}  \\
\includegraphics[clip,width=0.74\hsize]{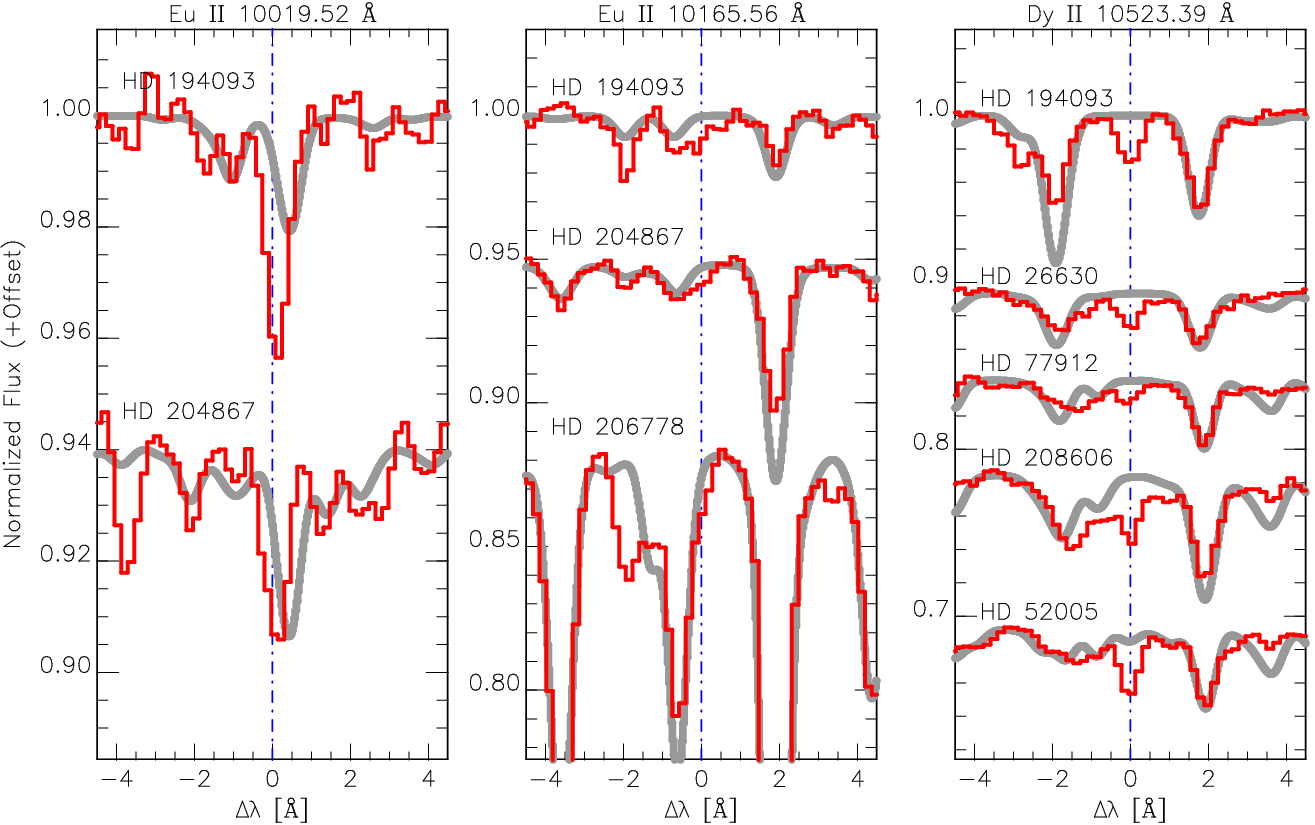} 
\end{tabular}

\newpage

\begin{center}
Here presented are individual plots for Figure~\ref{fig4} 
which are available as the online material. 
\end{center}
\begin{tabular}{cc}
\includegraphics[clip,width=0.42\hsize]{fig4a.eps} &
\includegraphics[clip,width=0.42\hsize]{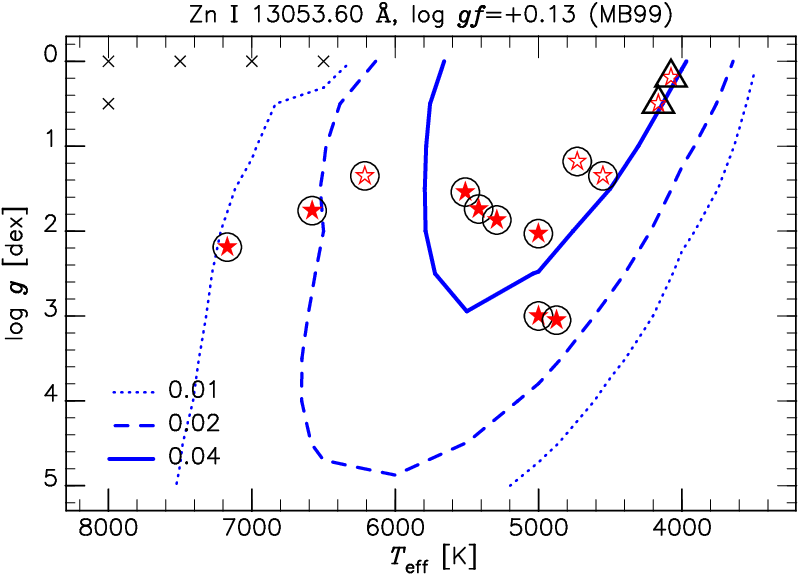} \\
\includegraphics[clip,width=0.42\hsize]{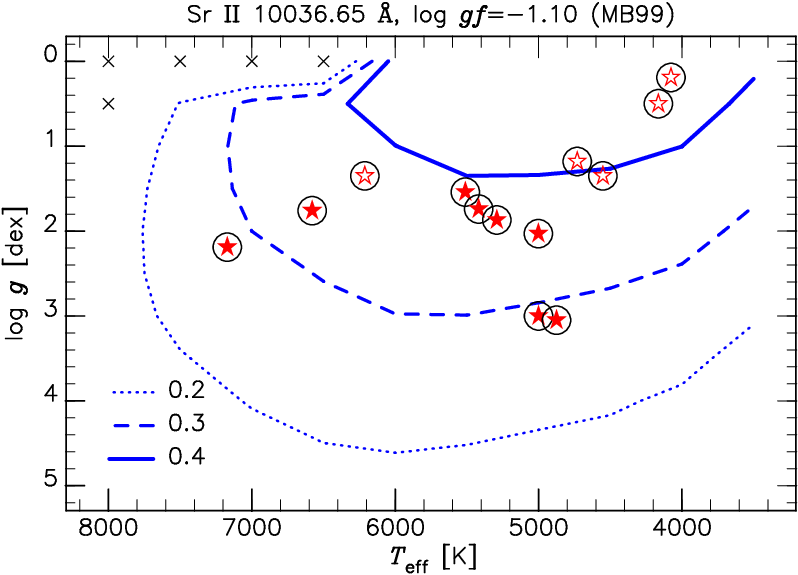} &
\includegraphics[clip,width=0.42\hsize]{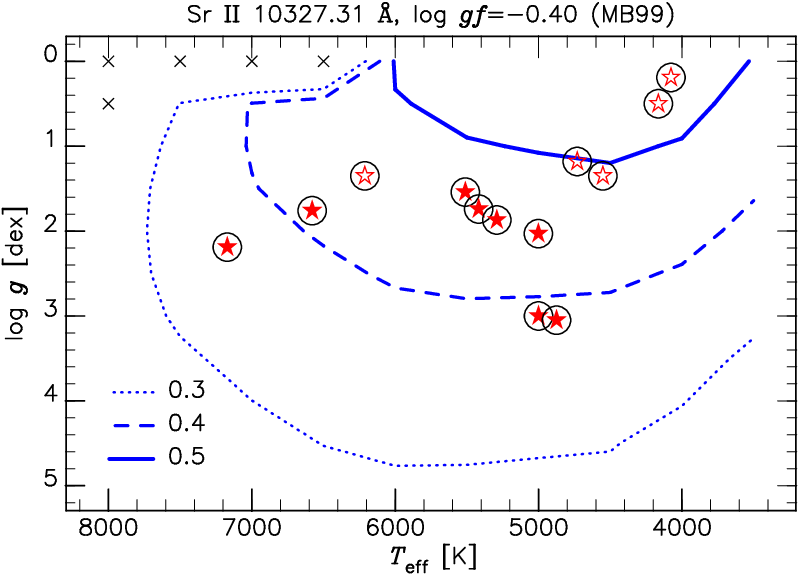} \\
\includegraphics[clip,width=0.42\hsize]{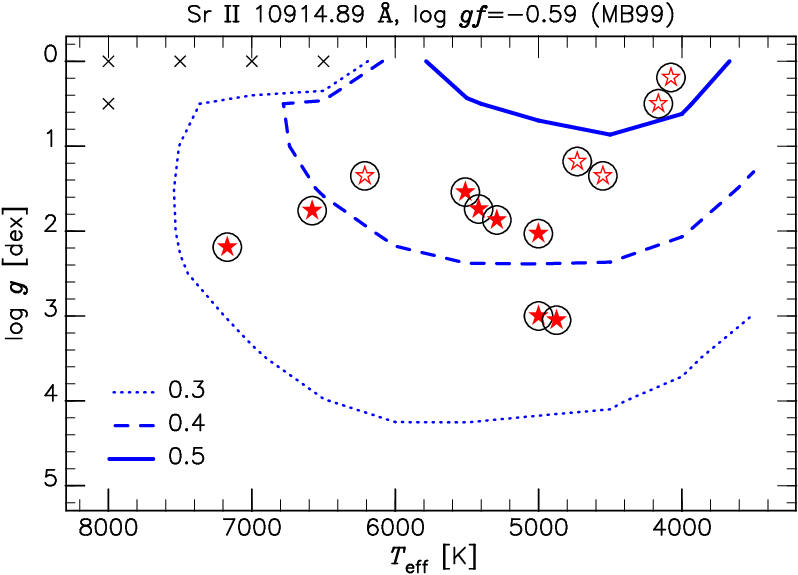} &
\includegraphics[clip,width=0.42\hsize]{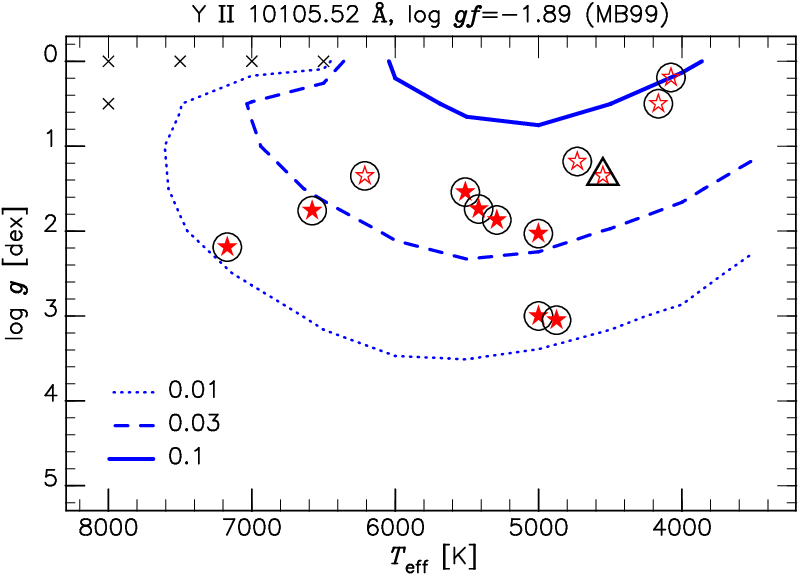} \\
\includegraphics[clip,width=0.42\hsize]{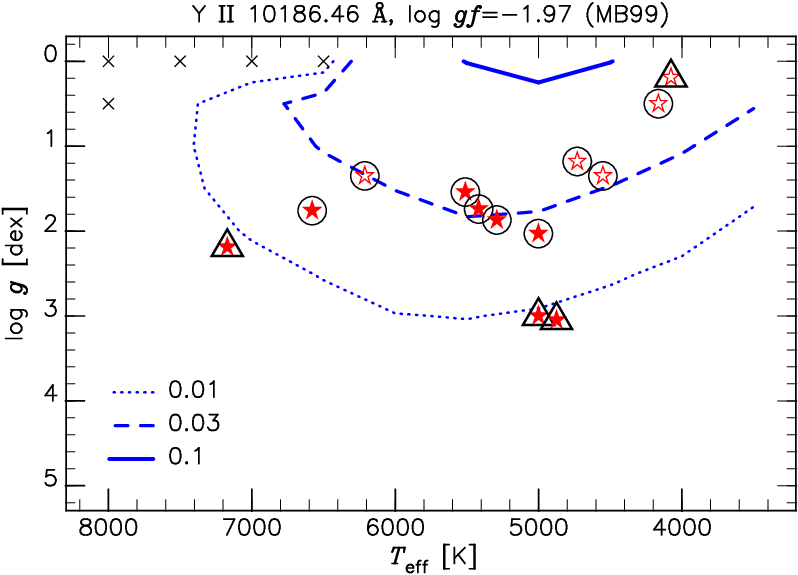} &
\includegraphics[clip,width=0.42\hsize]{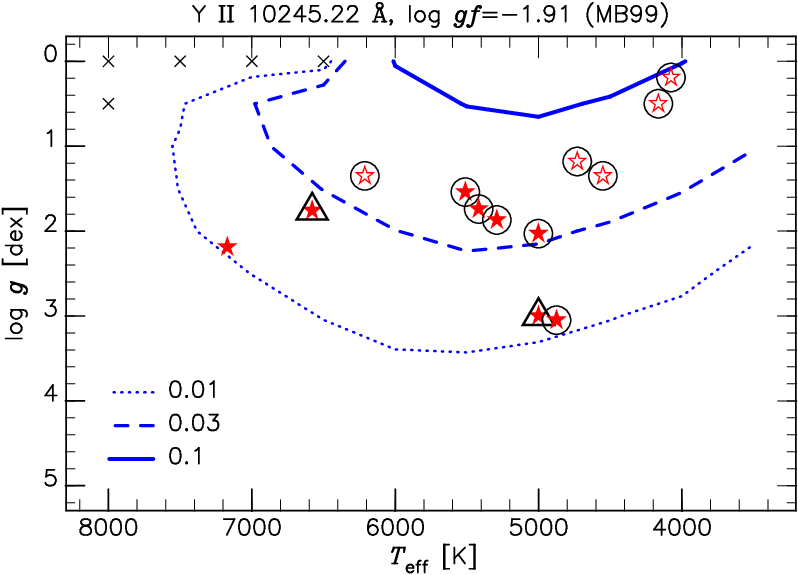} 
\end{tabular}

\newpage

\begin{center}
Here presented are individual plots for Figure~\ref{fig4} 
which are available as the online material. 
\end{center}
\begin{tabular}{cc}
\includegraphics[clip,width=0.42\hsize]{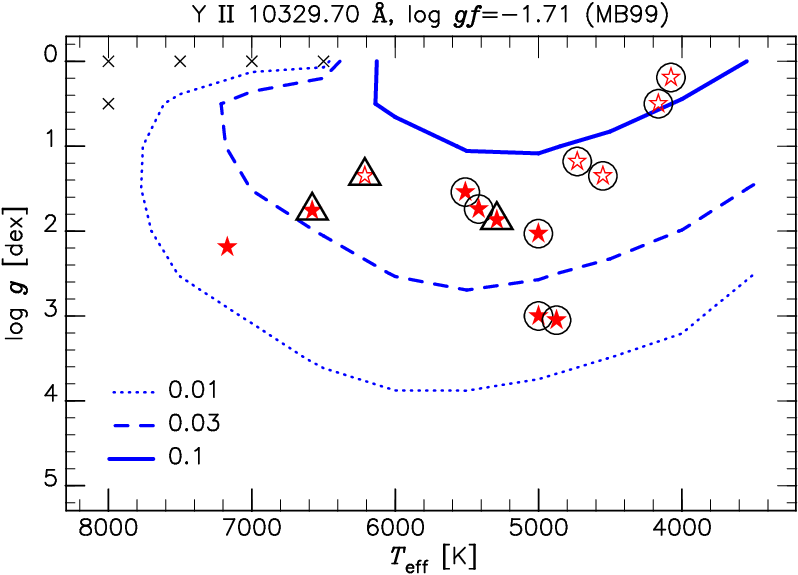} &
\includegraphics[clip,width=0.42\hsize]{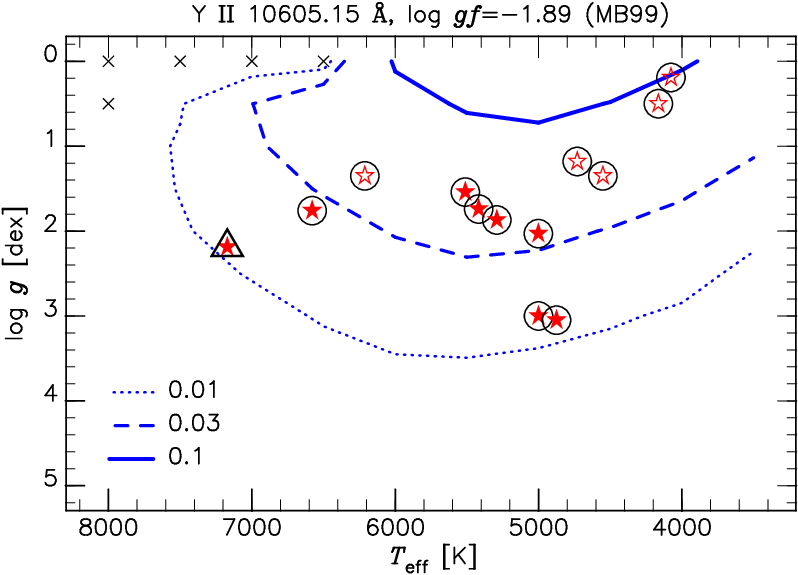} \\
\includegraphics[clip,width=0.42\hsize]{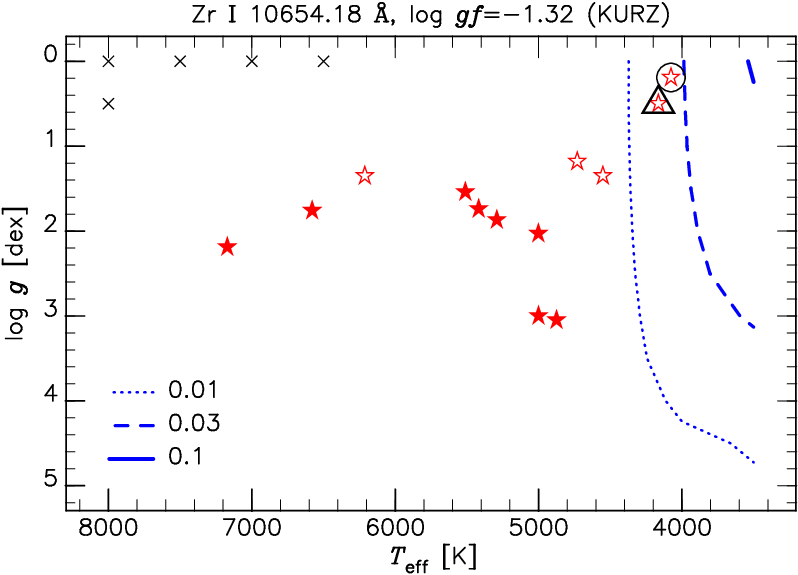} &
\includegraphics[clip,width=0.42\hsize]{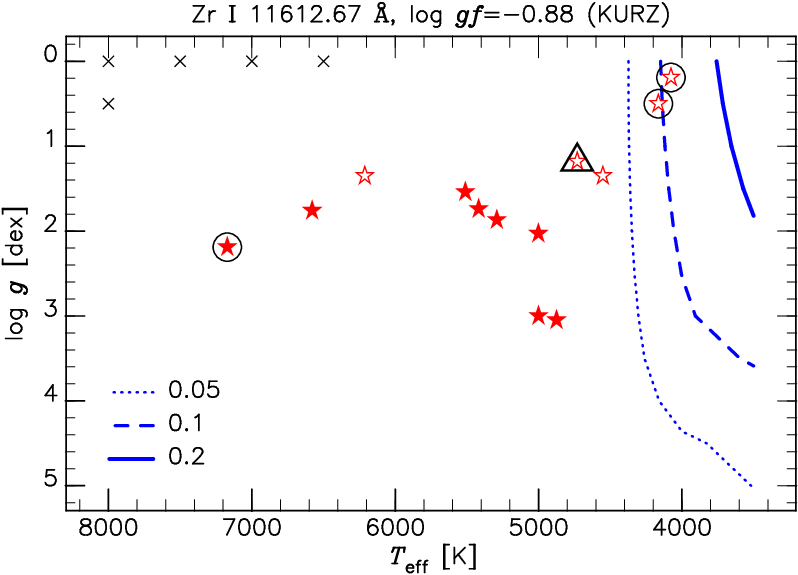} \\
\includegraphics[clip,width=0.42\hsize]{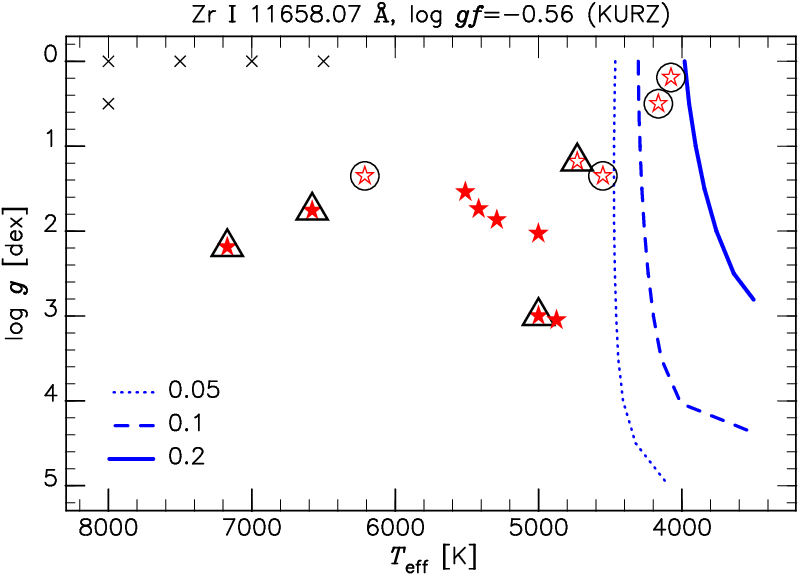} &
\includegraphics[clip,width=0.42\hsize]{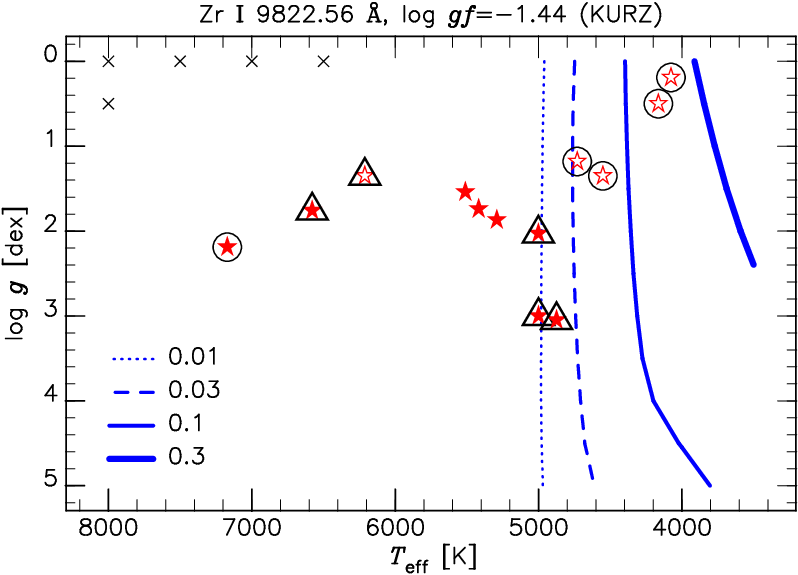} \\
\includegraphics[clip,width=0.42\hsize]{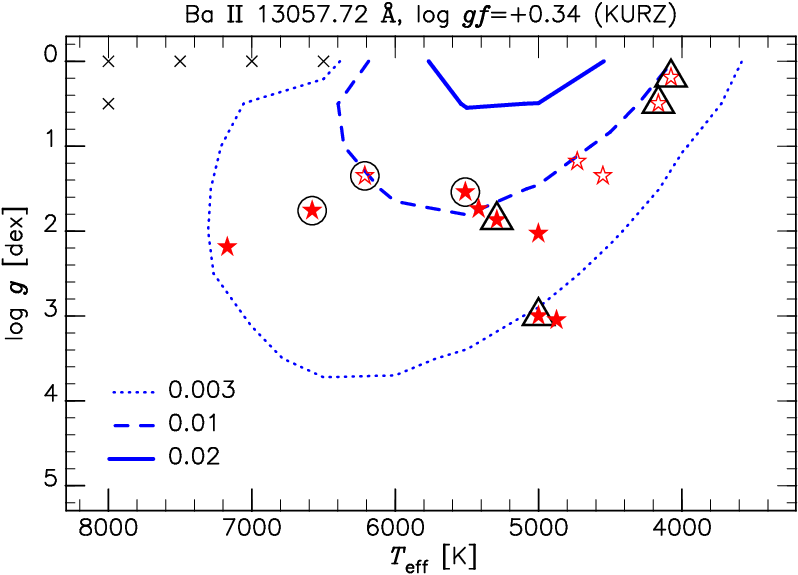} &
\includegraphics[clip,width=0.42\hsize]{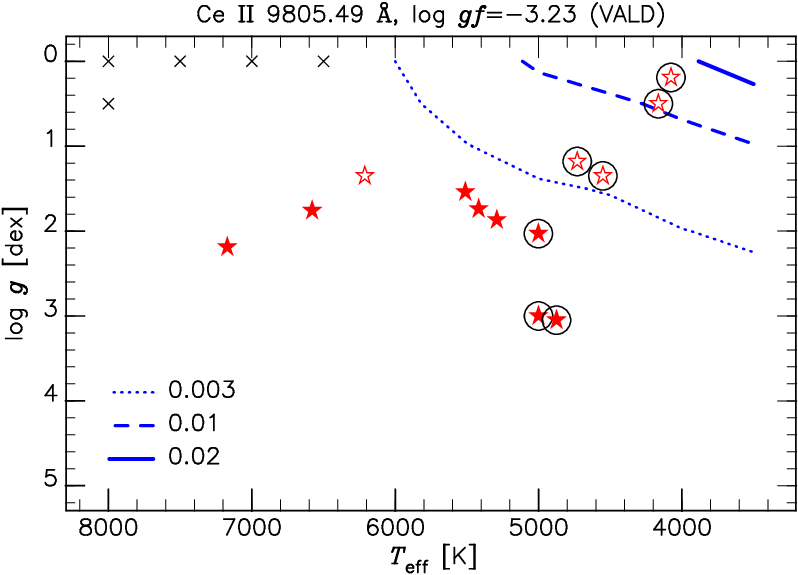} 
\end{tabular}

\newpage

\begin{center}
Here presented are individual plots for Figure~\ref{fig4} 
which are available as the online material. 
\end{center}
\begin{tabular}{cc}
\includegraphics[clip,width=0.42\hsize]{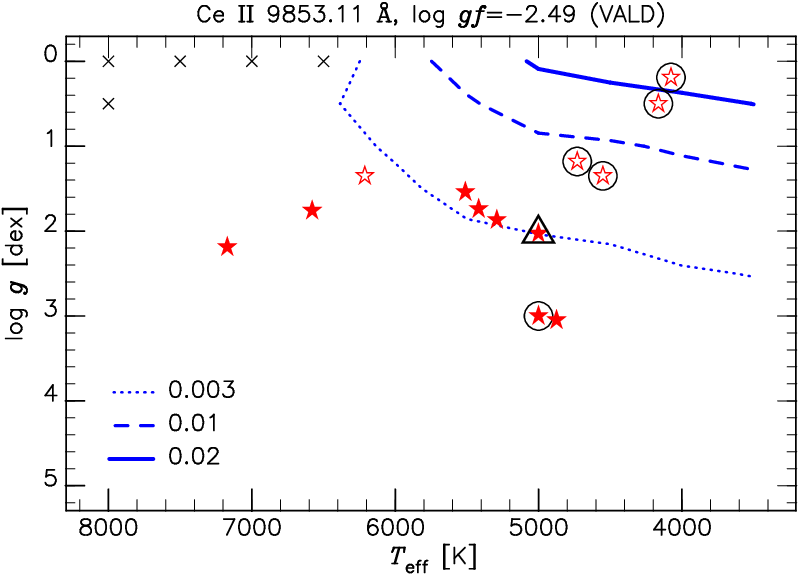} &
\includegraphics[clip,width=0.42\hsize]{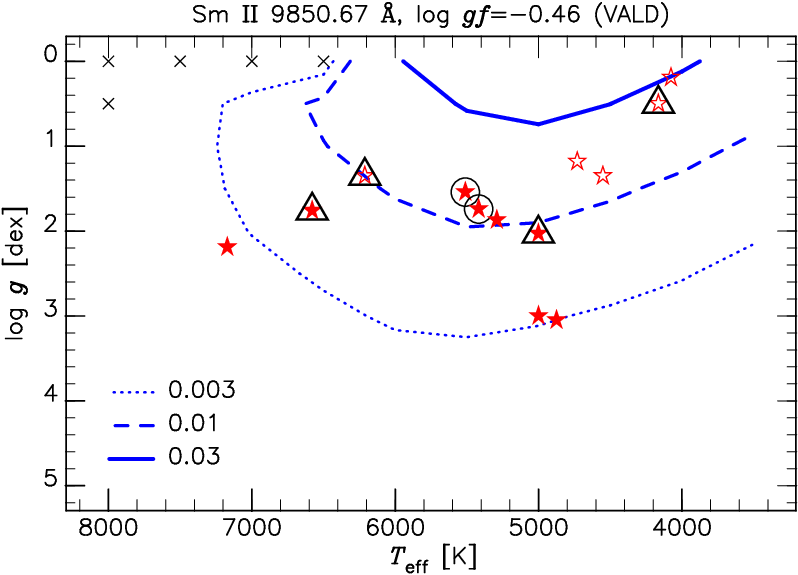} \\
\includegraphics[clip,width=0.42\hsize]{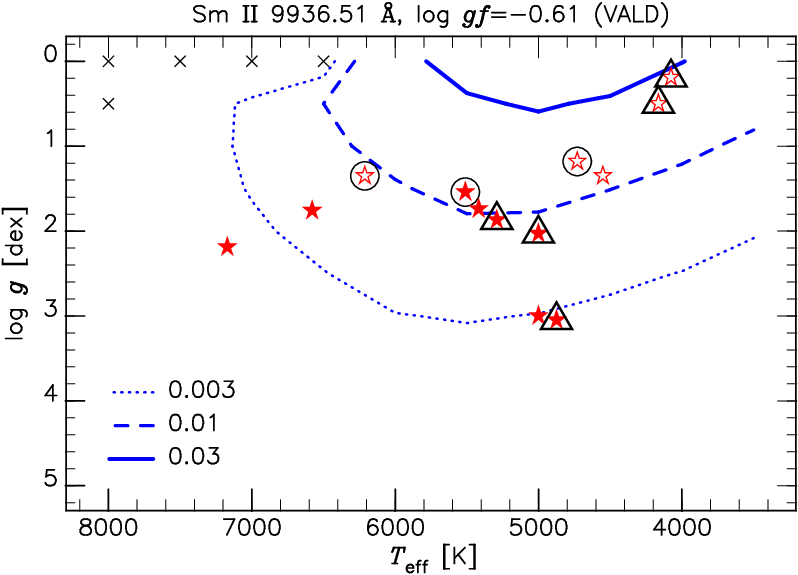} &
\includegraphics[clip,width=0.42\hsize]{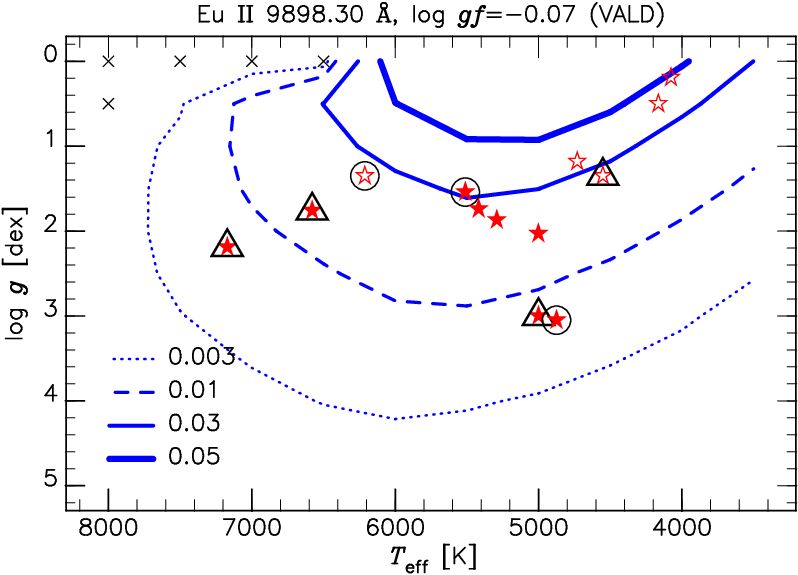} \\
\includegraphics[clip,width=0.42\hsize]{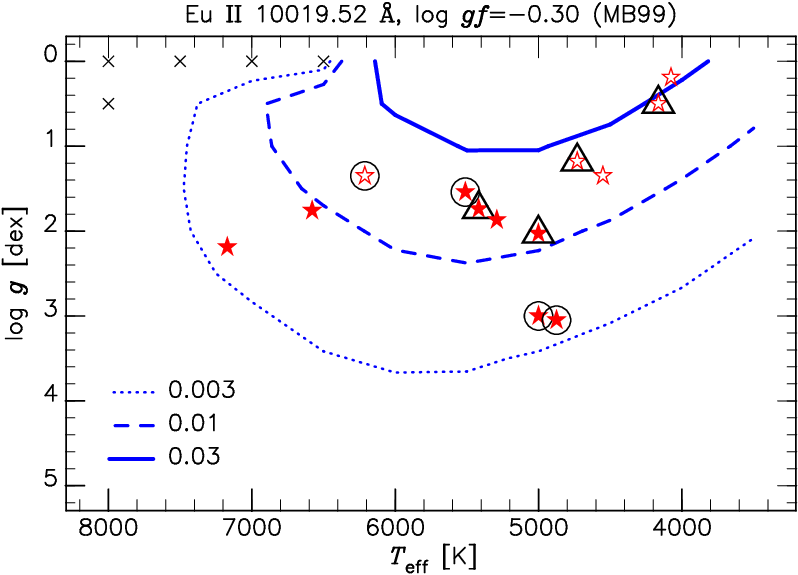} &
\includegraphics[clip,width=0.42\hsize]{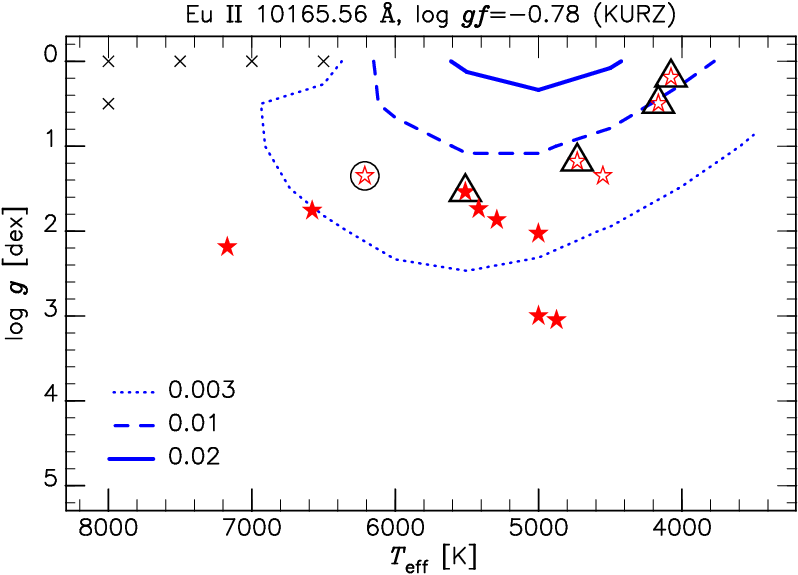} \\
\includegraphics[clip,width=0.42\hsize]{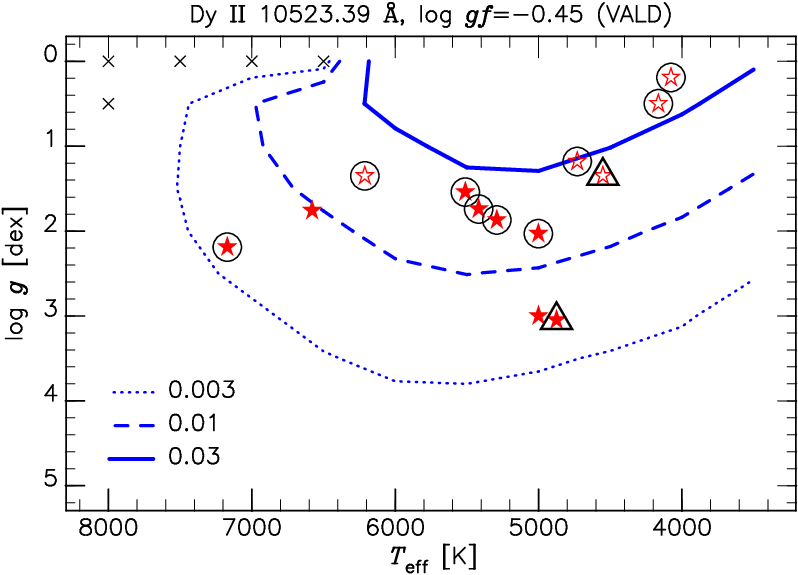} &
\end{tabular}

\newpage

\begin{center}
Here presented are individual plots for Figure~\ref{fig5} 
which are available as the online material. 
\end{center}
\begin{tabular}{c}
\includegraphics[clip,width=0.98\hsize]{fig5a.eps}  \\
\includegraphics[clip,width=0.98\hsize]{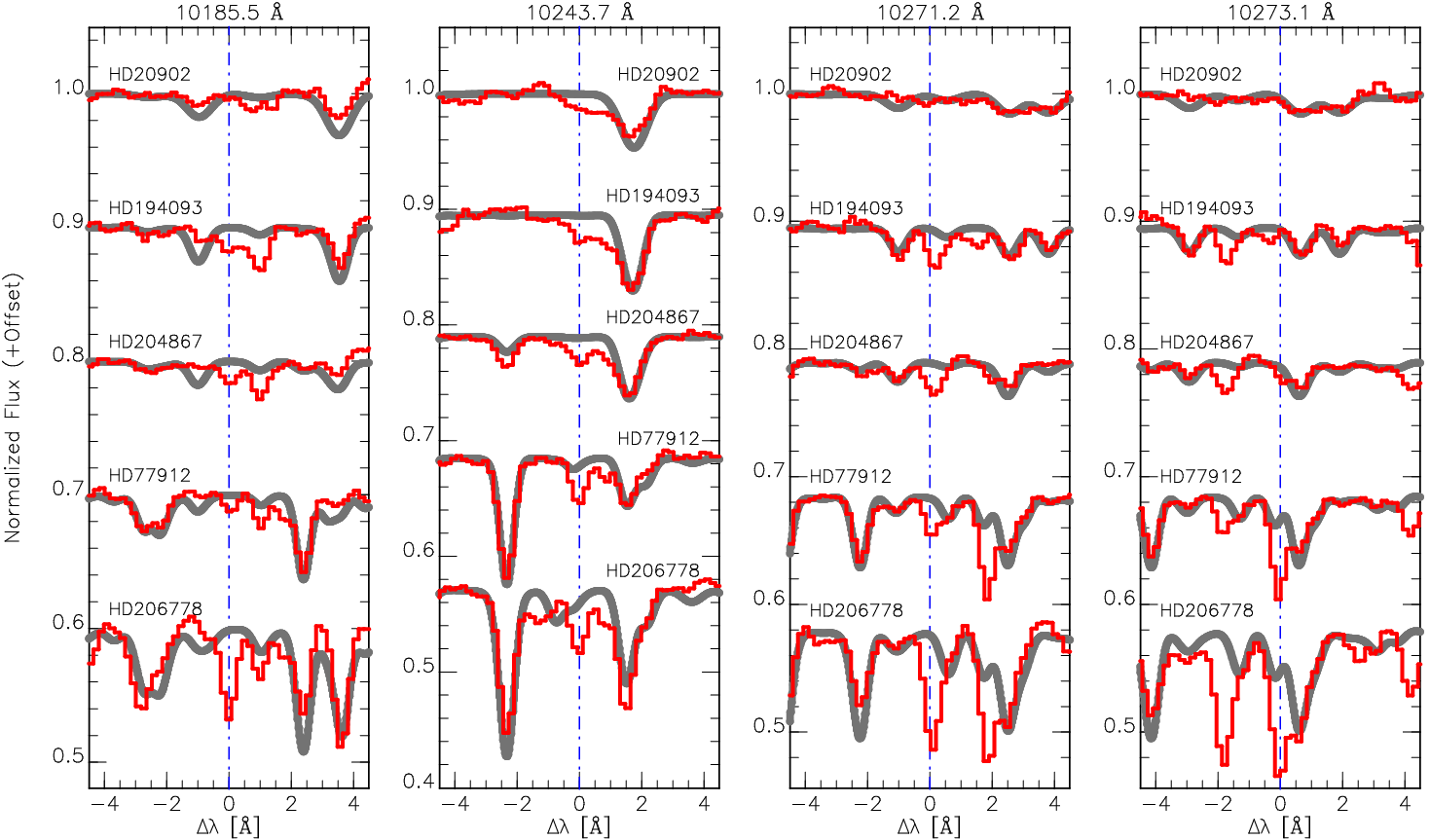} 
\end{tabular}

\newpage

\begin{center}
Here presented are individual plots for Figure~\ref{fig5} 
which are available as the online material. 
\end{center}
\begin{tabular}{c}
\includegraphics[clip,width=0.98\hsize]{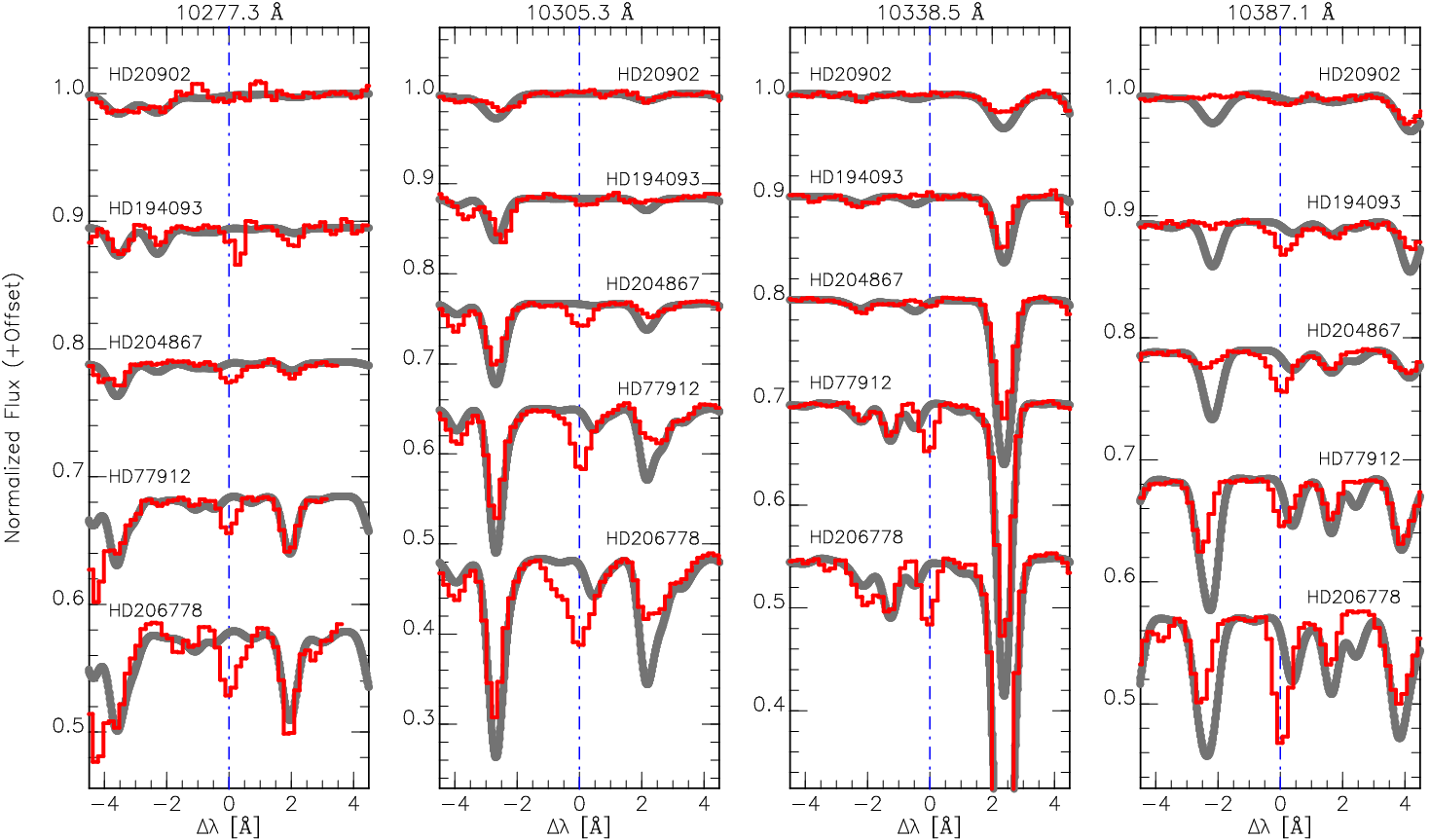}  \\
\includegraphics[clip,width=0.98\hsize]{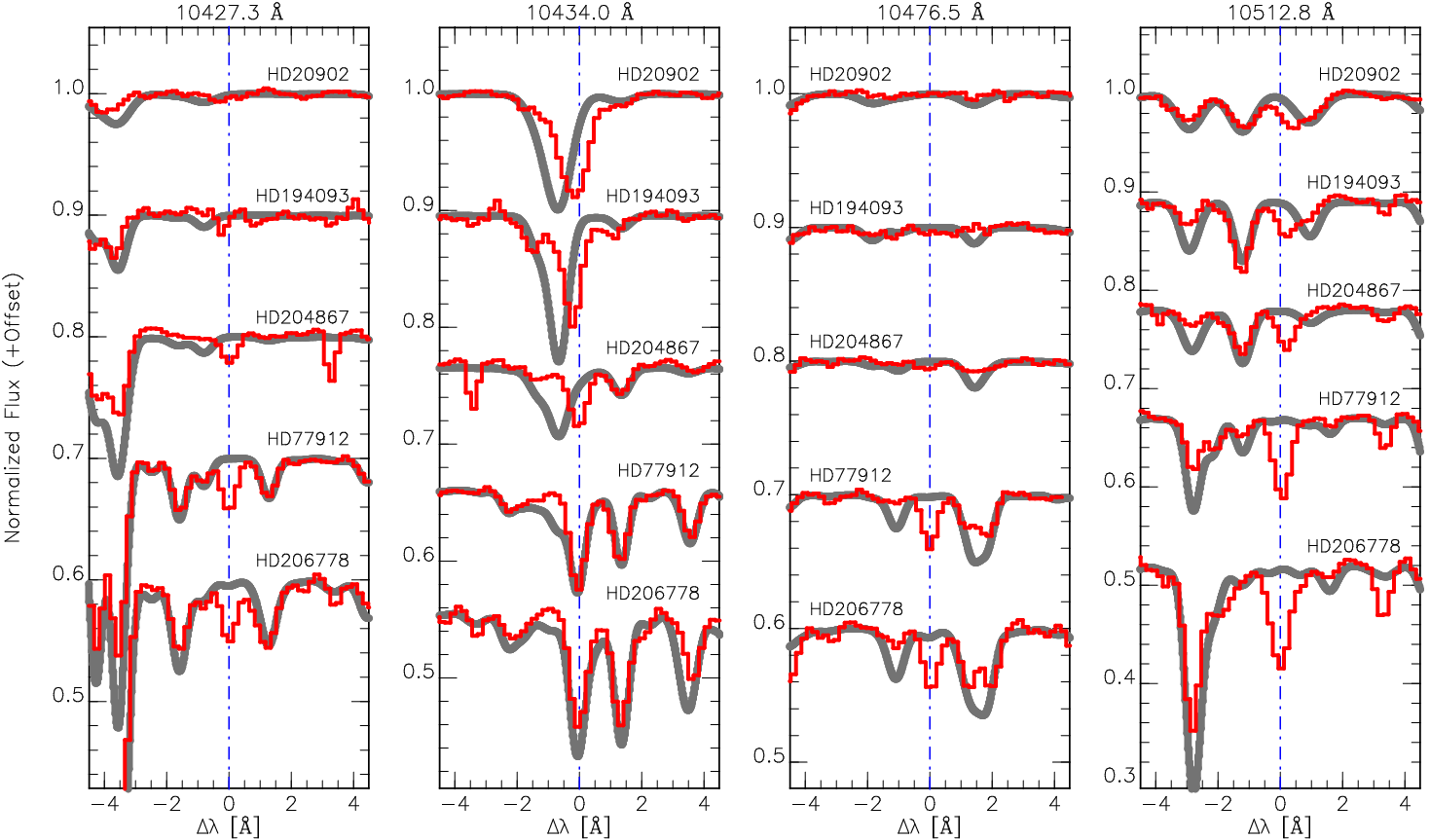} 
\end{tabular}

\newpage

\begin{center}
Here presented are individual plots for Figure~\ref{fig5} 
which are available as the online material. 
\end{center}
\begin{tabular}{c}
\includegraphics[clip,width=0.98\hsize]{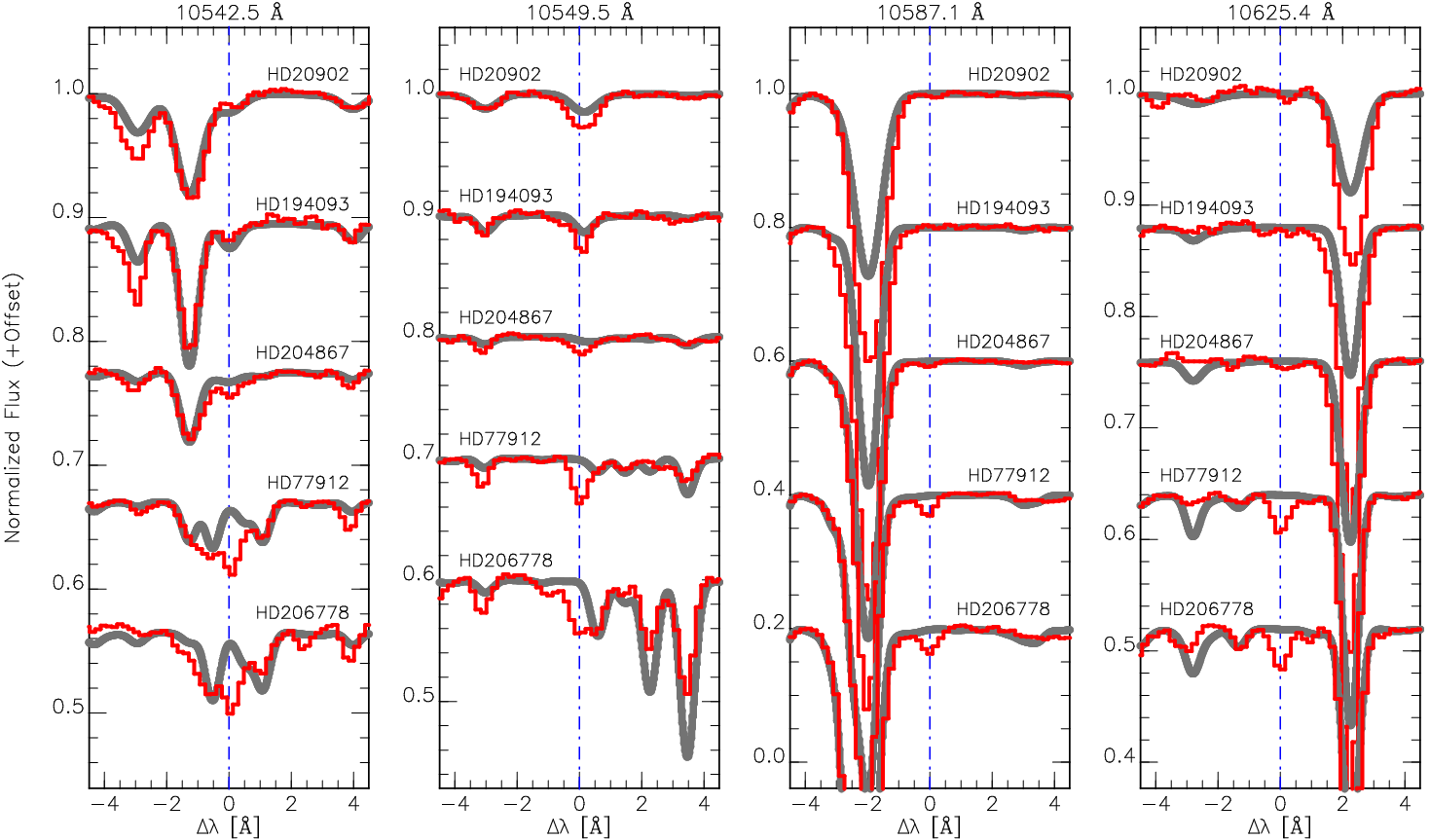}  \\
\includegraphics[clip,width=0.98\hsize]{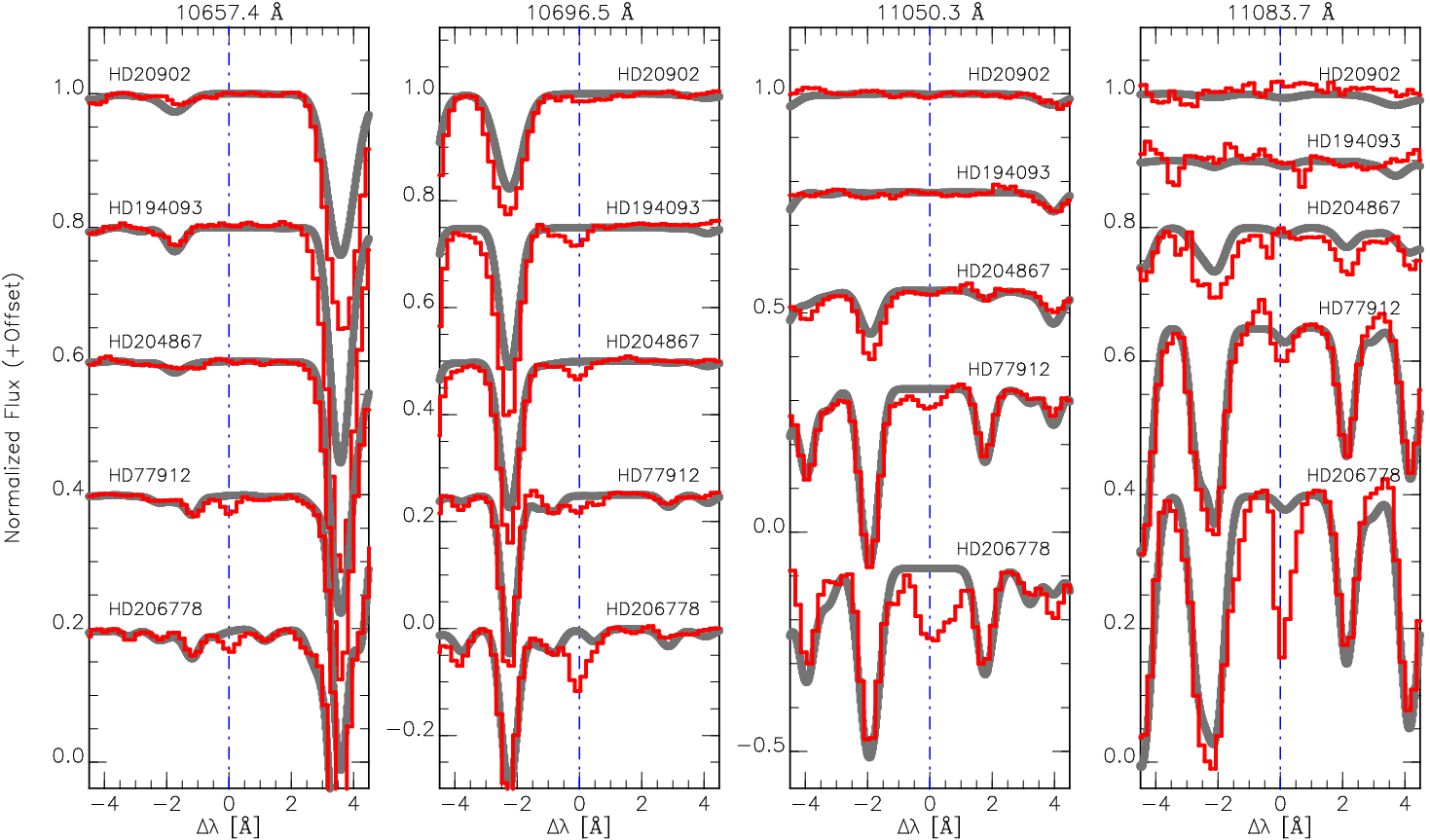} 
\end{tabular}

\newpage

\begin{center}
Here presented are individual plots for Figure~\ref{fig5} 
which are available as the online material. 
\end{center}
\begin{tabular}{c}
\includegraphics[clip,width=0.98\hsize]{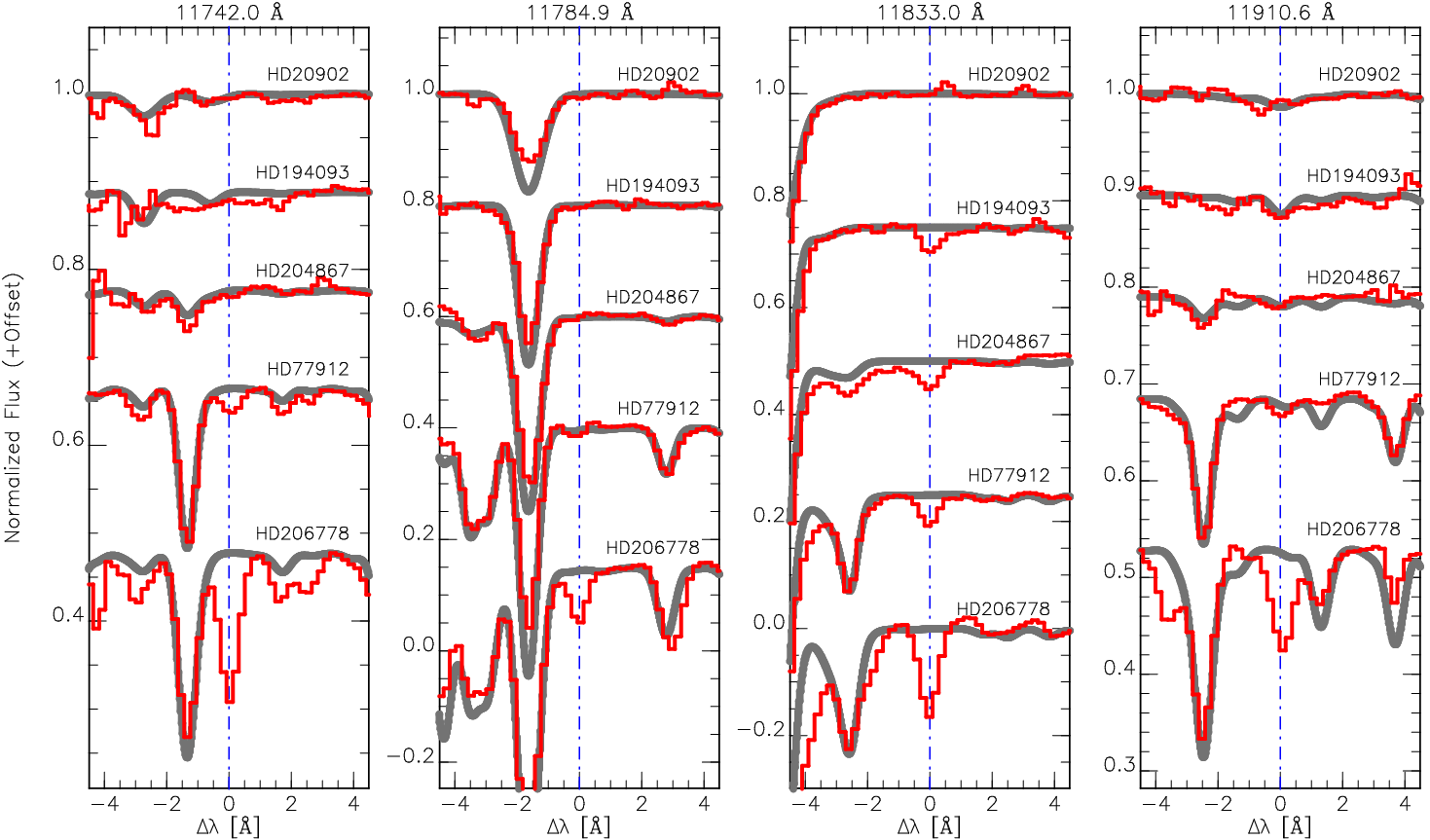}  \\
\includegraphics[clip,width=0.98\hsize]{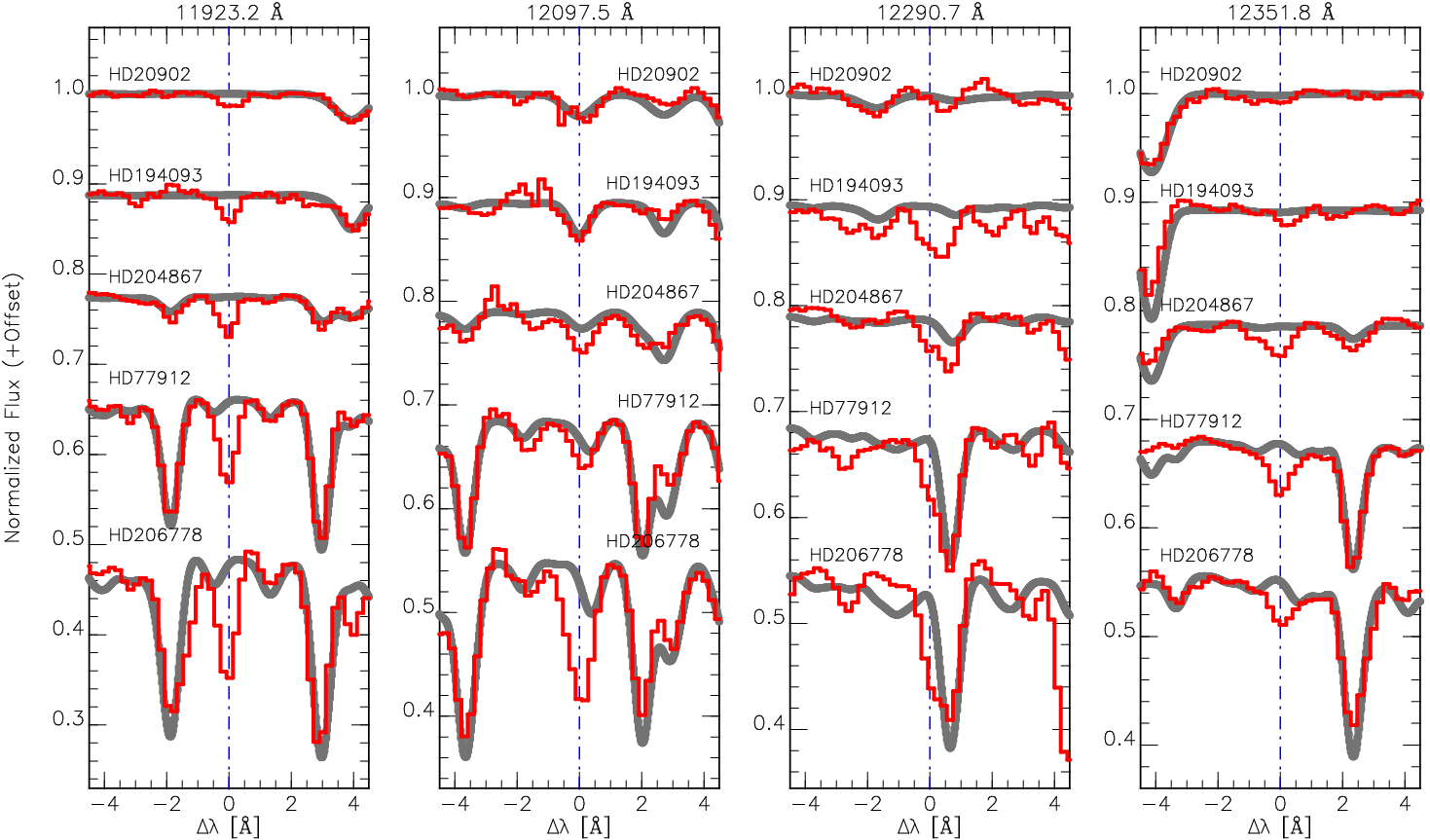} 
\end{tabular}

\newpage

\begin{center}
Here presented are individual plots for Figure~\ref{fig5} 
which are available as the online material. 
\end{center}
\begin{tabular}{c}
\includegraphics[clip,width=0.98\hsize]{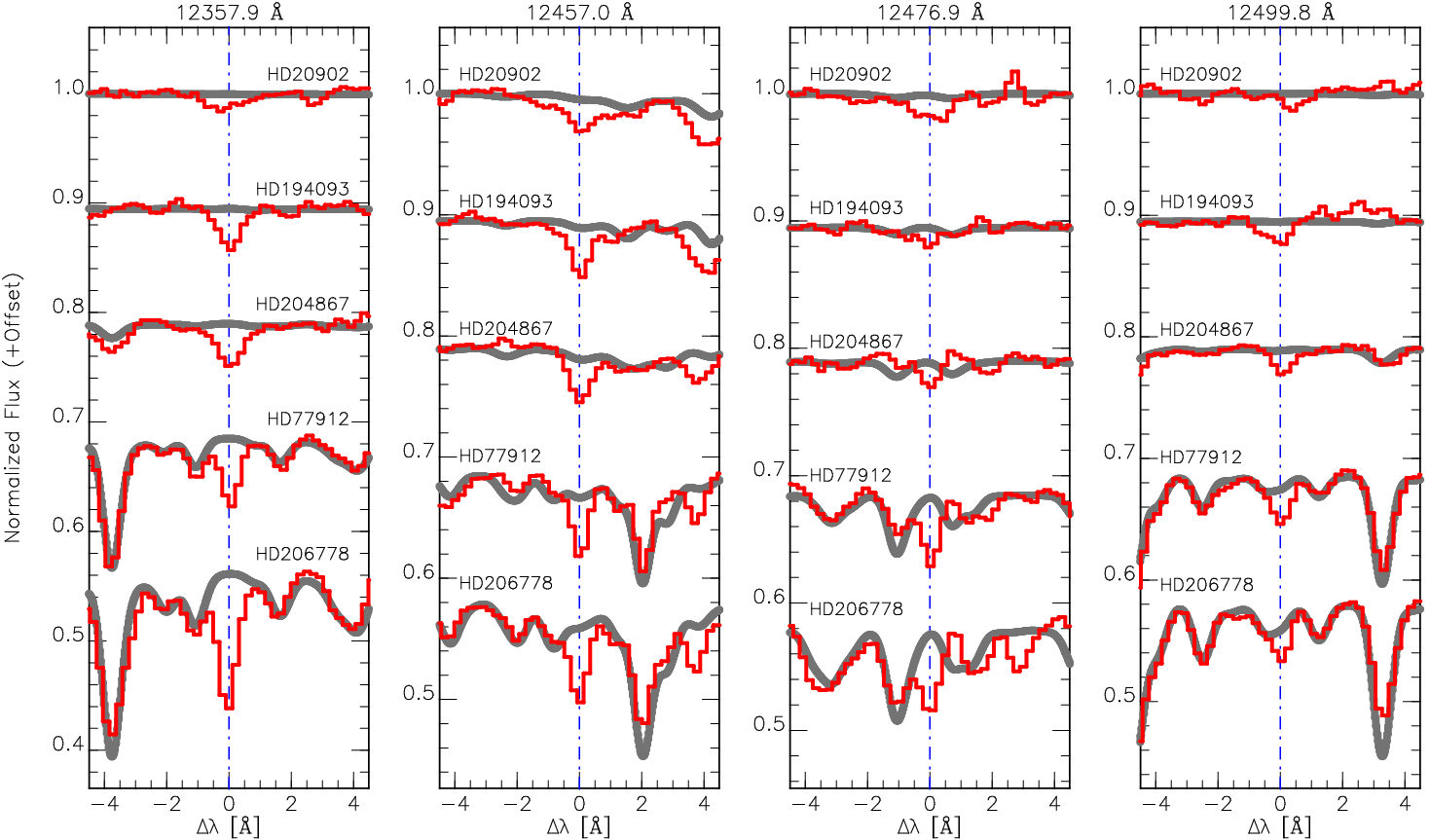}  \\
\includegraphics[clip,width=0.98\hsize]{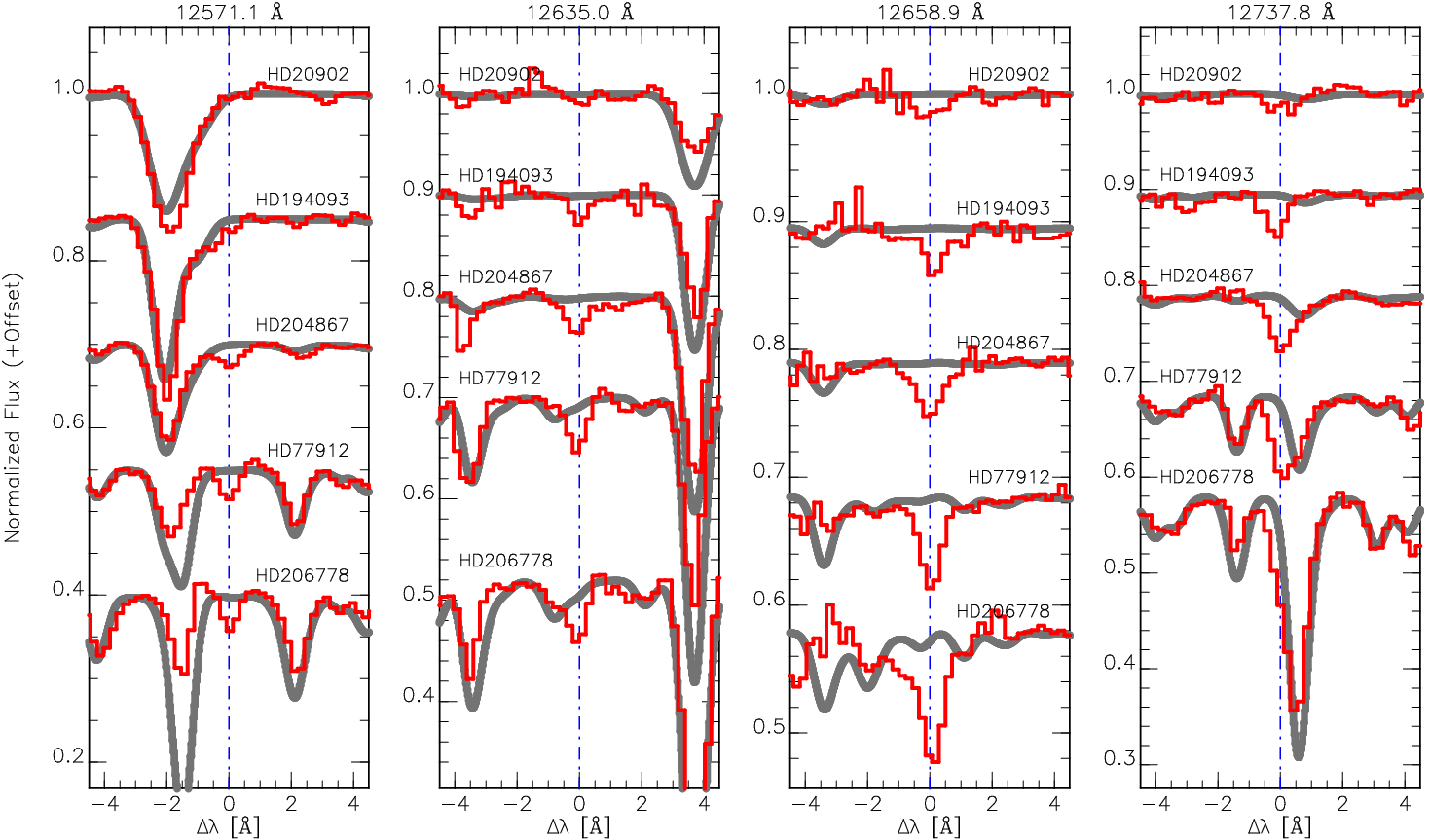}  
\end{tabular}

\newpage

\begin{center}
Here presented are individual plots for Figure~\ref{fig5} 
which are available as the online material. 
\end{center}
\begin{tabular}{c}
\includegraphics[clip,width=0.98\hsize]{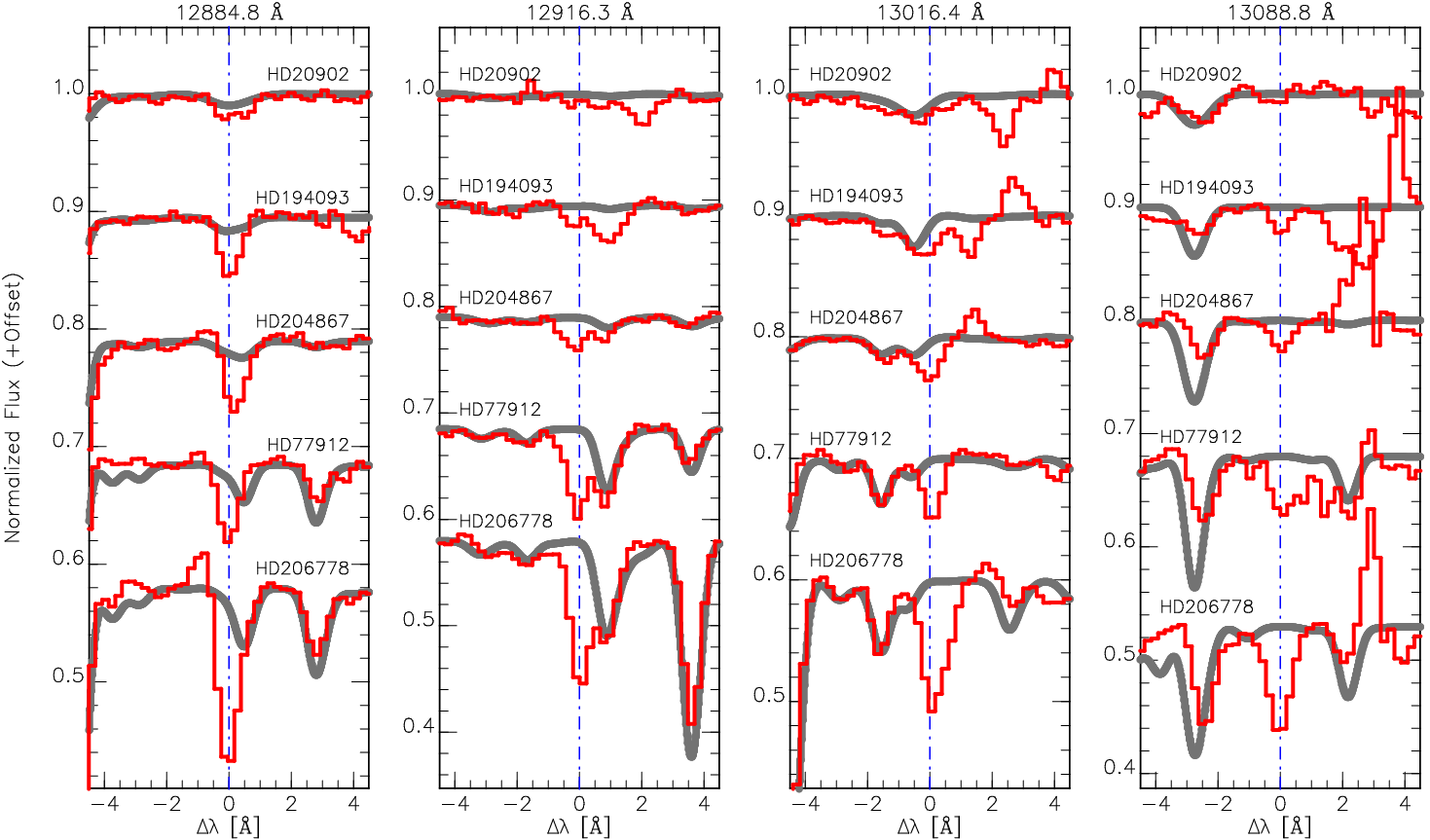}  
\end{tabular}

\end{document}

%% file: table1.tex
\begin{deluxetable*}{cccrrrrccrr}[!tb]
\tablecaption{Observed objects and their parameters
\label{tab1}}
\tablehead{
\multirow{2}{*}{HD} & \multirow{2}{*}{Name} & \multirow{2}{*}{Sp.~Type} &
\colhead{$\Teff$} & \colhead{$\log g$} &
\colhead{[Fe/H]} & \colhead{$\xi$} & \multirow{2}{*}{Ref.} &
\colhead{Obs.~Date} & \colhead{$\vbroad$} & \multicolumn{1}{c}{\multirow{2}{*}{$\gamma$}} \\ 
\colhead{} & \colhead{} & \colhead{} & \colhead{(K)} & 
\colhead{(dex)} & \colhead{(dex)} & \colhead{(\kms)} &
\colhead{} & \colhead{(UT)} & \colhead{(\kms)} & \colhead{}
}
\startdata
 25291 & HR\,1242       & F0\,II & 7171 & 2.19 & $+0.03$ & 1.67 & 1 & 2015.10.28 & 12.65 & 1.103 \\ 
 20902 & $\alpha$~Per   & F5\,Ib & 6579 & 1.76 & $+0.15$ & 3.65 & 1 & 2015.10.26 & 25.61 & 1.694 \\ 
194093 & $\gamma$~Cyg   & F8\,Ib & 6212 & 1.35 & $+0.05$ & 4.02 & 1 & 2015.10.26 & 16.64 & 1.386 \\ 
204867 & $\beta$~Aqr    & G0\,Ib & 5511 & 1.54 & $+0.03$ & 3.39 & 1 & 2015.10.26 & 17.60 & 1.535 \\ 
 26630 & $\mu$~Per      & G0\,Ib & 5418 & 1.74 & $+0.09$ & 3.02 & 1 & 2015.10.28 & 17.93 & 1.362 \\ 
159181 & $\beta$~Dra    & G2\,Ib--IIa & 5291 & 1.87 & $+0.15$ & 2.72 & 1 & 2016.02.03 & 18.10 & 1.198 \\ 
 77912 & HR\,3612       & G7\,IIa & 5001 & 2.03 & $+0.12$ & 2.16 & 1 & 2016.03.21 & 14.27 & 1.012 \\ 
 27697 & $\delta$~Tau   & G9.5\,III & 5000 & 3.00 & $+0.08$ & 1.50 & 2 & 2015.10.25 & 12.31 & 0.957 \\ 
 19787 & $\delta$~Ari   & G9.5\,IIIb & 4875 & 3.05 & $+0.09$ & 1.68 & 2 & 2016.03.11 & 12.07 & 0.917 \\ 
208606 & HR\,8374       & G8\,Ib & 4731 & 1.18 & $+0.25$ & 3.49 & 1 & 2015.10.31 & 17.05 & 0.896 \\ 
  9900 & HR\,461        & K0\,II & 4552 & 1.35 & $+0.19$ & 2.46 & 1 & 2015.10.31 & 13.51 & 0.815 \\ 
206778 & $\epsilon$~Peg & K2\,Ib--II & 4165 & 0.50 & $-0.01$ & 2.96 & 1 & 2015.10.31 & 14.73 & 1.409 \\ 
 52005 & HR\,2615       & K3\,Ib & 4077 & 0.19 & $-0.07$ & 2.76 & 1 & 2015.10.28 & 14.18 & 1.557 \\ 
\enddata
\tablecomments{Parameters of each target are taken from one of the two references, Ref.~1=\citet{Luck-2014} and Ref.~2=\citet{Hekker-2007}. In the last two columns, $\vbroad$ indicates the broadening width including the instrumental profile which corresponds to {$\sim$}10.7\,{\kms}, and $\gamma$ indicates the factor we used to convert the depths (see Section~\ref{sec:measurements}). Spectral types are taken from the SIMBAD database.}
\end{deluxetable*}

%% file: table2.tex
\begin{deluxetable}{ccr@{\extracolsep{1em}}l@{\extracolsep{0pt}}l@{\extracolsep{0pt}}l}[!tb]
\tablecaption{Absorption lines identified in the observed spectra
\label{tab2}}
\tablehead{
\multirow{2}{*}{Species} & \multicolumn{1}{c}{$\airwave$} & \multicolumn{1}{c}{EP} & \multicolumn{3}{c}{$\log gf$ (dex)} \\ 
\cline{4-6} 
\colhead{} & \colhead{(\AA)} & \colhead{(eV)} & \colhead{KURZ} & \colhead{VALD} & \colhead{MB99} 
}
\startdata
\ion{Zn}{1} & 11054.26 & 5.796 & $-0.20$ & $-0.30$ & $-0.50$ \\ 
\ion{Zn}{1} & 13053.60$^{a}$ & 6.655 & $+0.39$ & $+0.34$ & $+0.13$ \\ 
\ion{Sr}{2} & 10036.65 & 1.805 & $-1.22$ & $-1.31$ & $-1.10$ \\ 
\ion{Sr}{2} & 10327.31 & 1.839 & $-0.25$ & $-0.35$ & $-0.40$ \\ 
\ion{Sr}{2} & 10914.89 & 1.805 & $-0.57$ & $-0.64$ & $-0.59$ \\ 
\ion{Y}{2} & 10105.52 & 1.721 & $-1.63$ & $-1.89$ & $-1.89$ \\ 
\ion{Y}{2} & 10186.46 & 1.839 & $-2.65$\tablenotemark{$c$} & $-2.65$\tablenotemark{$c$} & $-1.97$ \\ 
\ion{Y}{2} & 10245.22 & 1.738 & $-1.82$ & $-1.82$ & $-1.91$ \\ 
\ion{Y}{2} & 10329.70 & 1.748 & $-1.51$ & $-1.76$ & $-1.71$ \\ 
\ion{Y}{2} & 10605.15 & 1.738 & $-1.71$ & $-1.96$ & $-1.89$ \\ 
\ion{Zr}{1} & 9822.56 & 0.623 & $-1.44$ & $-1.20$ & \multicolumn{1}{c}{---} \\ 
\ion{Zr}{1} & 10654.18 & 1.582 & $-1.32$ & \multicolumn{1}{c}{---} & \multicolumn{1}{c}{---} \\ 
\ion{Zr}{1} & 11612.67 & 1.366 & $-0.88$ & \multicolumn{1}{c}{---} & \multicolumn{1}{c}{---} \\ 
\ion{Zr}{1} & 11658.07 & 1.396 & $-0.56$ & \multicolumn{1}{c}{---} & \multicolumn{1}{c}{---} \\ 
\ion{Ba}{2} & 13057.72$^{b}$ & 5.251 & $+0.34$\tablenotemark{$c$} & $+0.34$\tablenotemark{$c$} & \multicolumn{1}{c}{---} \\ 
\ion{Ce}{2} & 9805.49 & 0.322 & \multicolumn{1}{c}{---} & $-3.23$\tablenotemark{$c$} & \multicolumn{1}{c}{---} \\ 
\ion{Ce}{2} & 9853.11 & 0.704 & \multicolumn{1}{c}{---} & $-2.49$\tablenotemark{$c$} & \multicolumn{1}{c}{---} \\ 
\ion{Sm}{2} & 9850.67 & 1.971 & \multicolumn{1}{c}{---} & $-0.46$\tablenotemark{$c$} & \multicolumn{1}{c}{---} \\ 
\ion{Sm}{2} & 9936.51 & 1.890 & \multicolumn{1}{c}{---} & $-0.61$ & \multicolumn{1}{c}{---} \\ 
\ion{Eu}{2} & 9898.30 & 2.108 & $-0.76$\tablenotemark{$c$} & $-0.07$ & \multicolumn{1}{c}{---} \\ 
\ion{Eu}{2} & 10019.52 & 2.091 & $-0.66$\tablenotemark{$c$} & $-0.66$\tablenotemark{$c$} & $-0.30$ \\ 
\ion{Eu}{2} & 10165.56 & 2.108 & $-0.78$ & $-0.78$ & \multicolumn{1}{c}{---} \\ 
\ion{Dy}{2} & 10523.39 & 1.946 & \multicolumn{1}{c}{---} & $-0.45$ & \multicolumn{1}{c}{---} \\ 
\enddata
\tablenotetext{a}{The wavelength in the KURZ list, 13053.559\,{\AA}, is slightly different from the counterparts in the VALD, 13053.627\,{\AA}, and the MB99, 13053.64\,{\AA}.} 
\tablenotetext{b}{The wavelength in the KURZ list, 13057.716\,{\AA}, is shorter than the counterpart in the VALD, 13058.015\,{\AA}.} 
\tablenotetext{c}{~The measured depths differ from those predicted with the given $\log gf$ by {$\sim$}0.3~dex or more.}
\end{deluxetable}

%% file: table3.tex
\startlongtable
\begin{deluxetable}{rp{155mm}}
\tablecaption{Absorption lines not predicted by the synthetic spectra
\label{tab3}}
\tablehead{
\colhead{$\airwave$ (\AA)} & \colhead{Comments}
}
\startdata
\multicolumn{2}{c}{$Y$ band} \\
9994.9 &
Blended at $\Teff\lesssim 5000$\,K by a couple of CN lines (the strongest is at 9995.10\,{\AA}), but there is an unknown line which gets strongest at around 5500\,K.
\\
10071.8 &
Stronger towards the lower temperature.
\\
10133.5 &
Stronger towards the lower temperature.
Maybe this is the \ion{Fe}{1} line listed
in both KURZ and VALD
at a different wavelength, 10134.21\,{\AA}, 
that has no counterpart in the observed spectra (see text).
\\
10163.6 &
Synthetic spectra have almost no absorption at around 4500--5000\,K,
but a clear line was detected. \ion{Fe}{2}~10163.606
may be present in some objects, but it is expected
to be significant only at $> 5000$\,K.
\\
10185.5 &
Stronger towards the lower temperature.
\\
10243.7 &
Stronger towards the lower temperature.
\\
10271.2 &
Stronger towards the lower temperature.
\\
10273.1 &
Stronger towards the lower temperature.
There are two lines, CN~10272.93\,{\AA} and \ion{Ca}{1}~10273.68\,{\AA},
blending with this unknown line, but they cannot explain the absorption detected.
\\
10277.3 &
Stronger towards the lower temperature.
\\
10305.3 &
Stronger towards the lower temperature.
\\
10338.5 &
Stronger towards the lower temperature.
\\
10387.1 &
Stronger towards the lower temperature.
\\
10427.3 &
Stronger towards the lower temperature.
\\
10434.0 &
There are a couple of contaminating CN lines dominant at
$\gtrsim 5000$\,K, but this unknown line is stronger towards
the higher temperature.
It is probably \ion{C}{1}~10433.35 listed in KURZ
(but not in VALD) which is not seen in the observed spectra at the given wavelength.
\\
10476.5 &
Synthetic spectra have almost no absorption at $\Teff\gtrsim 4500$\,K,
but there is a clear line whose depth seems to show a peak at {$\sim$}5000\,K. 
This line may be a CN line (see text).
\\
10512.8 &
Stronger towards the lower temperature.
\\
10542.5 &
Synthetic spectra have almost no absorption at $\Teff\gtrsim 4500$\,K,
but there is a clear line whose depth seems to show a peak at 4500--5000\,K. 
\\
10549.5 &
Blended with \ion{N}{1}~10549.64 at $\Teff \gtrsim 6000$\,K
and with \ion{Cr}{1}~10550.095 at $\Teff\lesssim 4500$\,K,
but there is clearly a line,
unexpected in the synthetic spectra,
whose depth seems to show a peak at the intermediate $\Teff$ range.
\\
10587.1 &
Strongest at around 4500--5000\,K. 
\\
10625.4 &
Stronger towards the lower temperature.
Synthetic spectra have almost no absorption at this wavelength
at the entire $\Teff$ range.
The observed absorption resembles \ion{Fe}{1}~10622.592 
in the synthetic spectra although the latter is not confirmed
in the observed spectra.
\\
10657.4 &
Stronger towards the lower temperature.
\\
10696.5 &
Observed spectra seem to show two lines not visible
in the synthetic spectra.
One of them gets stronger towards the lower temperature,
while the other is seen at around 6000\,K.
\\
11050.3 &
Stronger towards the lower temperature and seen at only $\Teff \lesssim 5000$\,K.
\\
11083.7 &
Stronger towards the lower temperature and seen at only $\Teff \lesssim 5000$\,K.
\\
\tableline
\multicolumn{2}{c}{$J$ band} \\
11742.0 &
Stronger towards the lower temperature and seen at only $\Teff<5000$\,K.
\\
11784.9 &
Stronger towards the lower temperature.
\\
11833.0 &
Stronger towards the lower temperature.
\\
11910.6 &
Stronger towards the lower temperature and seen at
only $\Teff \lesssim 5000$\,K. At the higher $\Teff$,
in contrast, a weak line of \ion{C}{1} is visible at 11910.62\,{\AA}.
\\
11923.2 &
Stronger towards the lower temperature.
\\
12097.5 &
Stronger towards the lower temperature and seen at
only $\Teff\lesssim 5000$\,K. At the higher $\Teff$,
in contrast, a weak line of \ion{C}{1} is visible at 12097.49\,{\AA}.
\\
12290.7 &
Blended with CN~12291.332, but clearly present
in a wide range of $\Teff$ and stronger towards the lower temperature.
\\
12351.8 &
Strongest at around the intermediate temperatures, {$\sim$}5000\,K.
\\
12357.9 &
Stronger towards the lower temperature.
\\
12457.0 &
Seems to be stronger towards the lower temperature,
but the $\Teff$ dependency is weak.
\\
12476.9 &
Strongest at around 5000\,K.
\\
12499.8 &
Blended with a couple of CN lines at $\Teff \lesssim 5000$\,K,
but there is a line showing a different trend with a peak at {$\sim$}5000\,K.
\\
12571.1 &
The $\Teff$ trend seems flat for a wide $\Teff$ range, 4000--5500\,K.
\\
12635.0 &
Strongest at around the intermediate temperatures, {$\sim$}5000\,K.
\\
12658.9 &
The $\Teff$ trend seems consistent with that of 
the combination of two lines of \ion{Ni}{1} at 12655.38 and 12655.60\,{\AA}
listed in both KURZ and VALD (see text).
\\
12737.8 &
Blended with \ion{Ti}{1}~12738.383 together with other weaker lines,
but there is a clear line, at the shoulder of the \ion{Ti}{1} line,
which is unexpected in the synthetic spectra.
\\
12884.8 &
Stronger towards the lower temperature.
\\
12916.3 &
Stronger towards the lower temperature.
\\
13016.4 &
Stronger towards the lower temperature.
\\
13088.8 &
There are strong telluric absorption lines around this wavelength,
but there is a reasonably strong line which shows a well-behaved trend
with the temperature, getting stronger towards the lower temperature.
\\
\enddata
\end{deluxetable}